\documentclass[11pt]{article}
\usepackage{old-arrows}
\usepackage[scaled=.9]{helvet}
\usepackage{charter}
\usepackage{XoohmG}

\newcommand{\tW}{\mkern1mu^\wtd\mkern-4mu}
\newcommand{\FF}[2][n]{F^{\sss(#1)}_{#2}}
\newcommand{\XX}[2][n-1]{X^{\sss(#1)}_{#2}}
\newcommand{\pDN}[2][]{{\pD^{\mkern-3mu\raisemath{1pt}{#1}}%
                       _{\mkern-1mu\smash{#2}}}}
\newcommand{\pDs}[2][]{\pDN[\star#1]{#2}}
\def\zZ#1#2{\big(\ZZ_{#1}{:}\,{#2}\big)}
\long\def\oMit#1{}
\def\vev#1{\langle#1\rangle}
\def\Vev#1{\left\langle#1\right\rangle}

 \SfTitles
 \allowdisplaybreaks
 
 \omitBackreferences

\begin{document}

\thispagestyle{empty}
\setcounter{page}{0}
\renewcommand{\thefootnote}{\fnsymbol{footnote}}
\vglue5mm
\begin{center}
{\LARGE\sf\bfseries\boldmath
  Ricci-Flat Mirror Hypersurfaces\\[4pt]
  in Spaces of General Type%
  \footnote{\,Based on the invited talk presented at the
  {\em 11th Mathematical Physics Meeting,} Sept.\;2--6, 2024, Belgrade, Serbia}
}
\vspace{2mm}

{\sf\bfseries Tristan H\"{u}bsch}\\*[1mm]
{\small\it
      Department of Physics \&\ Astronomy,
      Howard University, Washington, DC 20059, USA}\\[-1mm]
{\tt  thubsch@howard.edu}
\vspace{2mm}

{\sf\bfseries ABSTRACT}\\[3mm]
\parbox{152mm}{\addtolength{\baselineskip}{-3pt}\parindent=2pc\noindent
Complex Ricci-flat (i.e., Calabi--Yau) hypersurfaces in spaces admitting a maximal (toric) $U(1)^n$ gauge symmetry of general type (encoded by certain non-convex and multi-layered multitopes) may degenerate, but can be smoothed by rational (Laurent) anticanonical sections. Nevertheless, the phases of the Gauged Linear Sigma Model and an increasing number of their classical and quantum data are just as computable as for their siblings encoded by reflexive polytopes, and they all have transposition mirror models. Showcasing such hypersurfaces in so-called Hirzebruch scrolls shows this class of constructions to be {\em\/infinitely vast,\/} yet amenable to standard and well-founded algebro-geometric methods of analysis.
\begin{flushright}\footnotesize\sl
``Science fascinates by reaping dividends of conjecture\\*[-1pt]
from such trifling investment as facts.''\\*[-1pt]
--- Samuel Clemens [paraphrase]
\end{flushright}
}
\begin{minipage}{.75\hsize}\small
  \baselineskip=10.5pt plus1pt minus 1pt
  \renewcommand{\cftbeforetoctitleskip}{\smallskipamount}
  \renewcommand{\cftaftertoctitle}{\vskip-3pt}
  \tocloftpagestyle{empty}
  \setcounter{tocdepth}{3} 
  \tableofcontents
\end{minipage}
\end{center}

\setcounter{footnote}{0}
\renewcommand{\thefootnote}{\arabic{footnote}}
\section{Introduction, Roadmap and Summary}
\label{s:IRS}
Algebraic methods of solving the century-old Einstein equations are themselves centenarian, and have in the ``sourceless'' ($T_{\m\n}\<= 0$, vacuum) case aptly reconstructed the first exact and best known solution, the Schwarz\-schild geometry, but have also revealed its 2-sheeted nature\cite{Kasner:1921Fin, rCF-BH}. Herein, we consider this time-tested approach applied to the small and compact spatial factor of spacetime, as required in ``Calabi--Yau compactification'' in string theory\cite{rCHSW, rBeast}. 
 The extended nature of strings allows them to propagate consistently even through singular and otherwise defective spacetimes, which enables transitions that change even spacetime topology\cite{rReidK0, rCYCI1, rGHC, rGHPT, rCGH1, rCGH2, rBeast, rSingS, Avram:1997rs, Avram:1995pu, Kreuzer:2000xy, rKS-CY, McNamara:2019rup}.
 In the last three decades, model construction and analysis focuses on {\em\/gauged linear sigma models\/} (GLSMs)\cite{rPhases, rPhasesMF, rMP0, Distler:1993mk, rMP1, Schafer-Nameki:2016cfr, Sharpe:2024dcd}, the underlying (complex algebraic) toric geometry of which is governed by the supersymmetry-complexified gauge group,
 $U(1;\IC)^n=(\IC^*)^n$, governs\cite{rKKMS-TE1, rD-TV, rO-TV, rF-TV, rGE-CCAG, rCLS-TV, rCK}.\footnote{The gauge symmetry,
 $U(1)^n\<\to U(1;\IC)^n\<= (\IC^*)^n$
 complexified by worldsheet supersymmetry, may admit some mild degeneration or boundary sources (target-space 'branes),  relaxing further the underlying framework. In turn, the worldsheet quantum field theory itself has a manifestly $T$-dual and mirror-symmetric formulation, insuring a foundation to rely on\cite{Freidel:2015pka, Berglund:2021hbo}.}
 This has produced the largest known database of {\em\/reflexive\/} polytopes, each encoding (semi) Fano ($c_1{\geqslant}0$) {\em\/complex algebraic\/} toric 4-folds and their families of anticanonical Calabi--Yau 3-fold hypersurfaces\cite{Kreuzer:2000xy}.
 This vast pool of constructions exhibits {\em\/mirror symmetry,} whereby the
 $(\IC^*)^n$-equivariant quantum cohomology rings of mirror-pairs of Calabi--Yau hypersurfaces, $(Z_f,Z_g)$, satisfy
 $H^q(Z_f,\wedge^p\,T^*)\approx H^q(Z_g,\wedge^p\,T)$%
\cite{rGP1, rBH, rBaty01, rBatyBor1, Kontsevich:1994Mir, rBH-LGO+EG, rBatyBor3, rCOK, Kontsevich:1995wk, rLB-VA+MM, rBatyBor2, rT-TransMM, rPC-DuaLGO, Krawitz:2009aa, rC+R-MirrBH, rC+R-LG|CY, Ebeling_2011, rLB-MirrBH, rEGZ-OrbEu-BHH, rABS-BHK3, rS-BirBHK, rK-BHK, rPC-birBHK, rA+P-BHK, rACG-BHK, rABS-BHCY3, rEGZ-Saito-BHK, rET-StrangeDisBHK, rF+K-BHK, rDKSS-altMM, Basalaev:2017aa, Recknagel:2017bsx, Comparin:2021sas, rEGZ-BHHT, Bott:2019lmb, Belavin:2020qyl, Belavin:2020xhs, Ebeling:2021Str, Parkhomenko:2022kju, Clawson:2023cmp, Belavin:2024mmd, Parkhomenko:2024mxq}.

\begin{figure}[htb]
 \centering
 \TikZ{[scale=1.33]\path[use as bounding box](-1,-.2)--(11,3.2);
       \path(0,1.45)node{\includegraphics[height=30mm]{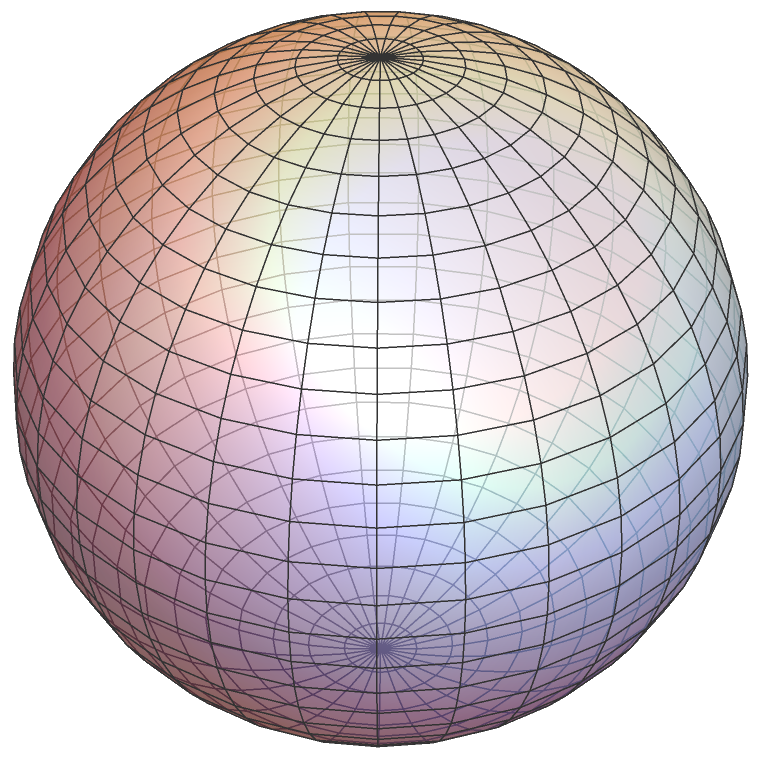}};
        \path(0,2.9)node{\large$\IP^1=S^2$}; \path(0,0)node{\small$g=0$};
        \fill[red](-.25,1.3)circle(.5mm); \fill[red](.25,1.5)circle(.5mm);
       \path(2.7,1.5)node{\includegraphics[height=30mm]{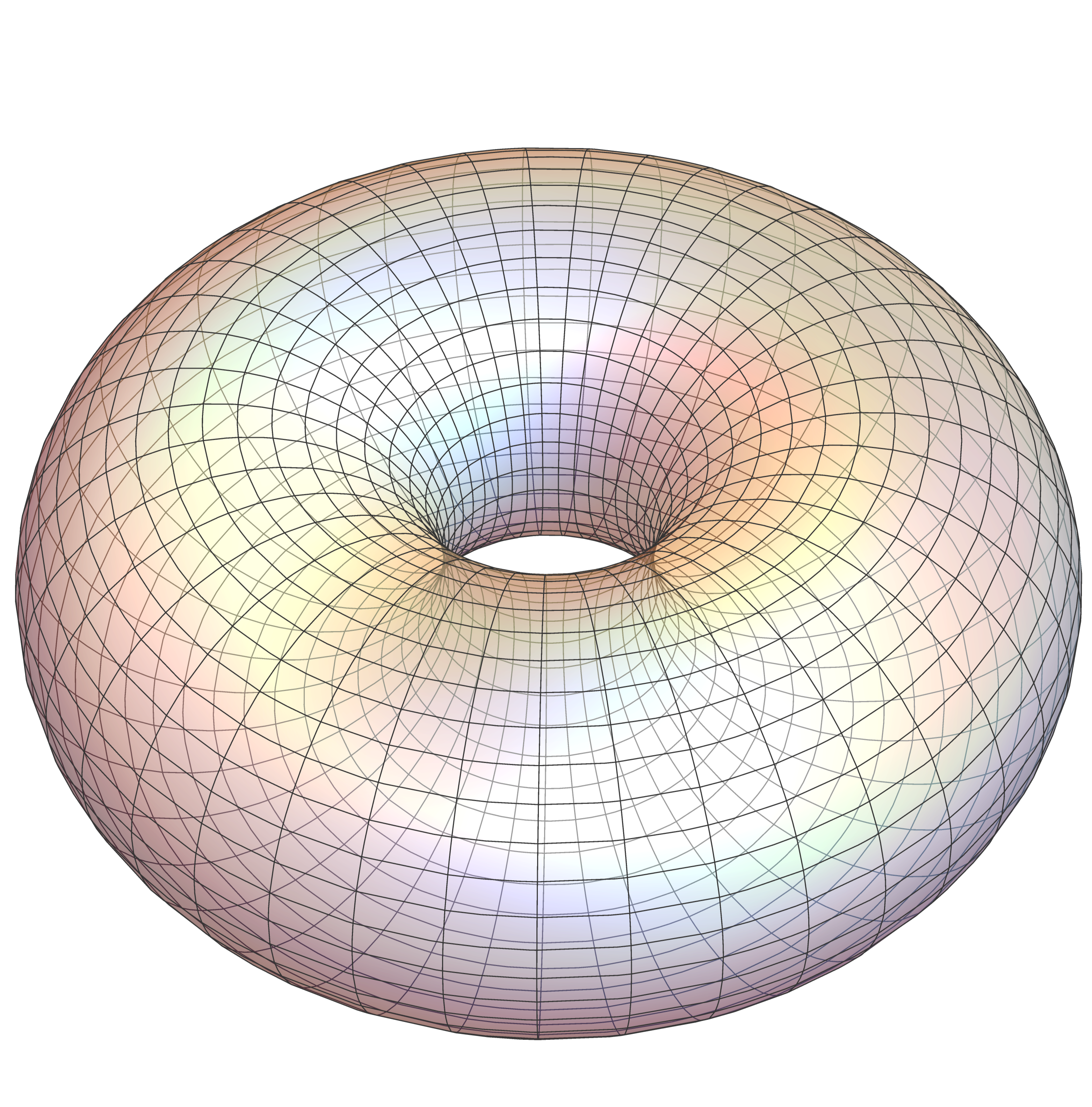}};
        \path(2.7,2.7)node{\large$T^2=S^1{\times}S^1$};
        \path(2.8,0)node{\small$g=1$};
        \fill[red](2.7,.9)circle(.5mm); \fill[red](3.2,1.1)circle(.5mm);
       \path(5,1.6)node{\includegraphics[height=30mm]{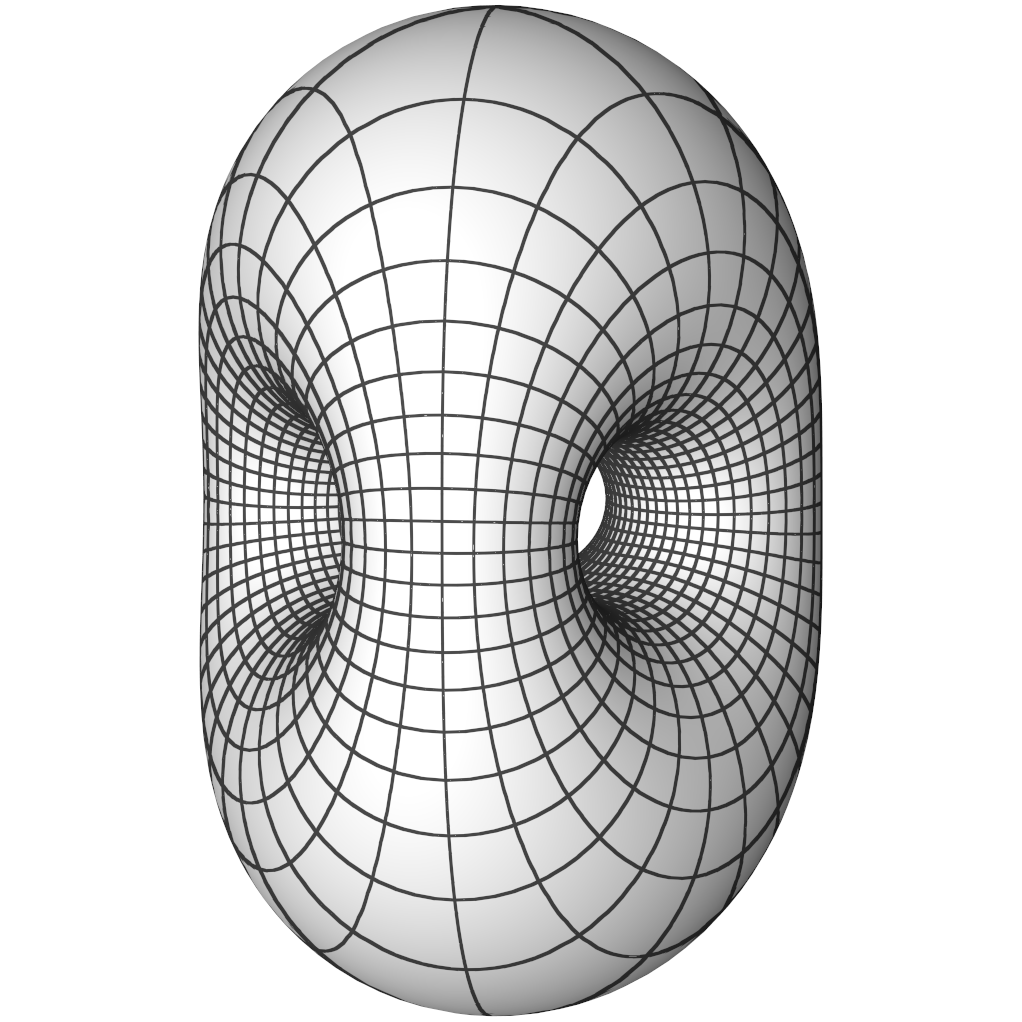}};
        \fill[red](4.65,.8)circle(.5mm); \fill[red](5.15,1)circle(.5mm);
        \path(5,0)node{\small$g=2$};
       \path(6.9,1.6)node{\includegraphics[height=30mm]{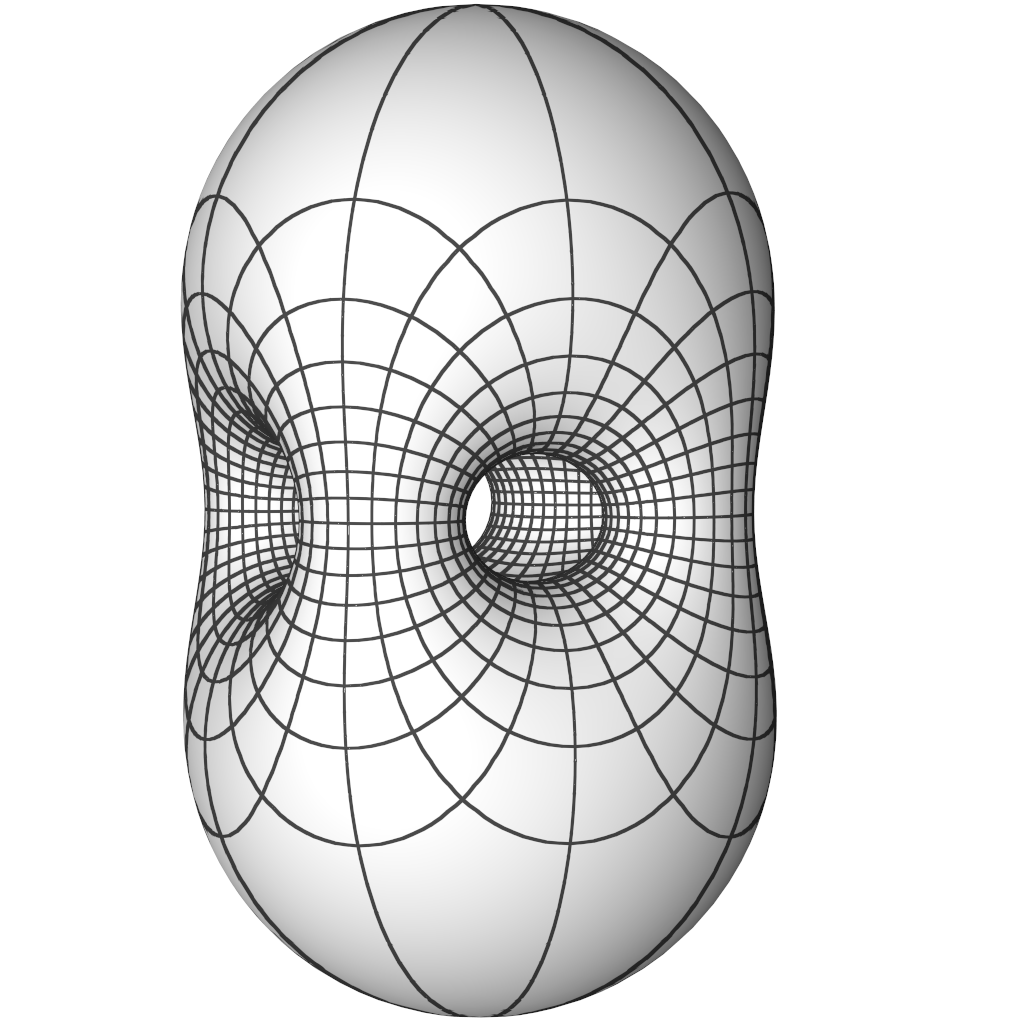}};
        \fill[red](6.6,.8)circle(.5mm); \fill[red](7.1,1)circle(.5mm);
       \path(6.9,0)node{\small$g=3$};
       \path(8.6,1.6)node{\includegraphics[height=30mm]{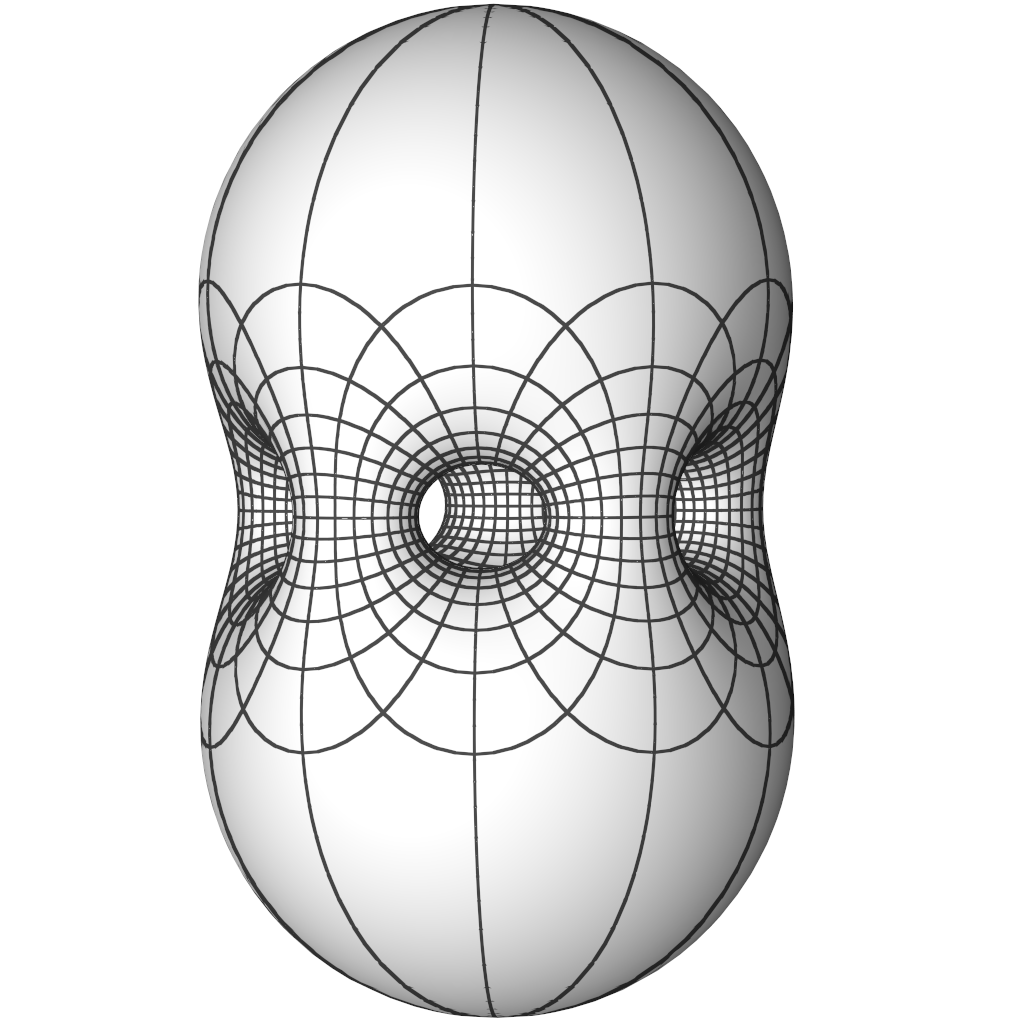}};
        \fill[red](8.3,.8)circle(.5mm); \fill[red](8.8,1)circle(.5mm);
        \path(8.6,0)node{\small$g=4$};
       \path(10.4,1.6)node{\includegraphics[height=30mm]{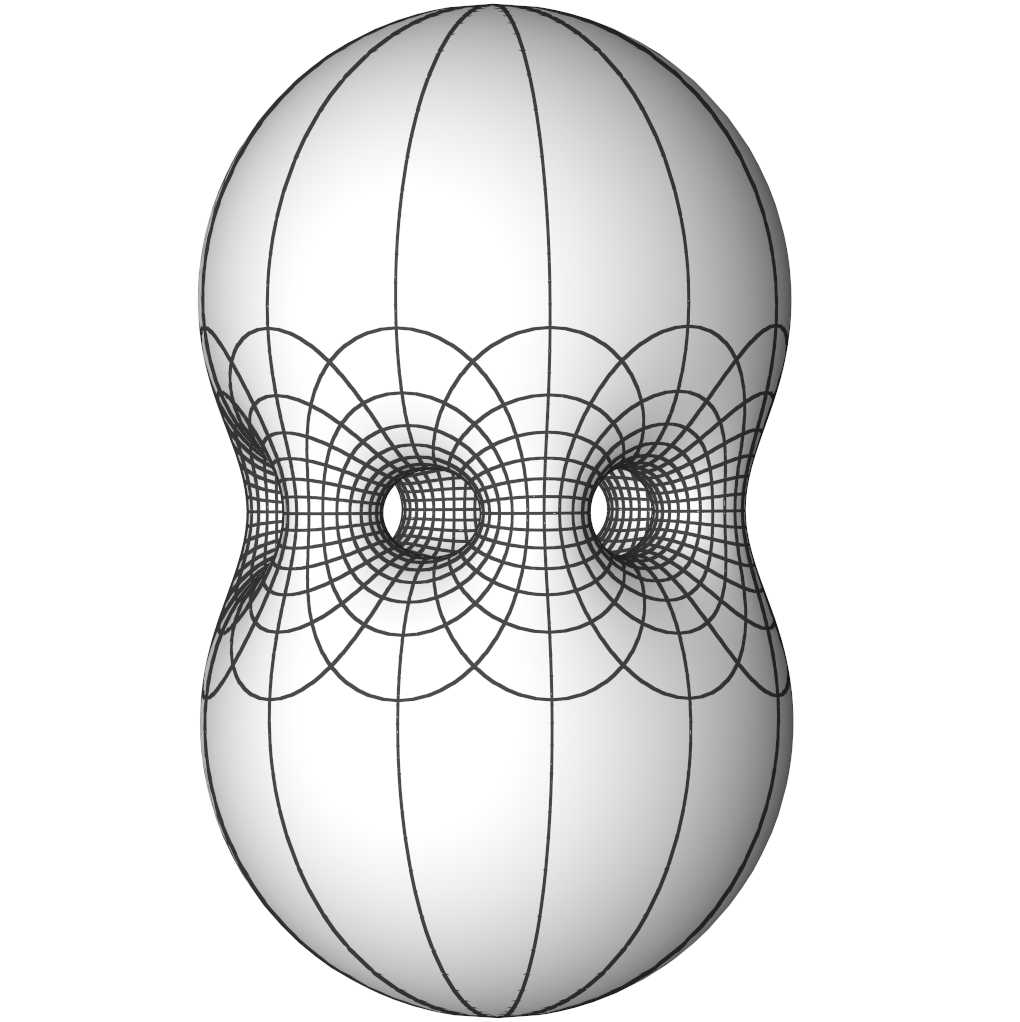}};
        \fill[red](10.1,.85)circle(.5mm); \fill[red](10.6,1.05)circle(.5mm);
        \path(10.4,0)node{\small$g=5$};
            }
\caption{Left-most, $g{=}0$: the 0-dimensional reduction,
 $\IP^4[5]\<\to \cdots\<\to \IP^1[2]$, Calabi-Yau 0-fold (two red points), the anticanonical hypersurface in $S^2{=}\IP^1$. This Calabi--Yau 0-fold easily embeds in increasingly higher-genus Riemann surfaces; those with $g\geqslant2$ are called ``of general type.''}
 \label{f:CY0}
\end{figure}
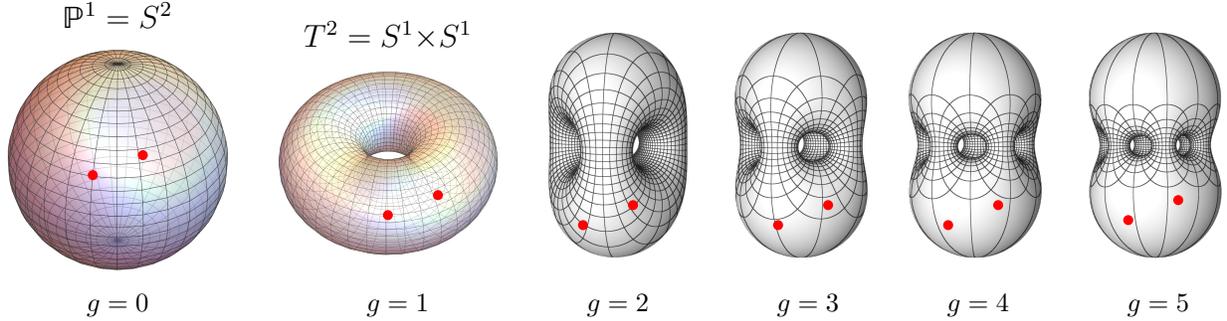

However, non-Fano ($c_1{\ngtr}0$) varieties also admit Calabi--Yau hypersurfaces\cite{rBH-gB, Berglund:2022dgb, Berglund:2024zuz} --- see Figure~\ref{f:CY0} for a 0-dimensional sketch,
and are found within deformation families of their (semi) Fano ($c_1{\geqslant}0$) variants\cite{rgCICY1, rBH-Fm, rGG-gCI, Berglund:2022dgb, Berglund:2024zuz}.
 Mirror symmetry then necessarily {\em\/extends\/} this web {\em\/even futher\/}: In addition to {\em\/reflexive\/} polytopes\cite{rBaty01, Kreuzer:2000xy}, ``{\em\/VEX multitopes\/}''\cite{rBH-gB, Berglund:2022dgb, Berglund:2024zuz} also include non-convex, possibly multi-layered multihedral but star-triangulable bodies, $\pDs{X}$, which span\footnote{A VEX multitope {\em\/spans\/} (is star-subdivided by) a multifan, $\pDs{}{\lat}\,\Sigma$, with a {\em\/primitive\/} (lattice coprime) central cone over every face of the multitope $\pDs{}$\cite{rGE-CCAG}. A $k$-{\em\/face\/} $\theta\in\partial\pDs{}$ is a closed, contiguous, multihedral, $k$-dimensional component in $\partial\pDs{}$, an adjacency-generated {\em\/poset\/} of faces; standard definitions\cite{rD-TV,rO-TV,rF-TV,rGE-CCAG,rCLS-TV} can be adapted for non-convex, self-crossing, folded and otherwise multi-layered multitopes\cite{rM-MFans, Masuda:2000aa, rHM-MFs}.} {\em\/multifan{s}}, $\S$\cite{rM-MFans, Masuda:2000aa, rHM-MFs, Masuda:2006aa, rHM-EG+MF, rH-EG+MFs2, Nishimura:2006vs, Ishida:2013ab}; see also\cite{Davis:1991uz, Ishida:2013aa, buchstaber2014toric, Jang:2023aa}.
The geometry of corresponding {\em\/torus manifolds,}\footnote{A {\em\/torus manifold\/} is a compact, real $2n$-dimensional (smooth) manifold with a torus,- i.e., $(S^1)^n$-action with a nonempty set of isolated fixed points\cite{rM-MFans, rHM-MFs}.} $X_\S$, is intrinsically $(S^1)^n$-equivariant, matching the GLSM $U(1)^n$-gauge symmetry in string theory\cite{rPhases}.
 This match helps in exploring the nature of Calabi--Yau mirror models constructed within both Fano and non-Fano such torus manifolds. The extraordinary success in analyzing GLSMs and related structures\cite{rPhases, rPhasesMF, rMP0, Distler:1993mk, rMP1, Schafer-Nameki:2016cfr, Sharpe:2024dcd} is then seen as a consequence of enhancing the minimally necessary worldsheet {\em\/local\/} $(0,1)$-supersymmetry\cite{rUDSS08, rUDSS09} to a global $(0,2)$- and even $(2,2)$-supersymmetry, which also enhances the $U(1)^n$-gauge symmetry to the $U(1;\IC)^n\<=(\IC^*)^n$ toric action and induces target spacetime supersymmetry, which in turn obstructs the physically desirable positive (albeit minuscule) cosmological constant of the observed asymptotically de~Sitter spacetime.

\subsection{Organization}
\label{s:org}
The remainder of this introduction reviews the key construction of holomorphic equivalence-class sections on hypersurfaces used to construct {\em\/generalized complete intersections\/}\cite{rgCICY1, rBH-Fm, rGG-gCI}, the related class of Laurent hypersurfaces in torus manifolds, and their key consequences. This affords the extension of the overarching mirror-symmetric framework of Calabi--Yau triples, $\big(Z_f,(X\<\ssm Z_f),\cK^*_{\!X}\big)$ from (almost/semi) Fano to certain non-Fano torus manifolds $X$\cite{Berglund:2024zuz}\footnote{The term ``transpolar'' is meant as a portmanteau, combining ``transpose'' and ``polar'' --- but also employs the prefix ``trans-'' to mean ``beyond''\cite{rBH-gB}.}:
\begin{equation}
\setbox9\hbox{$=:$}
 \vC{\begin{tikzpicture}[scale=1.6, 
                         every node/.style={inner sep=0,outer sep=.75mm}]
      \path[use as bounding box](0,-1)--(8.5,1.4);
      \node(K)  at(1.4,1.1) {$\G(\cK^*_X)$\,
            \rlap{$\Iff\sfa\pDN{X}$}};
      \node(G)  at(.7,1.1)  {$f\in$};
      \node(X)  at(1.4,.3)
          {$X$\,\rlap{$\Iff\S_X\smt\pDs{X}$}};
      \node(Z)  at(1.4,-.6) {\llap{$X[c_1]\ni$\,}$Z_f$};
      \node(*K) at(7,1.1)  {\llap{$\sfa\pDs{X}\Iff$}\,
            $\G(\smash{\cK^*_Y})$};
      \node(*G) at(7.7,1.1)   {$\ni g$};
      \node(*X) at(7,.3)   {\llap{$\pDN{X}\lat\S_Y
            \Iff$}\,$Y$};
      \node(*Z) at(7,-.6)  {$Z_g$\rlap{$
            \in Y[c_1]$}};
      \draw[semithick, ->](K)--(X);
      \draw[semithick, <-{Hooks[left]}](X)--node[right=2pt]
            {\scriptsize$f{=}0$}(Z);
      \draw[semithick, ->](*K)--(*X);
      \draw[semithick, <-{Hooks[right]}](*X)--node[left=2pt]
            {\scriptsize$g{=}0$}(*Z);
      \draw[blue, thick, densely dotted, -stealth](G)--node[below, rotate=-68]
            {\scriptsize def.\ section\quad~~}(Z);
      \draw[blue, thick, densely dotted, -stealth](*G)--node[below, rotate=68]
            {\scriptsize\quad def.\ section}(*Z);
      \draw[purple, thick, Stealth-Stealth](2.7,1.1) to
            node[above]{\scriptsize({\em\/trans\/})polar}
            node[below]{\scriptsize\cite{rBH-gB, Berglund:2022dgb, 
            Berglund:2024zuz}} (5.6,1.1);
      \draw[purple, thick, Stealth-Stealth](3,.3) to
            node[above]{\scriptsize({\em\/trans\/})polar}
            node[below]{\scriptsize\cite{rBH-gB, Berglund:2022dgb, 
            Berglund:2024zuz}} (5.45,.3);
      \draw[purple, thick, Stealth-Stealth](Z)to
            node[above]{\scriptsize transposition (``BHK'') mirrors}
            node[below]{\scriptsize \cite{rBH, rBH-LGO+EG, Krawitz:2009aa,
            rLB-MirrBH, rACG-BHK, rF+K-BHK, Belavin:2020xhs} \&
            \cite{rBH-gB, Berglund:2022dgb, Berglund:2024zuz}} (*Z);
   \end{tikzpicture}}
 \label{e:gMM}
\end{equation}
This framework is encoded by ({\em\/trans\/})polar pairs of multitopes, $(\pDs{X},\pDN{X})$, and consists of:
\begin{enumerate}[itemsep=-1pt, topsep=0pt]

 \item a deformation family of $Z_f:=\{f(x)=0,\:f\in\G(\cK^*_{\!X})\}$
    Calabi--Yau $n$-fold hypersurfaces,

 \item in a torus manifold $X$ --- wherein $(X\ssm Z_f)$ are
     non-compact Calabi--Yau $(n{+}1)$-folds\cite{rTY1, rTY2},

 \item and the anticanonical sheaf, $\cK^*_{\!X}$,
     itself a non-compact Calabi--Yau $(n{+}2)$-fold.

\end{enumerate}
as well as their mirror/transpolar images,
 $\big(Z_g,(Y\<\ssm Z_g),\cK^*_Y\big)$.
All K\"ahler manifolds admit a {\em\/compatible triple\/}: a
complex structure ($J_i{}^k$) compatible with a symplectic structure 
($\w_{kj}$) so that $g_{ij}\!:=\!J_i{}^k\,\w_{kj}$ is a Riemannian metric --- the corresponding K\"ahler metric. Generalizing the framework~\eqref{e:gMM} beyond Fano embedding spaces $X,Y$, allows the compatible triples, 
 $(J,\w,g_{\sss K})$ on $Z_f\<\subset X$ and/or on
 $Z_g\<\subset Y$, to degenerate.
 The entire mirror-symmetric framework~\eqref{e:gMM} is parametrized by a {\em\/joint moduli space\/} that admits a (trivial up to Bridgeland stability\cite{rD+G-DBrM}) foliation, $(\sM,\sW)$, such that:
\begin{enumerate}[resume, itemsep=-1pt, topsep=0pt]

 \item 
 $\sM$ parametrizes the choices of $f\<\in \G(\cK^*_X)$, and so (some of) the complex structure deformations of $Z_f$, as well as of $(X\<\ssm Z_f)$.
 Analogously, $\sW$ parametrizes $g\<\in \G(\cK^*_Y)$ and (some of) the complex structure deformations of $Z_g$ and of $(Y\<\ssm Z_g)$.

 \item 
 By mirror symmetry, $\sM$ also parametrizes (some of) the (complexified) K{\"a}hler class variations of $Z_g$ and of $(Y\<\ssm Z_g)$, while $\sW$ also parametrizes (some of) the (complexified) K{\"a}hler class variations of $Z_f$ and of $(X\<\ssm Z_f)$.

 \item\label{o:CKS}
 Varying (the cohomology class of) the K\"ahler metric while holding $J$ fixed is equivalent to varying the symplectic form, $\w_{kj}$, so the {\em\/joint moduli space,} $(\sM,\sW)$, parametrizes this {\em\/compatible triple,} $(J,\w,g_{\sss K})$, on both $Z_f\<\subset X$ and $Z_g\<\subset Y$, mirror-symmetrically.

 \item 
 The (left-right) mirror mapping~\eqref{e:gMM} is encoded by the swap:
\begin{equation}
  (\pDs{X},\pDN{X})
  \fif{~\text{({\em trans\/})polar}~}
  (\pDN{X},\pDs{X}) =: (\pDs{Y},\pDN{Y}),
\end{equation}
which then suggests that the non-compact complement
 $(Y\<\ssm Z_g)$ is also the mirror of $(X\<\ssm Z_f)$, and that the total space of $\cK^*_X$ is the mirror of $\cK^*_Y$, which may be less non-trivial, since the total space of $\cK^*_X$ contracts to $X$. This mirror mapping refers to non-compact Calabi--Yau manifolds, and is sometimes referred to as ``local mirror symmetry'' in physics literature.
\end{enumerate}

 With these seven observations, the roadmap diagram~\eqref{e:gMM} lends itself to a simpler but sweepingly generalizing (and conjectural!) re-statement:
\begin{equation}
\setbox9\hbox{$=:$}
 \vC{\begin{tikzpicture}
      [xscale=1.6, yscale=1.3, every node/.style={inner sep=0,outer sep=.75mm}]
      \path[use as bounding box](-.3,-.8)--(7.1,1.3);
      \node(K)  at(2,1.1) {$\cK^*_X$};
      \node(G)  at(.4,1.1){$f$\rlap{$\in\!\G(\cK^*_X)\Leftarrow$}};
      \node(X)  at(2,.3)  {$(X~$\rlap{\kern-4pt${\ssm}Z_f)$}};
      \node(Z)  at(2,-.6) {\llap{$X[c_1]\ni$\,}$Z_f$};
      \node(*K) at(5,1.1) {$\cK^*_Y$};
      \node(*G) at(6.6,1.1){\llap{$\To\G(\cK^*_Y)\!\ni$}$g$};
      \node(*X) at(5,.3)   {$(Y$
            \rlap{\!\!${\ssm}Z_g)$}};
      \node(*Z) at(5,-.6)  {$Z_g$
            \rlap{$\in Y[c_1]$}};
      \draw[semithick, ->](K)--(X);
      \draw[semithick, <-{Hooks[left]}](X)--node[right=2pt]
            {\scriptsize$f{=}0$}(Z);
      \draw[semithick, ->](*K)--(*X);
      \draw[semithick, <-{Hooks[right]}](*X)--node[left=2pt]
            {\scriptsize$g{=}0$}(*Z);
      \draw[blue, thick, densely dotted, -stealth](G)--node[below, rotate=-42]
            {\scriptsize def.\ section\quad~~}(Z);
      \draw[blue, thick, densely dotted, -stealth](*G)--node[below, rotate=42]
            {\scriptsize\quad def.\ section}(*Z);
      \draw[purple, thick, Stealth-Stealth](K) to
            node[above]{\scriptsize mirror*} (*K);
      \draw[purple, thick, Stealth-Stealth](2.7,.3) to
            node[above]{\scriptsize mirror*} (*X);
      \draw[purple, thick, Stealth-Stealth](Z)to
            node[above]{\scriptsize mirror} (*Z);
      \path(0,.6)node[rotate=90]{\scriptsize non-compact};
      \path(-.15,-.6)node[right]{\scriptsize compact};
   \end{tikzpicture}}
 \label{e:gMM0}
\end{equation}
which seems reasonable to expect to hold generally, in the sense:
\label{sC:0}
\begin{conj}\label{C:0}
For every Calabi--Yau hypersurface, $Z_f\<\in X[c_1]$ there exists a space $Y$ with a suitable anticanonical section, $g\in\G(\cK^*_Y)$, so $Z_g\<\in Y[c_1]$ is mirror to $Z_f$ and the rest of the mirror-symmetric framework~\eqref{e:gMM0} holds.
\end{conj}
\noindent{\bf Notes:}
 ({\small\bf1})~Extending from hypersurfaces to their complete intersections (by iteration) and higher-rank zero-loci (by splitting principle and/or resolutions) should be straightforward.\footnote{The immediate-embedding torus manifolds $X$ and $Y$ for the diagram~\eqref{e:gMM0} are then simply any (convenient) one of the combinatorially available penultimate intersections.}
 ({\small\bf2})~As mirror symmetry is known to swap the roles of complex and symplectic geometry\cite{rCK, Kontsevich:1995wk, Kontsevich:1994Mir, rSYZ-Mirr, rHV-MS}, the objects in the diagram~\eqref{e:gMM0} defined in standard complex algebro-geometric way, need a precisely mirror-symmetric symplectic geometry reformulation to make the mirror symmetry manifest.
 ({\small\bf3})~The non-compact mirror$^*$-relations may well need some qualification: Compact Calabi--Yau spaces (e.g., $Z_g$) may be found as hypersurfaces in various different embedding spaces, say $Z_g\<\subset Y$ and $Z_{g'}\<\subset Y'$, implying that embedding-complementary non-compact Calabi--Yau spaces, $(Y\<\ssm Z_g)$ and $(Y'\<\ssm Z_{g'})$, are multiple mirrors* of $(X\<\ssm Z_f)$; the same applies to
 $\cK^*_Y$ and $\cK^*_{Y'}$ being multiple mirrors* of $\cK^*_X$.
 ({\small\bf4})~Comparing with its straightforward physics origins\cite{rGP1, rBH, rJPS}, the mathematical statement of (homological) mirror symmetry, ``the derived category of coherent sheaves on $Z_f$ is isomorphic as a triangulated category to the derived Fukaya category on $Z_g$'' {\em\/should\/} have also a more direct (elementary?) formulation.
\\
\centerline{---$\,*\!*\!*\,$---}

Reviewing a showcasing infinite sequence of {\em\/generalized complete intersection Calabi--Yau\/} (``gCICY'') $n$-folds\cite{rgCICY1, rBH-Fm, rBH-gB, rGG-gCI, Berglund:2022dgb, Berglund:2024zuz}, Section~\ref{s:FmXm} presents the key construction of Laurent defining equations in gCICYs.
 A coordinate-level reinterpretation of such Calabi-Yau $n$-folds finds them in Section~\ref{s:TR} as toric hypersurfaces in {\em\/distinct\/} (complex, algebraic) toric varieties {\em\/within the same explicit deformation family\/} of (g)CICYs.

Section~\ref{s:DFP} then shows that:
 ({\small\bf1})~many if not all toric varieties may be realized as specific complete intersections in products of projective spaces, and
 ({\small\bf2})~Laurent deformations of (algebraically) ``un-smoothable'' Tyurin-degenerate Calabi-Yau $n$-folds\cite{Berglund:2022dgb} are smooth limits of regular deformations within the same explicit deformation family of (diffeomorphic) Calabi--Yau $n$-folds.
 This gathered evidence is used in Section~\ref{s:CSI} to characterize the class of torus manifolds, $X$ and $Y$, in which the mirror-symmetric framework~\eqref{e:gMM} exists.
 Section~\ref{s:Coda} then collects several conclusions of the foregoing analysis about the class of usable toric embedding spaces, $X$ and $Y$, most of which are of general type ($c_1{\ngeqslant}0$) and which are unitary torus manifolds\cite{rM-MFans, Masuda:2000aa, rHM-MFs, Masuda:2006aa, rHM-EG+MF, rH-EG+MFs2} or surgically related to them; see also\cite{Nishimura:2006vs, Ishida:2013ab, Davis:1991uz, Ishida:2013aa, buchstaber2014toric, Jang:2023aa}.
 The key conclusions are formulated as conjectures (\ref{C:0}, \ref{C:1}, \ref{C:2}, \ref{C:01dS}, \ref{C:VEX}, \ref{C:pCS}, \ref{C:3}), intending to motivate a more rigorous study, leading to a better understanding of this immensely vast new class of constructions, aiming ultimately to prove these conjectures or their more precise revision, perhaps by suitably generalizing the vertex-algebra proof of transposition mirror symmetry\cite{rLB-MirrBH}, which should also provide a rigorous foundation for the superconformal field theory and GLSM counterpart.

\subsection{Motivation}
\label{s:mot}
We seek a specific type of solutions to the ``source-less'' Einstein equations,
\begin{equation}
  R_{\m\n}-\tfrac12g_{\m\n}\,R = \tfrac{8\pi G}{c^4}\,T_{\m\n},
  \qquad\text{where}\qquad
  T_{\m\n}=0~~\Iff~~R_{\m\n}=0.
 \label{e:EE0}
\end{equation}
Without any matter distribution, this describes {\em\/empty space,} i.e., the vacuum. Ricci-flatness is also the lowest-order condition for quantum stability of the worldsheet quantum field theory underlying string theory\cite{rF79b, rF79a}. Identical to the Calabi--Yau condition ($c_1=0$), it is synonymous with certain necessary anomaly cancellations in this underlying worldsheet quantum field theory\cite{rBeast, rJPS, rGSW2, rBBS}, and also emerges from the arguably non-perturbative (oriented) loop-space analysis of string theory\cite{rFrGaZu86, rBowRaj87, rBowRaj87a, rBowRaj87b, Oh:1987sq, rHHRR-sDiffS1, Pilch:1987eb, rBowRaj88, Bowick:1988nj, Bowick:1990wt}.
 With such plentiful motivation, we continue the by now fantastically prolific quest to construct Ricci-flat, i.e., Calabi--Yau geometries, and foremost the space-like compact factors\cite{rH-CY0, rGHCYCI, rCYCI1, rCLS-WCP4, rKreSka95, SKARKE_1996, rKreSka98, Kreuzer:2000xy, rKS-CY}.
 
 To a macroscopic observer in the cosmically large $\IR^{1,3}$-like spacetime, the geometry of extra dimensions is observable primarily through the invariant characteristics of the compact 6-dimensional factor, $Z$, such as the number of topologically nontrivial {\em\/handle-bodies\/} and corresponding {\em\/cycles,\/} counted by the Betti/Hodge numbers $b_2=h^{11}$ and $b_3=2(h^{21}{+}1)$, and the Euler characteristic, $\chi=2(h^{11}{-}h^{21})$. For a candidate $Z$ with desirable $h^{11},h^{21}$, one proceeds computing the various Yukawa couplings\footnote{These are the Bridgeland-stable ring structure constants for the relevant quantum cohomology groups\cite{rD+G-DBrM}, or at least their classical limits, with quantum corrections included perturbatively\cite{rBeast, rCK}: $H^{1,1}(Z)=H^1(Z,T^*_Z)$, $H^{2,1}(Z)=H^1(Z,T_Z)$ and then also $H^1(Z,\mathrm{End}T_Z)$ with their mixed coupling included\cite{rBeast}, as well as their deformations and higher-rank stable bundle generalizations, $T_Z\to V_Z$.}, physically normalized by the Zamolodchikov metric on the relevant moduli spaces --- to which the essentially algebro-geometric methods of construction and analysis are well adapted. Given the well-publicized embarrassment of riches estimated in $10^{700}$-tuples of moles ($10^{23}$) of models\cite{Constantin:2018xkj}, only the advent of various machine learning methods makes a search meaningful\cite{He:2018jtw, Butbaia:2024tje, Constantin:2024yxh, Berglund:2024uqv, Constantin:2024yaz, Berglund:2024psp, Fraser-Taliente:2024etl}. These machine learning methods also make it possible to estimate other curvature details\cite{Berglund:2022gvm} in the search for ways to enable a realistic {\em\/de~Sitter uplift\/}\cite{Bento:2021nbb, Bento:2023De-, McAllister:2023vgy}, and so seek a superstring model that does recover all the observed features of our Universe.

\subsection{Computational Framework}
\label{s:CoF}
In general, one seeks to construct the Calabi--Yau space, $Z$, as a complete intersection of (algebraically constrained, $f_a(x)=0$) hypersurfaces in some well understood ``ambient'' space, $X$:
\begin{equation}
  Z_f := {\ttt\bigcap_a}
   \big(Z_{f_a}:=\{f_a(x)=0\}\big)\subset X.
 \label{e:CI}
\end{equation}
On so-constructed $Z_f$, functions are naturally defined as {\em\/equivalence classes,}
\begin{equation}
  \varrho:~
  \phi_Z(z) \simeq \big[ \phi_X(x)~\big(\mathrm{mod}\:\oplus_a\!f_a(x)\big) \big],
 \label{e:CIf}
\end{equation}
since all $f_a(x)$ vanish on $Z\subset X$. Coordinates themselves being functions, their differentials are then straightforwardly defined as the analogous equivalence classes (oft cited as the {\em\/adjunction formula\:1\/}\cite{rBeast}),
\begin{equation}
  \varrho:~
  \rd z \simeq \big[ \rd x~\big(\mathrm{mod}\: \oplus_a \rd f_a(x)\big) \big],
 \label{e:CIa1}
\end{equation}
extending the algebraic definition~\eqref{e:CIf} to calculus~\eqref{e:CIa1}. For such a complete intersection to be smooth and the projection/restriction $\varrho$ in~\eqref{e:CIf}--\eqref{e:CIa1} complete, the system of algebraic constraints must be {\em\/transverse,}
\begin{equation}
   \mathrm{Sing}(Z):=
    \{ {\ttt\bigwedge_a}\, \rd f_a=0\}\cap\{f_a=0\} ~\not\subset X.
 \label{e:CIt}
\end{equation}
The determinant of~\eqref{e:CIa1} results in the {\em\/adjunction formula\:2\/}\cite{rBeast},
\begin{subequations}
 \label{e:CIa2}
\begin{alignat}9
   (\cK^*_Z:=\det T_Z) &= (\cK^*_X:=\det T_X)
       \otimes\big[{\ttt\prod_a}\, \cO\big(\deg(f_a)\big)\big],\\
   \text{i.e.,}\quad
   c(Z) &= c(X) \big/ {\textstyle\prod_a} c(Z_{f_a}),
\end{alignat}
\end{subequations}
where $c(X)=1{+}c_1(X){+}\cdots{+}c_{\dim X}(X)$ is the (total) Chern class of $X$. Finally, the vanishing of the first Chern class, cited as {\em\/the Calabi--Yau condition,}
\begin{equation}
  c_1(Z) = c_1(X) - {\ttt\sum_a}\,c_1(Z_{f_a})\: \overset!= 0,
 \label{e:CYc}
\end{equation}
requires the total degree of the constraint system to equal that of the {\em\/anticanonical\/} class, $\cK^*_X$, insuring $\cK^*_Z$ to be trivial --- and the holomorphic volume form on $Z$ to be covariantly constant. The foregoing then guarantees that concrete computations on $X$ reduce to solving straightforward (if tedious) $U(n_r{+}1)$-tensor equations when $X$ is a product of projective spaces\cite{rBeast}, and analogous generalizations when $X$ is a toric variety\cite{Blumenhagen:2010pv, Blumenhagen:2010ed, Jurke:2011nq}. To this end, by now there exist
 \texttt{Mathematica}, \texttt{SAGE}, \texttt{Macaulay2}, \texttt{Magma}
and other computer-aided implementations, which however require adaptation and generalization to the {\em\/general type\/} (non-Fano) $X$ discussed herein.

As implied by the discussion of~\eqref{e:gMM}, the foregoing (essentially algebraic) computational framework applies directly to {\em\/compact\/} Calabi--Yau subspaces, $Z_f\subset X$, in some sufficiently known embedding space,\footnote{Toric varieties include as special cases products of weighted projective spaces, ordinary (isotropic) projective spaces being special cases thereof.} $X$. It is then noteworthy that the framework~\eqref{e:gMM} --- {\em\/automatically\/} --- includes also two distinct and closely related {\em\/non-compact\/} Calabi--Yau manifolds: the $(X\<\ssm Z_f)$ complement, and the anticanonical bundle $\cK^*_X$; see~\eqref{e:gMM0}.

\section{The Showcasing Sequence}
\label{s:FmXm}
The key novel results are simply showcased by the sequence of deformation families of Calabi--Yau 3-folds:
\begin{equation}
  \XX[3]m\in
  \K[{r||c|c}{\IP^4_x&1&4\\[4pt] \IP^1_y&m&2{-}m}]_{-168}^{2,86}\qquad
  \bigg\{
  \begin{array}{@{}r@{\,=\,}lr@{\,=\,}l}
   p(x,y) & 0, & \deg[p]&\pM{1\\m},\\[4pt]
   q(x,y) & 0, & \deg[q]&\pM{4\\2-m},\\
  \end{array}
 \label{e:gCIXm}
\end{equation}
with a $b_2\<= 2\<= h^{11}$-dimensional space of (complexified) K\"ahler classes and a $\tfrac12b_2{-}1\<= 86\<= h^{21}$-dimensional space of complex structures.

\subsection{Calabi--Yau Hypersurfaces in Hirzebruch Scrolls}
\label{s:CYnFm}
Consider first the family of deg-$(1,m)$ constrained hypersurfaces, explicitly parametrized by the $\e_{a,\ell}$ coefficients:
\begin{equation}
   p_\e(x,y) = x_0y_0\!^m+x_1y_1\!^m
   +{\ttt \sum_{a=2}^4\sum_{\ell=1}^{m-1}}
           \e_{a,\ell}\,x_a\,y_0\!^{m-\ell}y_1\!^\ell
   ~\overset!=0,
 \label{e:HFme}
\end{equation}
where the $\e_{a,\ell}=0$ ``central'' model, $p_0(x,y)=0$, was introduced by Hirzebruch\cite{rH-Fm}. Since the gradient
 $\vd p_\e(x,y)=(y_0^m,y_1^m,\dots)$ cannot vanish as $(y_0,y_1)\neq(0,0)$ on $\IP^1_y$, the first hypersurface, the Hirzebruch scrolls
 $\FF[4]{m}:=\{p_\e(x,y)=0\}\subset\IP^4{\times}\IP^1$, are smooth for all $\e_{a,\ell}$. With the $A=\IP^4_x{\times}\IP^1_y$ K{\"a}hler forms, $J_i$, generating $H^2(\FF[4]m,\ZZ)$, the Chern class is ($J_1\!^5\<= 0\<= J_2\!^2$)
\begin{equation}
  c(\FF[4]m)
  =\frac{(1{+}J_1)^5(1{+}J_2)^2}{(1{+}J_1{+}mJ_2)}
  =(1{+}J_1)^3(1{+}J_1{-}mJ_2)(1{+}J_2)^2,
\end{equation}
so that all various Chern class evaluations on corresponding powers of
 $(aJ_1{+}bJ_2)$ depend on $b$ only via the linear combination $4b{+}ma$, so that $H^*(\FF[4]{m+4k},\ZZ)$ is related to $H^*(\FF[4]m,\ZZ)$ by the simple integral basis change:
\begin{equation}
  \bigg[\begin{matrix}J_1\\ J_2\end{matrix}\bigg]_m
  \too{\:\approx\:} 
  \bigg[\begin{matrix}1 &-k\\ 0 &~~1\end{matrix}\bigg]\!\!
  \bigg[\begin{matrix}J_1\\ J_2\end{matrix}\bigg]_{m+4k}
  ~~\xRightarrow{\text{\cite{rBH-Fm}~\;}}\quad
  \FF[4]{m+4k}\mathop{\approx_\IR}\FF[4]m,~~k\in\ZZ.
\end{equation}
This in turn implies that the sequence~\eqref{e:gCIXm} contains only four distinct diffeomorphism classes of Calabi--Yau 3-folds\cite{rWall}:
\begin{equation}
  \XX[3]0{\in}\ssK[{r||c|c}{\IP^4_x&1&4\\[4pt] \IP^1_y&0&2}],~~
  \XX[3]1{\in}\ssK[{r||c|c}{\IP^4_x&1&4\\[4pt] \IP^1_y&1&1}],~~
  \XX[3]2{\in}\ssK[{r||c|c}{\IP^4_x&1&4\\[4pt] \IP^1_y&2&0}],~~
  \XX[3]3{\in}\ssK[{r||c|c}{\IP^4_x&1&~~4\\[4pt] \IP^1_y&3&-1}],
 \label{e:Gang4}
\end{equation}
where the first three families consist of regular complete intersections, but the fourth diffeomorphism class is a {\em\/generalized complete intersection\/}\cite{rgCICY1, rBH-Fm, rGG-gCI}. The sequence also implies that $\XX[3]4{\approx_\IR}\XX[3]0$, $\XX[3]5{\approx_\IR}\XX[3]1$, etc., and this {\em\/diffeomorphism\/} (isomorphism as smooth manifolds) is explored below; in particular, $\XX{k}$ and $\XX{k\,(\mathrm{mod}\,n)}$ differ as {\em\/complex\/} and {\em\/symplectic\/} manifolds.

The more immediate question, ``{\em\/How can a $\deg\!\text{-}\pM{~\>4\\-1}$ section possibly be complex-analytic?\/}'' is answered in several different ways in Refs.\cite{rgCICY1, rBH-Fm, rGG-gCI}. Most importantly: there {\em\/cannot\/} exist such complex-analytic sections on $A=\IP^4{\times}\IP^1$, but they {\em\/can\/} (and do, 105 of them) exist on $\FF[4]3\subset A$ --- and solely because of~\eqref{e:CIf}. While\cite{rgCICY1, rBH-Fm, rGG-gCI} proffer general algorithms for finding such sections, we illustrate the construction by considering Hirzebruch's original $\e_{a,\ell}=0$ case of~\eqref{e:HFme}:
\begin{subequations}
 \label{e:quiQ}
\begin{alignat}9
  p_0(x,y) &= x_0y_0\!^3+x_1y_1\!^3\quad\To\quad
  q_0(x,y) = c(x)\Big(\frac{x_0y_0}{y_1\!^2}-\frac{x_1y_1}{y_0\!^2}\Big),
  \label{e:tune}
\intertext{where $c(x)$ is a cubic over $\IP^4_x$. The equivalence class
 $[q\;(\mathrm{mod}\:p)]$ is then}
 q^{\sss(\lambda)}_0(x,y)
  &\simeq q_0(x,y) +\lambda\frac{c(x)}{(y_0y_1)^2}\,p_0(x,y),\\
  &\quad=\Bigg\{
     \begin{array}
           {@{\;}l@{~~\text{with}~}r@{\,}l@{~~\text{where}~}r@{\>\neq\>}l}
        c(x)\big({-}2\frac{x_1y_1}{y_0\!^2}\big), &\lambda&={-}1, &y_0 &0,
          \\[2mm]
        c(x)\big({+}2\frac{x_0y_0}{y_1\!^2}\big), &\lambda&={+}1, &y_1 &0;\\
     \end{array}\label{e:quiq}\\
 \text{and}~~
  &q^{\sss(+1)}_0(x,y){-}q^{\sss(-1)}_0(x,y)
   =2\frac{c(x)}{(y_0y_1)^2}\, p_0(x,y),~=0\;~\text{on}~\FF[4]3,
  \label{e:quiv}
\end{alignat}
\end{subequations}
since $\FF[4]3\!:=\!\{p_0(x,y)\<= 0\}$.
 This chart-wise definition~\eqref{e:quiQ} provides a complex analytic choice~\eqref{e:quiq} on each hemisphere of $\IP^1_y$, the difference between them vanishing on
 the ``first'' hypersurface, $\FF[4]3$;~\eqref{e:quiv}.
This reminds of the Wu-Yang construction of the magnetic monopole\cite{rWY-MM}.
 Understanding thus $q_0(x,y)$ to represent this {\em\/equivalence class\/} of complex-analytic sections defines
\begin{equation}
   \XX[3]3:=\{p_0(x,y)=0=q_0(x,y)\} ~\subset\IP^4_x\<\times \IP^1_y
\end{equation}
to be a {\em\/generalized complete intersection,} ``in a scheme-theoretic sense''\cite{rGG-gCI}. At the same time, the Calabi--Yau 3-fold $\XX[3]3$ is also a perfectly {\em\/regular\/} hypersurface {\em\/within\/} the Hirzebruch scroll $\FF[4]3$, which  in turn may be described equivalently in several ways other than as a hypersurface itself~\eqref{e:HFme}. Those alternative descriptions can be used for complementary and independent verifications, and to this end we provide below a 1--1 ``translation'' into the {\em\/toric geometry\/} framework.
 
As noted originally\cite{rgCICY1, rBH-Fm, rGG-gCI} and implied in~\eqref{e:quiQ}, the choice of $q_0(x,y)$ is specially ``tuned'' to the choice of $p_0(x,y)$. Deforming $p_0(x,y)\leadsto p_\e(x,y)$ as given in~\eqref{e:HFme} implies a \eqref{e:quiQ}-corresponding change in $q_0(x,y)\leadsto q_\e(x,y)$\cite{rBH-Fm, rGG-gCI}. The analysis of such variations, showcased by the explicit deformation families of Calabi--Yau 3-folds diffeomorphic to the sequence~\eqref{e:Gang4} and their mirror models \`a la~\eqref{e:gMM}, provides the key novelty discussed herein; see in particular the concrete result~\eqref{e:qGen} below.

\subsection{Hallmark Submanifolds in Hirzebruch Scrolls}
\label{s:SnFm}
The same construction~\eqref{e:quiQ} also provides for another hallmark submanifold of $\FF{m}$, the so-called {\em\/directrix\/}\cite{rGrHa} of degree $\pM{~\:1\\-m}$:
\begin{alignat}9
  \Fs_0^{\sss(\lambda)}(x,y)
   &:= \Big(\frac{x_0}{y_1\!^m}-\frac{x_1}{y_0\!^m}\Big)
   +\frac{\lambda}{(y_0y_1)^m}\big(x_0y_0\!^m+x_1y_1\!^m\big),
  \label{e:drx}\\[2mm]
  [\Fs^{-1}(0)]^n &=
  \K[{r||c|ccc}{\IP^n_x&1&~~1&\cdots&~~1\\[3pt]
                \IP^1_y&m& -m&\cdots& -m\\}]_
  {\TikZ{\path[use as bounding box](0,0)--(0,-.9);
         \path(-1.3,-.75)node{$\underbrace{\mkern95mu}_{n~\text{columns}}$};}}
  = 1(m) +n({-}m) = -(n{-}1)m.
 \label{e:sInt}
\end{alignat}
where the zero-locus $\Fs^{-1}(0):=\{\Fs(x,y)=0\}\subset\FF{m}$ is a holomorphic and irreducible subspace in $\FF{m}$, the existence of which and its maximally negative self-intersections~\eqref{e:sInt}, prove that
 $\{p_0(x,y)=0\}$ is indeed the $m$-twisted Hirzebruch scroll.

Comparing~\eqref{e:tune} with~\eqref{e:drx} shows
$q_0(x,y)=c(x){\cdot}(y_0y_1){\cdot}\Fs(x,y)$, relating the Calabi--Yau hypersurfaces, $\XX{m}\subset\FF{m}$\cite{rBH-gB} to the hallmark {\em\/directrix,} $\Fs^{-1}(0)\subset\FF{m}$\cite{rGrHa}.
 The astute reader will have noticed that the definition of $q_0(x,y)$ in~\eqref{e:tune} provides for only ${\scriptstyle\binom{3{+}4}{4}}\<= 35$ of the 105 sections, accounted for by the cubics $c(x)$ on $\IP^4_x$. The full complement is easily reproduced generalizing this pattern, and serves to further highlight the pivotal role of the directrix:
\begin{equation}
   q^{\sss(\lambda)}_0(x,y)
    =\big(\, c^{00}(x)y_0\!^2{+}c^{01}(x)y_0y_1{+}c^{11}(x)y_1\!^2 \,\big)
      \!\cdot \Fs^{\sss(\lambda)}_0(x,y),
 \label{e:qGen}
\end{equation}
now duly parametrized by three independent cubics, $c^{00}(x)$, $c^{01}(x)$ and $c^{11}(x)$, adding up to $3{\cdot}35\<= 105=\dim H^0(\FF[4]3,\cK^*)$. This provides additional and complementary reasoning to the various other construction methods exhibited earlier\cite{rBH-Fm, rgCICY1, rGG-gCI}.

This result additionally motivates the following\cite{rBH-gB}:
\label{sc:s}
\begin{cons}\label{c:s}
On a degree-$\pM{1\\m}$ hypersurface
 $\FF{m;\e}\!:=\!\{p_\e(x,y){=}0\}\,
   {\subset}\,\IP^n{\times}\IP^1$ akin to~\eqref{e:HFme}, construct the
deg-$\mkern1mu\pM{-1\\m{-}r_0{-}r_1}$ holomorphic equivalence classes of sections
\begin{equation}
  \Fs_\e^{\sss(\lambda)}(x,y):=
   \Big[\mathop{\mathrm{Flip}}_{y_0}
         \Big(\frac1{y_0\!^{r_0}\,y_1\!^{r_1}}p_\e(x,y)\Big)
    ~\big(\mathrm{mod}\:p_\e(x,y) \big)\Big],
 \label{e:gDrX}
\end{equation}
represented by the Laurent polynomials containing both $y_0$- and $y_1$-denomi\-na\-tors but no $y_0\,y_1$-mixed ones, for $r_0{+}r_1\<= 2m,\,2m{-}1,\,\cdots,\,0$. The ``\/$\mathrm{Flip}_{y_i}\!\!$'' operator changes the relative sign of the rational monomials with $y_i$-denomi\-na\-tors.
 For algebraically independent such Laurent polynomial sections, restrict to the subset with maximally negative degrees that are not overall $(y_0,y_1)$-multiples of each other.
\end{cons}
 The zero-locus of each so-obtained section~\eqref{e:gDrX} is a hallmark holomorphic submanifold (a {\em\/directrix\/}) of the hypersurface
 $\FF{m;\e}$, and the so-obtained list of {\em\/directrices\/} characterizes $\FF{m;\e}$ itself. Anticanonical sections~\eqref{e:qGen}, with which one defines Calabi--Yau hypersurfaces in non-Fano varieties, are such linear combinations of regular multiples of directrices.
 We now turn to describing the ``translation'' of the foregoing discussion of generalized hypersurfaces and complete intersections into the framework of toric geometry.

\section{Toric Rendition and Smoothing}
\label{s:TR}
For simplicity, we return to the ``central'' ($\e_{a,\ell}\<= 0$) hypersurface~\eqref{e:HFme}, 
\begin{equation}
   \FF{m;0} := \{p_0(x,y)\<= x_0y_0\!^m{+}x_1y_1\!^m\<= 0\}
   \subset \IP^4_x{\times}\IP^1_y,
 \label{e:p0}
\end{equation}
and then explore the effects of varying $\e_{a,\ell}$.

The familiar homogeneous coordinates of the embedding space, 
 $\IP^n_x{\times}\IP^1_y$, may be freely changed, and focusing on~\eqref{e:p0} we choose:
\begin{subequations}
 \label{e:x2psx}
\begin{alignat}9
  (x_0,x_1,x_2,\cdots;y_0,y_1) &\to (p_0,\Fs,x_2,\cdots;y_0,y_1),\\
  \det\big[{\ttt\frac{\vd(p_0,\,\Fs,\,x_2,\,\cdots;\,y_0,\,y_1)}
                 {\vd(x_0,\,x_1,\,x_2,\cdots;\,y_0,\,y_1)}}\big] 
  &=\textit{const.}
\end{alignat}
\end{subequations}
This leaves $(\Fs,x_2,\cdots;y_0,y_1)$ as coordinates on the $\{p_0(x,y)=0\}$ constrained hypersurface, which inherit the $\IP^n_x{\times}\IP^1_y$ bi-degrees:
\begin{equation}
  \begin{array}{@{}c|c@{~}c@{~}c@{~}c@{~}c@{~}c@{}}
     &X_1{=}\Fs &X_2{=}x_2 &\cdots &X_n{=}x_n &X_{n+1}{=}y_0 &X_{n+2}{=}y_1 \\ 
 \toprule\nGlu{-2pt}
 \deg_{\IP^n}\!=: Q^1 &~~1  &  1  & \cdots &  1  &  0      &  0  \\ 
 \deg_{\IP^1}\!=: Q^2 & -m  &  0  & \cdots &  0  &  1      &  1  \\ 
  \end{array}
 \label{e:Mori}
\end{equation}
Interpreting each row of bi--degrees~\eqref{e:Mori} as components of two Mori
 row-$(n{+}2)$-vectors defines its $n$-dimensional null-space:
 $\sum_{i=1}^{n+2}\n^i\,Q^a(X_i)=0$ for $a=1,2$.
The vertical stack of the $n$ so-obtained row-$(n{+}2)$-vectors then defines
 $(n{+}2)$ column-$n$-vectors $\skew{-1}\vec\n\mkern1mu^i$, up to
 $\mathrm{GL}(n)$ basis transformations, such as:
\begin{equation}
  \begin{array}{@{}r|c@{~}c@{~}c@{~}c@{~}c@{~}c@{~}c@{~}l@{}}
     &X_1 &X_2 &X_3 &\cdots &X_n &X_{n+1} &X_{n+2} \\*[-2pt]
 \cmidrule[.6pt](r{1pt}){1-8}\nGlu{-2pt}
 \multirow5*{$\skew{-1}\vec\n\mkern1mu^i\left.\rule{0pt}{35pt}\right\{$}
  & -1 & 1 & 0 & \cdots & 0 & 0 & -m &
 \multirow5*{$\left\}\rule{0pt}{35pt}\right.\S_{\smash{\FF{m}}}$}\\
  & -1 & 0 & 1 & \cdots & 0 & 0 & -m \\[-4pt]
  &~~\vdots &\vdots &\vdots & \ddots &\vdots &\vdots &~~\vdots \\
  & -1 & 0 & 0 & \cdots & 1 & 0 & -m \\
  &~~0 & 0 & 0 & \cdots & 0 & 1 & -1 \\
 \cmidrule[.6pt](r{1pt}){1-8}
 Q^1 &~~1  &  1  &  1  & \cdots &  1  &  0      &  0  \\ 
 Q^2 & -m  &  0  &  0  & \cdots &  0  &  1      &  1  \\ 
  \end{array}
 \label{e:nFmQ}
\end{equation}
 The column-$n$-vectors $\vec\n\mkern1mu^i$ generate the {\em\/fan\/} of cones $\S_{\smash{\FF{m}}}$, the (nearest lattice) bases of which form the facets of the {\em\/spanning polytope,} 
 $\pDs{\FF{m}}$; we write $\S_{\smash{\FF{m}}}\lat\pDs{\FF{m}}$.
 By corresponding each $n$-dimensional cone in $\S_{\smash{\FF{m}}}$ to an 
 $\IC^n$-like chart and gluing them as prescribed by the cone intersections within
 $\S_{\smash{\FF{m}}}$\cite{rD-TV, rO-TV, rF-TV, rGE-CCAG, rCLS-TV, rMP0, rCK}, these combinatorial structures fully specify the toric rendition of the Hirzebruch scrolls, $\FF{m}$\cite{Berglund:2022dgb}.
 The column-2-vectors $\vec{Q}(X_i)$ together with the negative of their sum,
 $-\sum_i\vec{Q}(X_i)\<= \binom{-n}{m{-}2}$, generate the {\em\/secondary fan\/}:
\begin{equation}
  \vC{\TikZ{[thick]
       \path[use as bounding box](-2.2,-3)--(1.2,1);
       \corner{(0,0)}{0}{90}{1}{red};
       \corner{(0,0)}{90}{225}{1}{yellow};
       \corner{(0,0)}{225}{360}{1}{green!50!blue};
       \draw[-stealth](0,0)--(1,0);
       \draw[-stealth](0,0)--(0,1);
       \draw[-stealth](0,0)--(-2,-2);
       \filldraw[fill=white](0,0)circle(.5mm);
       \path(0,.5)node[left]{$\S''(\FF[2]0)$};
            }}
 \quad
  \vC{\TikZ{[thick]
       \path[use as bounding box](-2.2,-3)--(1.2,1);
       \corner{(0,0)}{0}{90}{1}{red};
       \corner{(0,0)}{90}{180+atan(1/2)}{1}{yellow};
       \corner{(0,0)}{180+atan(1/2)}{315}{1}{green};
       \corner{(0,0)}{315}{360}{1}{blue};
       \draw[-stealth](0,0)--(1,0);
       \draw[-stealth](0,0)--(0,1);
       \draw[-stealth](0,0)--(1,-1);
       \draw[-stealth](0,0)--(-2,-1);
       \filldraw[fill=white](0,0)circle(.5mm);
       \path(0,.5)node[left]{$\S''(\FF[2]1)$};
            }}
 \quad
  \vC{\TikZ{[thick]
       \path[use as bounding box](-2.2,-3)--(1.2,1);
       \corner{(0,0)}{0}{90}{1}{red};
       \corner{(0,0)}{90}{180}{1}{yellow};
       \corner{(0,0)}{180}{360-atan(2)}{1}{green};
       \corner{(0,0)}{360-atan(2)}{360}{1}{blue};
       \draw[-stealth](0,0)--(1,0);
       \draw[-stealth](0,0)--(0,1);
       \draw[-stealth](0,0)--(1,-2);
       \draw[-stealth](0,0)--(-2,0);
       \filldraw[fill=white](0,0)circle(.5mm);
       \path(0,-.5)node[left]{$\S''(\FF[2]2)$};
            }}
 \quad
  \vC{\TikZ{[thick]
       \path[use as bounding box](-2.2,-3)--(1.2,1);
       \corner{(0,0)}{0}{90}{1}{red};
       \corner{(0,0)}{90}{180-atan(1/2)}{1}{yellow};
       \corner{(0,0)}{180-atan(1/2)}{360-atan(3)}{1}{green};
       \corner{(0,0)}{360-atan(3)}{360}{1}{blue};
       \draw[-stealth](0,0)--(1,0);
       \draw[-stealth](0,0)--(0,1);
       \draw[-stealth](0,0)--(1,-3);
       \draw[-stealth](0,0)--(-2,1);
       \filldraw[fill=white](0,0)circle(.5mm);
       \path(0,-.5)node[left]{$\S''(\FF[2]3)$};
            }}
 \quad\text{\large etc.}
 \label{e:2ndFn}
\end{equation}
Depicted here for $n\<= 2$, the sequence makes it obvious that
 $\S''(\FF{m}) \not\approx \S''(\FF{m\,(\mathrm{mod}\,n)})$, although all
 $\FF{m}$ and $\FF{m\,(\mathrm{mod}\,n)}$ are {\em\/diffeomorphic,} i.e., are all the same as real, smooth manifolds.

\subsection{Anticanonicals}
\label{s:Ac}
The bi-degrees listed in~\eqref{e:Mori} imply that
 $c_1(\FF{m})=\sum_i\deg(X_i)=\pM{n\\2-m}$,
so that anticanonical sections of the $\e_{a,\ell}=0$ ``central'' Hirzebruch scroll~\eqref{e:tune} (with which to define its Calabi--Yau hypersurfaces) are of the form
\begin{equation}
  \G(\cK^*_{\smash{\FF{m}}}) = \bigoplus\nolimits_{k=0}^n (X_1)^k\,
  (X_2{\oplus}\cdots{\oplus}X_n)^{n-k}\, (X_{n+1}{\oplus}X_{n+2})^{2+m(k-1)}.
 \label{e:K(nFm)}
\end{equation}
For $m\geqslant3$, the last factor acquires a negative exponent when $k\<= 0$. So, {\em\/standard practice\/} in complex algebraic toric geometry uses the full complement~\eqref{e:K(nFm)} when $m\<= 0,\,1,\,2$, but limits $k\<> 0$ when $m\geqslant3$. The $m=1,2,3$ sequence of monomial plots shown in Figure~\ref{f:F1-2-3} exhibit the evident tendency of the right-hand side column of monomials,
\begin{figure}[htb]
 \centering
  \TikZ{[xscale=1.1, yscale=.7]\path[use as bounding box](-2,-2.5)--(2,5);
          \draw[blue, thick, densely dotted, -stealth]
              (-1,-1)--(-1,2)--(1,0)--(1,-1)--cycle;
          \path(-1,2)node{\footnotesize$X^{2030}$};
          \path(-1,1)node{\footnotesize$X^{2021}$};
          \path(-1,0)node{\footnotesize$X^{2012}$};
          \path(-1,-1)node{\footnotesize$X^{2003}$};
          \path(0,1)node{\footnotesize$X^{1120}$};
          \path(0,0)node{\footnotesize$X^{1111}$};
          \path(0,-1)node{\footnotesize$X^{1102}$};
          \path(1,0)node{\footnotesize$X^{0210}$};
          \path(1,-1)node{\footnotesize$X^{0201}$};
          \path(0,-2)node{\small$\G
                           \big(\cK^*_{\smash{\FF[2]1}} \<= 
                                 \cK^*_{\smash{\FF[2]{(1,0)}}}\big)$};
          \path(1,2.3)node{\footnotesize
                           $\deg\cK^*_{\smash{\FF[2]1}}=\pM{2\\1}$};
          \path(1,3.7)node{\footnotesize
                           $\begin{array}[t]
                               {@{}r@{\:}|@{\:}c@{~}c@{~}c@{~}c@{}}
                             &X_1 &X_2 &X_3 &X_4\\ \toprule\nGlu{-2pt}
                            Q^1 &~~1& 1 & 0 & 0 \\
                            Q^2 &-1 & 0 & 1 & 1 \\
                           \end{array}$};
            }
  \qquad\qquad
  \TikZ{[xscale=1.1, yscale=.7]\path[use as bounding box](-1.7,-2.5)--(2,5);
          \draw[blue, thick, densely dotted, -stealth]
              (-1,-1)--(-1,3)--(1,-1)--cycle;
          \path(-1,3)node{\footnotesize$X^{2040}$};
          \path(-1,2)node{\footnotesize$X^{2031}$};
          \path(-1,1)node{\footnotesize$X^{2022}$};
          \path(-1,0)node{\footnotesize$X^{2013}$};
          \path(-1,-1)node{\footnotesize$X^{2004}$};
          \path(0,1)node{\footnotesize$X^{1102}$};
          \path(0,0)node{\footnotesize$X^{1111}$};
          \path(0,-1)node{\footnotesize$X^{1120}$};
          \path[purple](1,-1)node{\footnotesize$X^{0200}$};
          \path(.2,-2)node{\small$\G
                            \big(\cK^*_{\smash{\FF[2]2}} \<= 
                                  \cK^*_{\smash{\FF[2]{(2,0)}}}\big)$};
          \path(-.2,2.3)node[right]
              {\footnotesize$\deg\cK^*_{\smash{\FF[2]2}}=\pM{2\\0}$};
          \path(-.33,3.7)node[right]
            {\footnotesize$\begin{array}[t]
                               {@{}r@{\:}|@{\:}c@{~}c@{~}c@{~}c@{}}
                             &X_1 &X_2 &X_3 &X_4\\ \toprule\nGlu{-2pt}
                            Q^1 &~~1& 1 & 0 & 0 \\
                            Q^2 &-2 & 0 & 1 & 1 \\
                           \end{array}$};
            }
  \qquad\qquad
  \TikZ{[xscale=1.1, yscale=.7]\path[use as bounding box](-2,-2.5)--(3,5);
          \draw[blue, thick, densely dotted]
              (.7,-1)--(-1,-1)--(-1,4)--(.8,-1.7);
          \draw[red, thick, densely dotted](1,-1.2)--++(0,-.6);
          \path(-1,4)node{\footnotesize$X^{2050}$};
          \path(-1,3)node{\footnotesize$X^{2041}$};
          \path(-1,2)node{\footnotesize$X^{2032}$};
          \path(-1,1)node{\footnotesize$X^{2023}$};
          \path(-1,0)node{\footnotesize$X^{2014}$};
          \path(-1,-1)node{\footnotesize$X^{2005}$};
          \path(0,1)node{\footnotesize$X^{1120}$};
          \path(0,0)node{\footnotesize$X^{1111}$};
          \path(0,-1)node{\footnotesize$X^{1102}$};
          \path[red](1.3,-1)node{\footnotesize$X_2\!^2/X_4$};
          \path[red](1.3,-2)node{\footnotesize$X_2\!^2/X_3$};
          \path(.7,-2)node[left]
              {\small$\G\big(\cK^*_{\smash{\FF[2]3}} \<= 
                                  \cK^*_{\smash{\FF[2]{(3,0)}}}\big)$};
          \path(-.1,2.3)node[right]
              {\footnotesize$\deg\cK^*_{\smash{\FF[2]3}}=\pM{~~2\\{-}1}$};
          \path(-.1,3.7)node[right]
            {\footnotesize$\begin{array}[t]
                               {@{}r@{\:}|@{\:}c@{~}c@{~}c@{~}c@{}}
                             &X_1 &X_2 &X_3 &X_4\\ \toprule\nGlu{-2pt}
                            Q^1 &~~1& 1 & 0 & 0 \\
                            Q^2 &-3 & 0 & 1 & 1 \\
                           \end{array}$};
            }
 \caption{The $m=1,2,3$ sequence of anticanonical monomials, $\G(\cK^*_{\smash{\FF[2]m}})$, where ``$X^{abcd}$'' stands for
 $X_1\!^aX_2\!^bX_3\!^cX_4\!^d$}
 \label{f:F1-2-3}
\end{figure}
as do the left-most and middle columns of $\G(\cK^*_{\smash{\FF[2]3}})$-monomials, making evident the downward (red-ink) stacking of
 $X_2\!^2/X_4$ and $X_2\!^2/X_3$ for $m\<= 3$.
This inevitably flip-folded (red-ink) completion of this plot for $m\geqslant3$ reminds of Gell-Mann's completion of the spin-$\tfrac32$ baryon decuplet, which led to the discovery of the $\Omega^-$-baryon and the confirmation of the quark model of hadrons.

The {\em\/standard practice\/} in (complex algebraic) toric geometry restricts to regular sections (omitting $X_2\!^2/X_3$ and $X_2\!^2/X_4$),
which comes at a steep price: All regular anticanonical sections factorize:
\begin{equation}
 \mathrm{reg}\big[\G(\cK^*_{\smash{\FF[2]3}})\big]
  =(X_1)\big( \underbrace{ X_1(X_3{\oplus}X_4)^5 \oplus X_2(X_3{\oplus}X_4)^2}
                        _{C(X)}\big),
 \label{e:regK*F3}
\end{equation}
as already seein in~\eqref{e:qGen}.
Therefore, all Calabi--Yau hypersurfaces constructed from regular anticanonical sections~\eqref{e:K(nFm)} are necessarily Tyurin-degene\-rate\cite{tyurin2003fano} for $m\geqslant3$: They reduce to the union of two ``lobes'':
 the {\em\/directrix,} $\cD:=\{X_1=0\}$, and
 the {\em\/complementrix,} $\cC:=\{C(X)=0\}$, with
\begin{equation}
  C(X) \in \bigoplus\nolimits_{k=1}^n (X_1)^{k-1}\,
  (X_2{\oplus}\cdots{\oplus}X_n)^{n-k}\, (X_{n+1}{\oplus}X_{n+2})^{2+m(k-1)}.
\end{equation}
Thus, all Calabi--Yau hypersurfaces,
 $(\XX{m}=\cD\cup\cC)\subset\FF{m}[c_1]$ for $m\geqslant3$, 
are always singular (at least) at the intersection of the two lobes,
\begin{equation}
   \mathrm{Sing}(\XX{m})=\cD\cap\cC \in
   \K[{r||c|cc}{\IP^n&1&~~1&n{-}1\\ \IP^1&m&{-}m&2\\}],
\end{equation}
which is itself a Calabi--Yau $(n{-}2)$-fold.
With only the $k>0$ sections~\eqref{e:K(nFm)} available to the standard practice in complex algebraic toric geometry, all these Tyurin-degenerate Calabi--Yau $(n{-}1)$-folds are deemed {\em\/unsmoothable\/}.

\subsection{Smoothing the Unsmoothable}
\label{s:SuS}
Only the $k\<= 0$ monomials listed in~\eqref{e:K(nFm)}, $(X_2{\oplus}\cdots{\oplus}X_n)^n\, (X_{n+1}{\oplus}X_{n+2})^{2-m}$, are independent of $X_1$ and so do not vanish along the directrix. Deforming a \eqref{e:regK*F3}-section by a $k\<= 0$ section then moves the so-deformed zero-locus away from
 $\mathrm{Sing}(\XX{m})=\cD\cap\cC$, and direct computation verifies that the generic so-deformed section is {\em\/transverse.}
 The price to pay for smoothing the ({\em\/algebraically\/}) unsmoothable hypersurface is the fact that the deformation monomials are {\em\/rational\/} for $m\<\geqslant 3$, so that one must address the putative pole locations, $(X_{n+1}{\oplus}X_{n+2})=0$.

\paragraph{Intrinsic Limit:}
In its simplest version\cite{Berglund:2022dgb}, consider specifying the zero-locus
\begin{equation}
  f(X) = X_1\!^2X_3\!^5+X_1\!^2X_4\!^5+X_2\!^2/X_4\overset!=0.
 \label{e:LZL}
\end{equation}
Its gradient,
 $\big( 2X_1X_3\!^5{+}2X_1X_4\!^5,\; \frac{2X_2}{X_4},\;
        5X_1\!^2X_3\!^4,\; 5X_1\!^2X_4\!^4-\frac{X_2\!^2}{X_4\!^2} \big)$,
can vanish only if $X_1{=}0{=}X_2$, which cannot happen\footnote{Having changed variables~\eqref{e:x2psx} from $(x_0,\,x_1,\,x_2)\in\IP^2$, which must not all vanish, to $(p_0,\Fs,x_2)$ where $\FF[2]m\!:=\!\{p_0{=}0\}$, implies that $\Fs{=}X_1$ and $x_2{=}X_2$ must not both vanish.} in $\FF[2]m$.
Away from the $X_4=0$ putative pole-locus, we may solve
\begin{equation}
\eqref{e:LZL}\To\quad
  X_2 = i\sqrt{(X_1\!^2X_3\!^5+X_1\!^2X_4\!^5)X_4}.
 \label{e:LZLs}
\end{equation}
Since the unqualified limit $\lim_{X_4\to0}X_2\!^2/X_4$ in~\eqref{e:LZL} is not well defined, the zero-locus, $\{f(X){=}0\}$, would appear to be undefined at its intersections with $\{X_4=0\}$. A \eqref{e:LZLs}-constrained, ``intrinsic limit'' however clearly {\em\/is\/} well defined:
Away from $\{X_4{=}0\}$, the zero-locus~\eqref{e:LZL} is well defined as usual in algebraic geometry, while the points intersecting $\{X_4{=}0\}$ are defined by \eqref{e:LZLs}-constraining $X_2$ in the $X_4\to0$ limit.
 This ``intrinsic limit'' reminds of L'Hopital's theorem, and in that sense veers outside the standard complex algebro-geometric framework.

\paragraph{Alternatives:}
There also exist at least two alternative ways of specifying the zero-locus~\eqref{e:LZL}, which are decidedly algebro-geometric, albeit considerably more involved and less well explored:
 Setting $X_1\to1$ (away from the directrix, $\{X_1{=}0\}$) and
 writing $f(X)=(X_3\!^5X_4+X_4\!^6+X_2\!^2)/X_4$ defines the {\em\/divisor classes\/}:
\begin{subequations}
 \label{e:DivRef}
\begin{alignat}9
 \big[f_{\sss R}^{-1}(0)\big]
  &\rlap{$=\big[\big\{\Fn(x)/\Fd(x)=0\big\}\big]
   =\big[\Fn^{-1}(0)\big]{-}\big[\Fd^{-1}(0)\big],$}
 \label{e:5=6-1}
\intertext{where}\nGlu{-4mm}
 \Fn(x)      &:= X_3\!^5X_4 +X_4\!^6 +X_2\!^2,&\qquad
 \Fn^{-1}(0) &\in\IP^2_{(3{:}1{:}1)}[6], \label{e:N6}\\
 \Fd(x)      &:= X_4,&\qquad
 \Fd^{-1}(0) &\in\IP^2_{(3{:}1{:}1)}[1] \label{e:D1}
\end{alignat}
\end{subequations}
Formal {\em\/differences\/} (as opposed to unions or intersections) of subvarieties were introduced a century ago as {\em\/virtual varieties\/} by F.~Severi~\cite{rSeveri-VirtVar}, and are by now known as {\em\/Weil divisors\/}~\cite{rF-TV,rGE-CCAG,rCLS-TV}. Note that
\begin{equation}
   c_1\big(\mathfrak{n}^{-1}(0)\big)\<= (3{+}1{+}1){-}6\<= {-}1\<< 0,
\end{equation}
so the $\{\Fn(x){=}0\}$ hypersurface and the divisor class it represents are {\em\/of general type\/}; in turn, 
$c_1(\Fd^{-1}(0)) = (3{+}1{+}1){-}1\<= 4\<> 0$ and $\{\Fd(x)\<= 0\}$ is Fano.

Separately, David Cox has found that a simple-loking fractional change of variables,\footnote{I wish to thank Hal Schenck for communicating this to me, which in turn motivated searching (and finding) generalizations for all $\FF{m}[c_1]$ hypersurfaces.} recasts the $X_1{=}1$ simplification of~\eqref{e:LZL}, the Calabi--Yau Laurent hypersurface
 $\{f(X)_{\sss X_1{=}1}\<= 0\}\<\in \IP^2_{(3{:}1{:}1)}[5]$
\begin{equation}
  (X_2,X_3,X_4)\mapsto(z_3\sqrt{z_2},z_1\!^2,z_2)
  ~\To~
  f(X)_{X_1=1}\to h(z) =z_1\!^{10}+z_2\!^5+z_3\!^2,
 \label{e:Cox}
\end{equation}
as (the branched double-cover of a $\ZZ_2$-quotient of) the regular hypersurface,
 $\{h(z)\<= 0\}\in\IP^2_{(1{:}2{:}5)}[10]$ which is an algebraic variety {\em\/of general type\/}: its 1st Chern class is negative, $c_1\big(h^{-1}(0)\big)=(1{+}2{+}5){-}10=-2$. Of course, the change of variables~\eqref{e:Cox} has a non-constant Jacobian and is not regular: Its inverse involves {\em\/division,}
$(z_1,z_2,z_3)\<\to (\pm\sqrt{X_3},X_4,X_2/\sqrt{X_4})$, which may be seen as introducing the rational monomial in~\eqref{e:LZL}.

Either way, the Laurent deformations~\eqref{e:K(nFm)} of ``unsmoothable'' Tyurin-degenerate hypersurfaces~\eqref{e:regK*F3}, such as~\eqref{e:LZL}, do specify {\em\/smooth\/} Calabi--Yau models, albeit rather more involved than the regular hypersurfaces of the standard practice in complex algebraic and toric geometry.

\section{Deformation Family Picture}
\label{s:DFP}
We now turn to explore the explicit deformation family~\eqref{e:HFme} of Hirzebruch scrolls embedded as hypersurfaces, e.g., and examine a few simple examples of $\FF[4]5\<\in \ssK[{r||c}{\IP^4&1\\ \IP^1&5}]$:
\begin{alignat}9
 \begin{array}[t]{r@{\,}l}
 p_0(x,y) &\underset{\CW{\:\To}}= x_0y_0\!^5 {+}x_1y_1\!^5,\\[-5pt]
 \Fs_0(x,y) &=\big[\big(\frac{x_0}{y_1\!^5}{-}\frac{x_1}{y_0\!^5}\big)
                   \big/\!\!\big/ p_0(x,y)\big];
 \end{array}
 ~&\To~~
  \begin{array}[t]{@{}c@{\:}|@{\:}c@{~}c@{~}c@{~}c@{~}c@{~}c@{}}
     &X_1 &X_2 &X_3 &X_4 &X_5 &X_6 \\
 \toprule\nGlu{-2pt}
 Q^1 &~~1  &  1  &  1 &  1  &  0      &  0  \\[-2pt]
 Q^2 & -5  &  0  &  0 &  0  &  1      &  1  \\ 
  \end{array}
 ~&&\To \FF[4]5, 
 \label{e:F5}
\intertext{where ``$A/\!\!/B$'' abbreviates ``$A~(\mathrm{mod}~B)$.'' Including some of the 
$\e_{a,\ell}$-deformations produces (absorbing $\e_{a,\ell}$'s in~\eqref{e:HFme} into the $\IP^4_x$-coordinates)\footnote{This corrects the typos in the definitions of $p_1(x,y)$ and $p_2(x,y)$ in\cite{Berglund:2022dgb}.}:}
 \begin{array}[t]{r@{\,}l}
 p_1(x,y) &\underset{\CW{\:\To}}
           = x_0y_0\!^5 {+}x_1y_1\!^5 {+}x_2y_0\!^4y_1,\\[-5pt]
 \Fs_{1,1}(x,y) &=\big[\big(\frac{x_0y_0}{y_1\!^5}
                            {+}\frac{x_2}{y_1\!^4}
                            {-}\frac{x_1}{y_0\!^4}\big)
                   \big/\!\!\big/ p_1(x,y)\big],\\[2pt]
 \Fs_{1,2}(x,y) &=\big[\big(\frac{x_0}{y_1}
                            {-}\frac{x_2}{y_0}
                            {-}\frac{x_1y_1\!^4}{y_0\!^5}\big)
                   \big/\!\!\big/ p_1(x,y)\big];
 \end{array}
 ~&\To~~
  \begin{array}[t]{@{}c@{\:}|@{\:}c@{~}c@{~}c@{~}c@{~}c@{~}c@{}}
     &X_1 &X_2 &X_3 &X_4 &X_5 &X_6 \\
 \toprule\nGlu{-2pt}
 Q^1 &~~1  &~~1  &  1 &  1  &  0      &  0  \\[-2pt]
 Q^2 & -4  & -1  &  0 &  0  &  1      &  1  \\ 
  \end{array}
 ~&&\To \FF[4]{\!\sss(4,1,0,0)}.
 \label{e:F41}\\[2mm]
 \begin{array}[t]{r@{\,}l}
 p_2(x,y) &\underset{\CW{\:\To}}
           = x_0y_0\!^5 {+}x_1y_1\!^5
                            {+}x_2y_0\!^3y_1\!^2,\\[-5pt]
 \Fs_{2,1}(x,y) &=\big[\big(\frac{x_0y_0\!^2}{y_1\!^5}
                            {+}\frac{x_2}{y_1\!^3}
                            {-}\frac{x_1}{y_0\!^3}\big)
                   \big/\!\!\big/ p_2(x,y)\big],\\[2pt]
 \Fs_{2,2}(x,y) &=\big[\big(\frac{x_0}{y_1\!^2}
                            {-}\frac{x_2}{y_0\!^2}
                            {-}\frac{x_1y_1\!^3}{y_0\!^5}\big)
                   \big/\!\!\big/ p_2(x,y)\big];
 \end{array}
 ~&\To~~
  \begin{array}[t]{@{}c@{\:}|@{\:}c@{~}c@{~}c@{~}c@{~}c@{~}c@{}}
     &X_1 &X_2 &X_3 &X_4 &X_5 &X_6 \\
 \toprule\nGlu{-2pt}
 Q^1 &~~1  &~~1  &  1 &  1  &  0      &  0  \\[-2pt]
 Q^2 & -3  & -2  &  0 &  0  &  1      &  1  \\
  \end{array}
 ~&&\To \FF[4]{\!\sss(3,2,0,0)}.
  \label{e:F32}\\[2mm]
 \begin{array}[t]{r@{\,}l}
 p_3(x,y) &\underset{\CW{\:\To}}= x_0y_0\!^5 {+}x_1y_1\!^5
                            {+}x_2y_0\!^3y_1\!^2 
                            {+}x_3y_0\!^2y_1\!^3,\\[-5pt]
 \Fs_{3,1}(x,y) &=\big[\big(\frac{x_0}{y_1\!^2}
                            {-}\frac{x_1 y_1\!^3}{y_0\!^5}
                            {-}\frac{x_2}{y_0\!^2} 
                            {-}\frac{x_3 y_1}{y_0\!^3}\big)
                   \big/\!\!\big/ p_3(x,y)\big],\\[2pt]
 \Fs_{3,2}(x,y) &=\big[\big(\frac{x_0 y_0\!^3}{y_1\!^5}
                            {+}\frac{x_2 y_0}{y_1\!^3}
                            {+}\frac{x_3}{y_1\!^2}
                            {-}\frac{x_1}{y_0\!^2}\big)
                   \big/\!\!\big/ p_3(x,y)\big],\\[2pt]
 \Fs_{3,3}(x,y) &=\big[\big(\frac{x_0 y_0\!^2}{y_1\!^3}
                            {+}\frac{x_2}{y_1}
                            {-}\frac{x_1 y_1\!^2}{y_0\!^3}
                            {-}\frac{x_3}{y_0}\big)
                   \big/\!\!\big/ p_3(x,y)\big];
 \end{array}
 ~&\To~~
  \begin{array}[t]{@{}c@{\:}|@{\:}c@{~}c@{~}c@{~}c@{~}c@{~}c@{}}
     &X_1 &X_2 &X_3 &X_4 &X_5 &X_6 \\[-2pt]
 \toprule\nGlu{-2pt}
 Q^1 &~~1  &~~1  &~~1 &  1  &  0      &  0  \\[-2pt]
 Q^2 & -2  & -2  & -1 &  0  &  1      &  1  \\
  \end{array}
 ~&&\To \FF[4]{\!\sss(2,2,1,0)}.
   \label{e:F221}
\end{alignat}%
The deformations~\eqref{e:F41}--\eqref{e:F221} represent distinct but still non-Fano toric varieties, with $c_1\propto\pM{~~4\\-3}$. However, the so-constructed network of deformations eventually leads to 
 $\FF[4]{\!\sss(2,1,1,1)}\<{\approx_\IR}\FF[4]{\!\sss(1,0,0,0)}=\FF[4]{\,1}$, which is Fano ($c_1>0$).
 
 This is simpler to see in the 3-dimensional reduction of~\eqref{e:F221}, obtained by setting $\IP^4{\supset}\{X_4{=}x_4{=}0\}\<= \IP^3$. This data then specifies
 $\FF[3]{\!\sss(2,2,1)}$, which is {\em\/diffeomorphic\/} to $\FF[3]{\!\sss(1,1,0)}$ --- and which is semi-Fano, $c_1\propto\pM{4\\0}\geqslant0$.
 This may be seen from the Segr\'e-like change of $\IP^3_x\<\times\IP^1_y$ coordinates:
\begin{subequations}
 \label{e:Chng}
\begin{equation}
  \big( x_0 y_0\!^3,\, x_1 y_1\!^3,\, x_2 y_0\!^2y_1,\, x_3 y_0y_1\!^2;\, 
               y_0,\, y_1 \big)
  =(\xi_0,\, \xi_1,\, \xi_2,\, \xi_3;\, \eta_0,\, \eta_1),
 \label{e:Segre}
\end{equation}
which maps
 $\ssK[{r||c}{\IP^3&1\\\IP^1&5}]\to\ssK[{r||c}{\IP^3&1\\\IP^1&2}]$ by changing the defining equation:
\begin{equation}
  p_3(x,y) \to 
  (\xi_0{+}\xi_3)\eta_0\!^2 {+}(\xi_1{+}\xi_2)\eta_1\!^2.
 \label{e:S5>2}
\end{equation}
\end{subequations}
However, the Segr\'e-like change of variables~\eqref{e:Segre} has a non-constant Jacobian,
$\det\!\big[\frac{\vd(\xi,\eta)}{\vd(x,y)}\big]\<= (y_0y_1)^6$, which vanishes at both ``poles,'' $y_0{=}0$ and $y_1{=}0$,
at each of which the hypersurface $\{p_3(x,y)\,{=}\,0\}$ has a 
 $\IP^2\,{\subset}\,\IP^3_x$ hyperplane.
 There, additional ({\em\/non-holomorphic\/} but smooth) partitions of unity/``bump-function'' modifications are required to locally ``repair'' the transformation~\eqref{e:S5>2} into a proper
 $\FF[3]{\!\sss(2,1,1)}\<{\approx_\IR}\FF[3]{\!\sss(1,0,0)}=\FF[3]1$ {\em\/diffeomorphism\/}. In the corresponding GLSM, this merely redefines the $U(1)^2$-gauge symmetry charge basis, $(Q^1,Q^2)\<\to(Q^1,Q^1{+}Q^2)$, identical to the
 $\GL(2,\ZZ)$ mapping
 $H^2(\FF[3]{\!\sss(2,1,1)},\ZZ)\to H^2(\FF[3]{\!\sss(1,0,0)},\ZZ)$ of the Chern (curvature) classes.

In two dimensions, similarly deforming Hirzebruch's original defining equation\cite{rH-Fm}, $p_0(x,y)=x_0y_0\!^3{+}x_1y_1\!^3$ as another simple example, results in:
\begin{equation}
  \K[{r||c}{\IP^2&1\\ \IP^1&3}]: \qquad
 \begin{array}{@{}r@{\,}l}
 p_1(x,y) &\underset{\CW{\:\To}}
           = x_0y_0\!^3 {+}x_1y_1\!^3
                            {+}x_2y_0\!^2y_1,\\[-5pt]
 \Fs_{1,1}(x,y) &=\big[\big(\frac{x_0 y_0}{y_1\!^3}
                            {-}\frac{x_1}{y_0\!^2}
                            {+}\frac{x_2}{y_1\!^2}\big)
                   \big/\!\!\big/ p_1\big],\\[2pt]
 \Fs_{1,2}(x,y) &=\big[\big(\frac{x_0}{y_1}
                            {-}\frac{x_1 y_1\!^2}{y_0\!^3}
                            {-}\frac{x_2}{y_0}\big)
                   \big/\!\!\big/ p_1\big].
 \end{array}
 \label{e:2F21a}
\end{equation}
This yields the toric rendition,
\begin{equation}
 \begin{matrix}
  \FF[2]{\!\sss(2,1)} \\[4mm]
  \begin{array}[t]{@{}c|c@{~}c@{~}c@{~}c@{~}c@{~}c@{}}
     &X_1 &X_2 &X_3 &X_4 \\
 \toprule\nGlu{-2pt}
 Q^1 &~~1  &~~1  &  0      &  0  \\[-2pt]
 Q^2 & -2  & -1  &  1      &  1  \\ 
  \end{array}
  \end{matrix}
 \To
 \vC{\TikZ{[xscale=.9, yscale=.7]\path[use as bounding box](-3,-1.2)--(3,2);
          \draw[blue, thick, densely dotted]
              (-2,2)--(2,0)--(2,-1)--(-2,-1)--cycle;
          \path(-2,2)node{\footnotesize$X_1\!^2X_3\!^3$};
          \path(-2,1)node{\footnotesize$X_1\!^2X_3\!^2X_4$};
          \path(-2,0)node{\footnotesize$X_1\!^2X_3X_4\!^2$};
          \path(-2,-1)node{\footnotesize$X_1\!^2X_4\!^3$};
          \path(0,1)node{\footnotesize$X_1X_2X_3\!^2$};
          \path(0,0)node{\footnotesize$X_1X_2X_3X_4$};
          \path(0,-1)node{\footnotesize$X_1X_2X_4\!^2$};
          \path(2,0)node{\footnotesize$X_2\!^2X_3$};
          \path(2,-1)node{\footnotesize$X_2\!^2X_4$};
          \path(-.5,2)node[right]{$\cK^*_{\smash{\FF[2]{(2,1)}}}
                                     =\cK^*_{\smash{\FF[2]1}}$};
            }}
 \label{e:2F21b}
\end{equation}
where the stack of anticanonical monomials~\eqref{e:2F21b} is clearly identical to $\G(\cK^*_{\smash{\FF[2]1}}{}_{\strut})$ displayed on the left of Figure~\ref{f:F1-2-3}.

The above computations explicitly demonstrate that the diffeomorphisms such as
 $\FF[2]{\!\sss(2,1)}\<{\approx_\IR}\FF[2]1$ and
 $\FF[3]{\!\sss(2,2,1)}\<{\approx_\IR}\FF[3]{\!\sss(1,1,0)}$
are directly relevant for providing sections with which to construct Calabi--Yau hypersurfaces.

\subsection{Two Conjectures}
\label{s:2C}
The foregoing analysis demonstrates that the explicitly $\e_{a,\ell}$-parametrized deformation family $\ssK[{r||c}{\IP^3&1\\\IP^1&m}]$ of smooth manifolds~\eqref{e:HFme} includes 
 ({\small\bf0})~the (non-Fano for $m\geqslant3$) Hirzebruch scroll $\FF{m}$ at the $\e_{a,\ell}=0$ center,
 ({\small\bf1})~all toric versions of Hirzebruch scrolls, 
 $\FF{\sss\ora{m}}$ where $\ora{m}$ is an $n$-tuple of nonnegative integral twists with a constant sum of components
 ($|m|\!:=\!\sum_i m_i\<= m$),
 ({\small\bf2})~hypersurfaces at generic 
 $\e$-locations within the same $\e$-deformation family,
 $\ssK[{r||c}{\IP^n&1\\ \IP^1&m}]$, that are {\em\/diffeomorphic\/} to their (semi-)Fano cousin, $\FF{\!\smash{m\,(\mathrm{mod}\,n)}}$.

This fact then prompts a sweeping extrapolation:
\label{sC:1}
\begin{conj}\label{C:1}
All toric varieties and so also their Calabi--Yau hypersurfaces\cite{Kreuzer:2000xy, rKS-CY} can be found, in a similar manner, as more or less particularly chosen (perhaps generalized) complete intersections in products of projective spaces\cite{rH-CY0, rGHCYCI, rCYCI1, rgCICY1, rBH-Fm, rGG-gCI}.
\end{conj}
All toric varieties encoded by convex, reflexive polytopes\cite{Kreuzer:2000xy, rKS-CY} {\em\/are\/} projective, i.e., can be embedded in some large enough projective space. That they ---along with their variants corresponding to the much more general VEX multitopes--- may in fact all be embedded as complete intersections in products of projective spaces is far from obvious, which raises the challenge of this sweeping conjecture.

Since generic-$\e$ members of the deformation family
 $\ssK[{r||c}{\IP^n&1\\ \IP^1&m}]$ admit regular smoothing of (purposefully chosen) Tyurin-degenerate Calabi--Yau hypersurfaces, while the $\e=0$ ``central'' Hirzebruch scroll has only ``un-smoothable'' Tyurin-degenerate Calabi--Yau hypersurfaces requiring Laurent deformations for smoothing, we conclude that:
\label{sC:2}
\begin{conj}\label{C:2}
Laurent smoothing deformations of ``un-smoothable'' Tyurin-degenerate Calabi--Yau $n$-folds\cite{rBH-gB} are smooth $\e\<\to 0$ limits of regular smoothing deformations of non-degenerate (and diffeomorphic) Calabi--Yau $n$-folds within the same explicit deformation family\cite{Berglund:2022dgb}.
\end{conj}
\noindent
These smoothing deformations clearly need not stay within the class and framework of (globally complex) {\em\/algebraic\/} constructions.

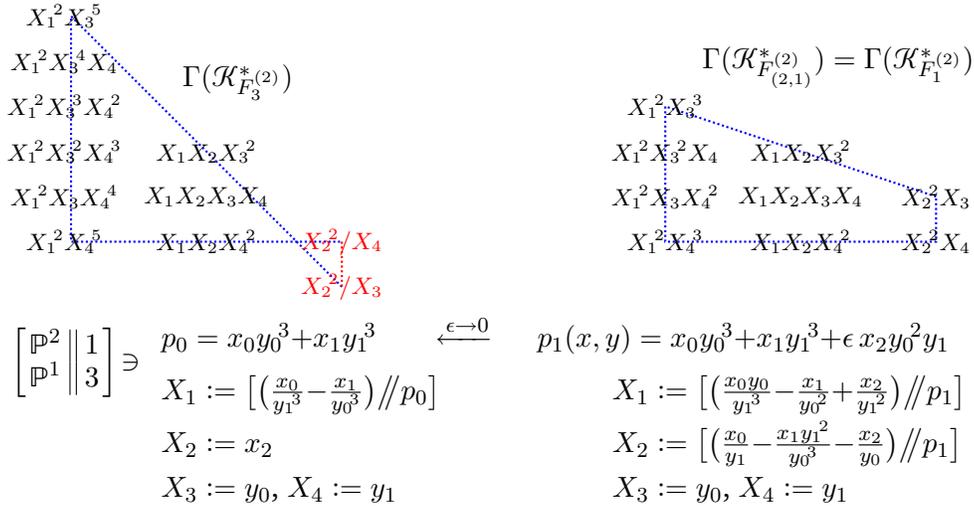
\begin{figure}[htb]
 \begin{center}\unitlength=1mm
  \begin{picture}(135,65)(-5,2)
   \put(0,17){$\K[{r||c}{\IP^2&1\\\IP^1&3}]\!\ni$}
   \put(20,20){$p_0=x_0y_0\!^3 {+}x_1y_1\!^3$}
   \put(20,13){$X_1:=\big[\big(\frac{x_0}{y_1\!^3}{-}\frac{x_1}{y_0\!^3}\big)
                           \big/\!\!\big/ p_0\big]$}
   \put(20,6){$X_2:= x_2$}
   \put(20,0){$X_3:= y_0$, $X_4:= y_1$}
   \put(8,25){\TikZ{[xscale=.9, yscale=.6]
          \path[use as bounding box](-2,-2.5)--(2,4.5);
          \draw[blue, thick, densely dotted](2,-2)--(-2,4)--(-2,-1)--(2,-1);
          \draw[red, thick, densely dotted](2,-1)--(2,-2);
          \path(-2.1,4)node{\footnotesize$X_1\!^2X_3\!^5$};
          \path(-2.1,3)node{\footnotesize$X_1\!^2X_3\!^4X_4$};
          \path(-2.1,2)node{\footnotesize$X_1\!^2X_3\!^3X_4\!^2$};
          \path(-2.1,1)node{\footnotesize$X_1\!^2X_3\!^2X_4\!^3$};
          \path(-2.1,0)node{\footnotesize$X_1\!^2X_3X_4\!^4$};
          \path(-2.1,-1)node{\footnotesize$X_1\!^2X_4\!^5$};
          \path(0,1)node{\footnotesize$X_1X_2X_3\!^2$};
          \path(0,0)node{\footnotesize$X_1X_2X_3X_4$};
          \path(0,-1)node{\footnotesize$X_1X_2X_4\!^2$};
          \path[red](2,-1)node{\footnotesize$X_2\!^2/X_4$};
          \path[red](2,-2)node{\footnotesize$X_2\!^2/X_3$};
          \path(-.5,2.6)node[right]{$\G(\cK^*_{\smash{\FF[2]3}})$};
            }}
   \put(70,20){\llap{$\fro{\e\to0}$\quad~~}%
               $p_1(x,y)=x_0y_0\!^3 {+}x_1y_1\!^3
                         {+}\e\,x_2y_0\!^2y_1$}
   \put(80,13){$X_1:=\big[\big(\frac{x_0 y_0}{y_1\!^3}
                            {-}\frac{x_1}{y_0\!^2}
                            {+}\frac{x_2}{y_1\!^2}\big)
                              \big/\!\!\big/ p_1\big]$}
   \put(80,6){$X_2:=\big[\big(\frac{x_0}{y_1}
                            {-}\frac{x_1 y_1\!^2}{y_0\!^3}
                            {-}\frac{x_2}{y_0}\big)\big/\!\!\big/ p_1\big]$}
   \put(80,0){$X_3:= y_0$, $X_4:= y_1$}
   \put(87,32.8){\TikZ{[xscale=.9, yscale=.6]
          \path[use as bounding box](-2,-1.2)--(2,2);
          \draw[blue, thick, densely dotted]
              (2,0)--(-2,2)--(-2,-1)--(2,-1)--cycle;
          \path(-2,2)node{\footnotesize$X_1\!^2X_3\!^3$};
          \path(-2,1)node{\footnotesize$X_1\!^2X_3\!^2X_4$};
          \path(-2,0)node{\footnotesize$X_1\!^2X_3X_4\!^2$};
          \path(-2,-1)node{\footnotesize$X_1\!^2X_4\!^3$};
          \path(0,1)node{\footnotesize$X_1X_2X_3\!^2$};
          \path(0,0)node{\footnotesize$X_1X_2X_3X_4$};
          \path(0,-1)node{\footnotesize$X_1X_2X_4\!^2$};
          \path(2,0)node{\footnotesize$X_2\!^2X_3$};
          \path(2,-1)node{\footnotesize$X_2\!^2X_4$};
          \path(-1.6,3)node[right]{$\G(\cK^*_{\smash{\FF[2]{(2,1)}}})
                                     =\G(\cK^*_{\smash{\FF[2]1}})$};
            }}
  \end{picture}
 \end{center}
 \caption{Change in the system of anticanonical sections used to define the Calabi--Yau hypersurface; the $\FF[2]3$-system on the left is over-complete, but factorizes without the rational monomials.}
 \label{f:2F3F1}
\end{figure}

As a particularly simple template for the latter conjecture, note that the deformation family of deg-$\pM{1\\3}$ hypersurfaces in $\IP^2{\times}\IP^1$ contains two particular Hirzebruch scrolls, which have rather different collections of anticanonical sections, collected from above in Figure~\ref{f:2F3F1}.
For $\e_{a,\ell}\neq0$, the deg-$\pM{1\\3}$ hypersurfaces in $\IP^2{\times}\IP^1$ defines the complex algebraic surface with the anticanonical monomials plotted in the upper-right section of Figure~\ref{f:2F3F1}: they are all regular, and their generic linear combination is transverse.
As $\e_{a,\ell}\to0$, the deg-$\pM{1\\3}$ hypersurfaces in $\IP^2{\times}\IP^1$ becomes the complex algebraic surface with the anticanonical monomials plotted in the upper-left section of Figure~\ref{f:2F3F1}: the generic linear combination of the regular monomials not only fails to be transverse but clearly factorizes; only the inclusion of the rational monomials recovers transversality. Since
 $\dim H^0(\FF[2]3,\cK^*)=9$, the displayed system of sections is over-complete, but also admits additional equivalence relations\cite{rCK}, allowing the rational monomials to ``take the place'' of some of the regular ones in constructing Calabi--Yau hypersurfaces. In this sense, the monomials 
 $X_2\!^2X_3$ and $X_2\!^2X_4$ on the Fano surface 
 $\FF[2]{\!\sss(2,1)}\mathop{\approx_{\IR}}\FF[2]1$ map, in the 
 $\e_{a,\ell}\<\to 0$ limit, to the rational monomials 
 $X_2\!^2/X_3$ and $X_2\!^2/X_4$ on $\FF[2]3$.
 
 More generally and for the purposes of constructing smooth Calabi--Yau hypersurfaces,
 the rational monomials on $\FF{m}$ for $m\geqslant3$
 play the role of certain regular monomials in $\FF{\sss\ora{m}}$, a sufficiently generic deformation of the ($\e_{a,\ell}\<= 0$) ``central'' $\FF{m}$ within the same deformation family,
 $\ssK[{r||c}{\IP^n&1\\ \IP^1&m}]$.

\subsection{Are These New?}
\label{s:new}
Given the huge database of 473\,800\,776 reflexive polytopes, many of which admit several distinct triangulations, each encoding a distinct toric variety and each of which defining a ({\em\/continuous\/}) deformation family of Calabi--Yau 3-folds\cite{Kreuzer:2000xy}, the obvious question is whether the above-described constructions offer anything new.

To this end, note that all (generalized) complete intersection Calabi--Yau 3-folds~\eqref{e:gCIXm} have the same Hodge numbers, but form four distinct {\em\/diffeomorphism\/} types, \eqref{e:Gang4}, i.e., four distinct {\em\/real (smooth) manifolds.} The first three of these are regular CICYs and so were long since known; the last one, $\XX[3]3$, does not appear to have been known. Furthermore, whereas all (classical) diffeomorphism invariants of $\XX[3]{m}$ and $\XX[3]{m{-}4}$ agree\cite{Berglund:2024zuz}, this equivalence fails for subtler properties of $H^{2k}(\XX[3]m,\ZZ)$ such as quantum cohomology and the K\"ahler/Fayet--Illiopoulos phase space\footnote{The distinction already shows at the ``semi-classical'' level of analysis\cite{rBH-gB}, but is even more evident upon including the worldsheet instanton effects\cite{rMP0}.} and so ought to be physically distinguishable.

On the construction side, the very need to use the L'Hopital rule-like ``intrinsic {\em\/limit,}'' as discussed around Eqs.~\eqref{e:LZL}--\eqref{e:LZLs}, indicates a significant departure from the otherwise three-decades standard {\em\/purely algebraic\/} constructions in the field, and a non-algebraic nature  of the so-smoothed Calabi--Yau hypersurfaces in $\FF{m}$ for $m\geqslant3$. Whereas the firmly algebro-geometric recasting of such models as discussed around~\eqref{e:DivRef}--\eqref{e:Cox} may prove sufficient in some sample cases or as regards some of the relevant aspects, further evidence noted below seems to corroborate the need to extend the framework beyond the standard complex algebraic toric geometry\cite{rD-TV, rO-TV, rF-TV, rGE-CCAG, rCLS-TV, rCK}, so as to include non-convex, self-crossing, folded and otherwise multi-layered (but still VEX\cite{rBH-gB, Berglund:2022dgb, Berglund:2024zuz}) multitopes\cite{rM-MFans, Masuda:2000aa, rHM-MFs, Masuda:2006aa, rHM-EG+MF, rH-EG+MFs2, Nishimura:2006vs, Ishida:2013ab, Davis:1991uz, Ishida:2013aa, buchstaber2014toric, Jang:2023aa}.

Indeed, multifans and multitopes (even the VEX ones) correspond to a class of so-called (unitary) {\em\/torus manifolds}\cite{rM-MFans, Masuda:2000aa, rHM-MFs, Masuda:2006aa} which do include complex algebraic toric varieties (when its multitope is an ordinary, convex polytope) but are incommensurately more general and abundant: Most of (real $2n$-dimensional) torus manifolds do not admit a global complex structure, nor are they algebraic varieties. However, they all admit a (maximal) action of $(S^1)^n$ --- which is in 1--1 correspondence with the $U(1)^n$ gauge symmetry of a corresponding GLSM. In torus manifolds that are not complex algebraic toric varieties, a complexified $(S^1)^n\<\to (\IC^*)^n$-action (as would be required in a GLSM with at least $(0,2)$-supersymmetry on the worldsheet) is not integrable or is otherwise obstructed.

This reminds of the following facts\cite{rJPS, rGSW2, rBBS}:
 ({\small\bf1})~Worldsheet $(0,1)$-super\-sym\-metry suffices to eliminate the stringy tachyon instability and guarantee the existence of a finite-energy ground state; for such {\em\/maximally generic\/} models, see\cite{rUDSS08, rUDSS09}.
 ({\small\bf2})~Worldsheet $(0,2)$-supersymmetry is required for target space supersymmetry and the complexification of the $(S^1)^n$ gauge symmetry to a complex-algebraic toric $(\IC^*)^n$-action.
 ({\small\bf3})~String theory compactification\cite{rCHSW, rBeast}, a priori, does {\em\/not require\/} the compact Calabi--Yau space be an algebraic variety, nor must it be a K\"ahler manifold\cite{Strominger:1986uh, Hull:1986kz}.
 ({\small\bf4})~Since a K\"ahler class is a non-degenerate symplectic 2-form twisted by a compatible complex structure, it follows that in spaces wherein a putative K\"ahler class is obstructed or otherwise defective, the symplectic and/or the complex structure must be analogously obstructed or defective.
 These observations seem to indicate:
\label{sC:01dS}
\begin{conj}\label{C:01dS}
GLSMs with worldsheet $(0,1)$-supersymmetry provide for consistent superstring models, which need not have target-space supersymmetry and so can include (asymptotically) de~Sitter spacetime geometries.
\end{conj}
\label{sC:VEX}
\begin{conj}\label{C:VEX}
Ground state spaces, $Z$, of such GLSMs may be modeled by ``toral'' (at least 
$(S^1)^n$-equivariant) hypersurfaces in torus manifolds, amongst which the by now well-studied (com\-plex-algebraic) toric constructions\cite{rPhases, rPhasesMF, rMP0, Distler:1993mk, rMP1, Schafer-Nameki:2016cfr, Sharpe:2024dcd} are special cases, without such defects obstructing (super)sym\-metry, and the complex and symplectic (and K\"ahler) structures of $Z$. 
\end{conj}

Torus manifolds are incommensurately more abundant than complex-algebraic toric varieties,\footnote{Dubbed ``frigid toric crystals'' by Dolgachev\cite[p.\,100]{rD-TV}, one may well picture toric varieties as isolated specks in a continuum of torus manifolds.} and so are then these ``toral'' hypersurfaces as compared to the complex-algebraic subvarieties of toric varieties. In turn, nothing in string theory a priori requires the vacua to be complex-{\em\/algebraic\/} varieties; our focus on those by now well familiar constructions is evidently correlated with computational accessibility afforded by the underlying worldsheet $(0,2)$- and even more so $(2,2)$-superconformal quantum field theories.

In apparent correspondence, the class of VEX multitopes (and embeddings of Calabi--Yau spaces in torus manifolds) is in fact infinitely large. It is thus actually {\em\/to be expected\/} that the GLSMs corresponding to Laurent defining polynomials such as~\eqref{e:LZL}, i.e., generic combination of monomials on the right-hand side of Figure~\ref{f:F1-2-3} cannot be ``made regular'' (and already known) Ricci-flat models by {\em\/regular\/} variable changes. (Cox's change of variables~\eqref{e:Cox} is neither regular nor are the resulting hypersurface Ricci-flat; there $\IP^2_{(1{:}2{:}5)}[10]$ has $c_1\<< 0$.) In turn, even with additional variables specifically introduced to recast, e.g.,~\eqref{e:LZL} as a regular polynomial, one ends up requiring $U(1)^n$-neutral variables, which introduces non-compact directions of ground states. While I am not aware of a rigorous proof, the next section presents indications that the inclusion of Laurent-deformed defining equations in non-Fano toric varieties corresponds to GLSMs that cannot be recast in that by now well-studied and familiar form\cite{rPhases, rPhasesMF, rMP0, Distler:1993mk, rMP1, Schafer-Nameki:2016cfr, Sharpe:2024dcd}, and so are indeed new.\footnote{This line of reasoning elaborates on a discussion with Ilarion Melnikov.} Finally, note that the semi-classical ``phase spaces'' $\S''(\FF{m})$ in~\eqref{e:2ndFn} being manifestly different for all different $m$ corroborates this, and implies that the corresponding GLSMs and string compactifications also differ, and in physically relevant ways.

\subsection{Extending Gauged Linear Sigma Models}
 \label{s:xGLSM}
The hallmark feature of the $m\<\geqslant 3$ models with the full complement of anticanonical sections~\eqref{e:K(nFm)} is the inclusion of rational monomials in the defining equations --- to which many standard methods of algebraic geometry and cohomological algebra are straightforward to adapt\cite{rgCICY1, rBH-Fm, rGG-gCI, rBH-gB, Berglund:2022dgb, Berglund:2024zuz}. For applications in string theory and its M- and F-theory extensions, it is desirable to find a world-sheet field theory model with such target spaces, and so compute semi-classical and quantum data.

For over three decades now, the standard vehicle to this end is Witten's gauged linear sigma model (GLSM)\cite{rPhases, rPhasesMF, rMP0, Distler:1993mk, rMP1, Schafer-Nameki:2016cfr, Sharpe:2024dcd}, where the space of ground-states is defined as the subset of the field space where a well-specified potential is minimized, at the intersection
\begin{equation}
   \bigg\{\!\!\underbrace{\frac{\partial W}{\partial x_i}}_{F\text{-terms}}
           \!\!=0\bigg\} \cap
   \bigg\{\!\underbrace{\sum\nolimits_iq^a_i\,|x_i|^2\!-\!r_a}
                    _{D\text{-terms}}=0\bigg\} \cap
   \bigg\{\!\underbrace{\sum\nolimits_{\smash{a,b,i}}
                      \bar\sigma_a\,\sigma_b\,q^a_i\,q^b_i\,|x_i|^2}
                     _{\sigma\text{-terms}}=0\bigg\}.
 \label{e:GndS}
\end{equation}
Here, $W(X)$ is the {\em\/superpotnential\/} function of chiral $(2,2)$-superfields\footnote{This brief discussion is limited to the simplest, $(2,2)$-supersymmetric case; $(0,2)$-generalizations are described in the literature\cite{Distler:1993mk, Schafer-Nameki:2016cfr, Sharpe:2024dcd}, while the ultimate, minimally $(0,1)$-supersymmetric extensions remain largely unexplored.}, $X_i$, with lowest component fields $x_i$. They couple to the twisted-chiral superfields, $\S_a$ (with lowest component fields $\s_a$), with $U(1)$-gauge charges $q^a_i$, identifiable with components of the Mori vectors such as~\eqref{e:Mori}.
 Being gauge-invariant, the superpotential $W(x)$ is quasi-homogeneous and the vanishing of $\pd{W}{x_i}$ implies that $W(x)=0$ also. The first of these factors defines the {\em\/base-locus\/} of the superpotential, the second factor assigns subsets of this base-locus to specific values of (Fayet-Iliopoulos parameters) $r_a$, while the third factor determines the $\s_a$ as depending on the $r_a$-specified subsets of the base-locus.

We are interested in single-hypersurface superpotentials, of the form
 $W(X)=X_0{\cdot}f(X_1,\dots)$ and with $f(X)$ a Laurent polynomial, where {\em\/some\/} of the monomials are rational, with {\em\/some\/} of the superfields (say $X_4$) in the denominator. The standard requirement for $W(X)$ itself to be chiral is straightforwardly satisfied:
\begin{alignat}9
  \bar{D}_{\dot\a}\,W(X)
   &=f(X_i)\big(\underbrace{\bar{D}_{\dot\a}\,X_0}_{=\,0}\big)
     +X_0\sum_{i>0} \pd{f}{X_i}
      \big(\underbrace{\bar{D}_{\dot\a}\,X_i}_{=\,0}\big),
 \label{e:DW=0}\\*[0mm]
   &= 0,\quad\textit{if}~~
         \big| f(X)\vphantom{\big|\,}\big|,~ \Big|X_0\pd{f}{X_i}\Big|<\infty,
 \label{e:BL<infty}
\end{alignat}
regardless of the possibly negative powers of some of the chiral superfields.
 As long as the {\em\/vacuum expectation value\/} (vev\footnote{\label{n:vev}The vev of the fermionic component fields in any superfield must vanish to preserve Lorentz symmetry, and the vevs of the auxiliary fields must vanish to preserve supersymmetry. It then follows that the vev of any chiral (and also twisted-chiral) superfield is the vev of its lowest component field, $\vev{X_i}=\vev{x_i}$.}) of the scalar component of the chiral superfield $X_4$ (appearing the denominator) does not identically vanish, the Laurent polynomial $W(X)$ has an analytic (in fact, finite) expansion in the fermionic coordinates.
 As defined in \SS\;\ref{s:SuS}, the ``intrinsic limit'' maintains the related conditions
\begin{equation}
 \big|\Vev{ f(X)\vphantom{\big|} }\big|,~
          \Big|\Big\langle X_0\pd{f}{X_i} \Big\rangle\Big|<\infty
 \label{e:vBL<infty}
\end{equation}
in all cases of interest, even when approaching putative poles of the rational monomials in the superpotential, such as $\vev{X_4}\to0$. Notably, this {\em\/chirality\/} requirement is a system of inequalities implied by the equality conditions, $f(x)=0=\vd_i\,f(x)$, defining the {\em\/base-locus\/} (see Ref.\cite{rGrHa}) of the {\em\/defining function\/} $f(x)$.
 Since $W(X)=X_0{\cdot}f(X_1,\dots)$, the vanishing of the $F$-terms (defining the absolute, supersymmetric ground states) in~\eqref{e:GndS} specifies
\begin{equation}
 \begin{array}{@{}r@{~}l@{}}
        f(x) &=0;\\[1mm] \displaystyle
        x_0\,\frac{\partial f}{\partial x_i} &=0,~ i=1,\cdots,n{+}2;\\
 \end{array}\Bigg\}~~
 :=~\text{base-locus of}~~W(x)=x_0\,{\cdot}\,f(x),
 \label{e:BLW}
\end{equation}
since the conditions in the left-hand side of~\eqref{e:BLW} imply the vanishing of $W(x)$ by homogeneity. There are two obvious (collections of) branches of solutions to~\eqref{e:BLW}:
\begin{enumerate}[itemsep=-1pt, topsep=0pt]
 \item The ``geometric phase(s)'' (branch~A), where $\vev{x_0}=0$, so that the gradient components $\vev{\pd{f}{x_i}}$ {\em\/need not\/} all vanish where 
 $\vev{f(x)}$ does; this is the $f(x)\<=0$ hypersurface in the complement of the base-locus of a transverse ``defining function,'' $f(x)$.
 \item The ``non-geometric phases(s)'' (branch~B), where $\vev{x_0}\neq0$, so that the gradient components $\vev{\pd{f}{x_i}}$ {\em\/must\/} all vanish where 
 $\vev{f(x)}$ does; this is the base-locus of a transverse $f(x)$.
\end{enumerate}
The above analysis may be extended to also accommodate non-transverse 
$f(x)$\cite{Hubsch:2002st}.
 As the ground states are defined by the vanishing of~\eqref{e:GndS} and that implies~\eqref{e:BLW}, the chirality condition~\eqref{e:BL<infty} is then automatically satisfied in the ground state. Furthermore, the chirality condition~\eqref{e:BL<infty} will continue to be satisfied also away from the ground state, as long as the background values are bounded by the inequalities~\eqref{e:vBL<infty}. The analogous finiteness condition~\eqref{e:BL<infty} then allows for a very wide (but correlated) range of (quantum) fluctuations of the superfields $X_0,X_1,\cdots$, and so defines the GLSM model as a (manifestly) supersymmetric quantum field theory in terms of a correlated background field expansions around the ground state; see \SS\;\ref{s:LQFT} below.
\\
\centerline{---$\,*\!*\!*\,$---}
 
 We now turn to discuss the space of ground states in a simple toy-model, the $X_1\<\to 1$ simplification of~\eqref{e:LZL}:
\begin{equation}
 W(X) := X_0\,{\cdot}\,f(X),\quad
 f(X):=X_3^5 +X_4^5 +\frac{X_2^2}{X_3}.
 \label{e:LaurV0}
\end{equation}
The $X_i$ have $q(X_0;X_2,X_3,X_4)\<= (-5;\,3,1,1)$ charges with respect to the single $U(1)$-gauge symmetry, which indicate that $(x_2,x_3,x_4)$ are homogeneous coordinates of the weighted projective space $x_i\in\IP^2_{(3:1:1)}$, including the standard MPCP-desingularization\footnote{The point $(1,0,0)\in\IP^2_{(3:1:1)}$ is a $\ZZ_3$ singularity and is smoothed by the ``maximal projective crepant partial'' (MPCP) desingularization\cite{rBaty01}; the so-desingularized space is biholomorphic to the Hirzebruch scroll $\FF[2]3$.}\cite{rBaty01} of the $\ZZ_3$-singular point 
 $(1,0,0)\in\IP^2_{(3:1:1)}$, whereupon $f(x)=0$ specifies the quintic (Calabi-Yau) hypersurface (2-torus), $\mathrm{Bl}^\uA[\IP^2_{(3:1:1)}][5]\approx_{\sss\IC}\!\FF[2]3$; the $X_1\<\to 1$ simplification effectively defers this desingularization.

\subsubsection{Ground States}
 \label{s:GndStoy}
For simplicity in discussing the ground states, let $x_i$ from now on denote the vev of the (scalar) lowest component field in the superfield $X_i$, not the (quantum) field itself.
 We also neglect the distinction between $\vev{x_i^{\,\a}}$ and $\vev{x_i}^\a$ (important though it may be in QFT in general), assume that they are equal for all intended purposes and simply write $x_i^{\,\a}$. Dropping the $\vev{{\cdots}}$-brackets from vevs also avoids confusion with other uses of these symbols, such as matrix elements as in the left-hand side of~\eqref{e:GLSM-M}, below.

The potential is a sum of positive-definite terms, each one of which~\eqref{e:GndS} must vanish in the supersymmetric ground states:
\begin{subequations}
 \label{e:GLSM-V}
\begin{alignat}9
 \frac{\vd W}{\vd x_0}\!&:\quad&
  0&\overset!=f(x)
   =x_3\!^5 +x_4\!^5 +\frac{x_2\!^2}{x_4},~~
  &&\To&\quad \{f(x)&=0\};
 \label{e:GLSM-G}\\
 \frac{\vd W}{\vd x_2}\!&:\quad&
  0&\overset!=
    x_0\,\Big(\frac{\vd f}{\vd x_2}\<= 2\,\frac{x_2}{x_4}\Big),
    &&\To&\quad \Big\{ \frac{x_2}{x_4}=0 \Big\}&\cup\{ x_0=0 \};
 \label{e:GLSM-2}\\
 \frac{\vd W}{\vd x_3}\!&:\quad&
  0&\overset!=
    x_0\,\Big(\frac{\vd f}{\vd x_3}\<= 5\,x_3\!^4\Big),
    &&\To&\quad \big\{ x_3\!^4=0 \big\}&\cup\{x_0=0\};
 \label{e:GLSM-3}\\
 \frac{\vd W}{\vd x_4}\!&:\quad&
  0&\overset!=
    x_0\,\Big(\frac{\vd f}{\vd x_4}\!
              =\!5\,x_4\!^4 -\frac{x_2\!^2}{x_4\!^2}\Big),
    &&\To&\quad \Big\{5\,x_4\!^4=\frac{x_2\!^2}{x_4\!^2}\Big\}&\cup\{x_0=0\};
 \label{e:GLSM-4}\\
 D\text{-term}&:~~&
  0&\makebox[0pt][l]{$\ddd\>\overset!=
    -e^2\big(3|x_2|^2 +|x_3|^2 +|x_4|^2 -5|x_0|^2 -r\big);$}
 \label{e:GLSM-D}\\
 \Vev{\s|Q^2(x)|\s}&:~~&
  0&\makebox[0pt][l]{$\ddd\>\overset!=
    |\s|^2\big(9|x_2|^2 +|x_3|^2 +|x_4|^2 +25|x_0|^2\big),$}
 \label{e:GLSM-M}
\end{alignat}
\end{subequations}
For the algebraic system~\eqref{e:GLSM-V}, and the first-stated branch of~\eqref{e:GLSM-2} and~\eqref{e:GLSM-4} in particular, to be well defined, we must rely on the {\em\/intrinsic limit\/} of \SS\;\ref{s:SuS}, which is how the present analysis veers outside the otherwise perfectly corresponding framework of (complex) algebraic geometry.
 The condition~\eqref{e:GLSM-M} immediately implies $\s=0$ since at least one of $x_0,x_3,x_4,x_2$ is nonzero in all the cases of interest\footnote{Exceptionally, the vev of $\s$ is not forced to vanish only at the origin 
 $(x_0,x_3,x_4,x_2)=(0,0,0,0)$ of {\em\/all\/} ``matter'' field-space; as in Refs.\cite{rPhases,rMP0} this special location will not be of interest herein.} and all the coefficients are positive.

In turn, the condition~\eqref{e:GLSM-D} also specifies two separate branches of solutions:
\paragraph{$r>0$:}
 The condition~\eqref{e:GLSM-D} implies:
\begin{equation}
   3|x_2|^2 +|x_3|^2 +|x_4|^2 = 5|x_0|^2+r>0,
 \label{e:CYph}
\end{equation}
which precludes $x_2,x_3,x_4$ from vanishing simultaneously.
 Projectivized by the (complexified) $U(1)$-gauge symmetry, this defines
 $\big\{\IC^4\<\ssm \{0\}\big\}/U(1;\IC) = \IP^2_{(3:1:1)}$,
 in which~\eqref{e:LaurV0} defines a quintic, $\IP^2_{(3:1:1)}[5]$.
Since $f(x)$ is homogeneous with respect to the
 $q(x_2,x_3,x_4)\<= (3,1,1)$-specified $U(1;\IC)$-gauge symmetry, its zero-locu is well defined away from its base-locus --- which is omitted from $\IP^2_{(3:1:1)}$.
With $(x_2,x_3,x_4)\<\neq (0,0,0)$, \eqref{e:GLSM-2}--\eqref{e:GLSM-4} force $x_0=0$. Then~\eqref{e:GLSM-M} implies that:
 ({\small\bf1})~the vev $\vev{\s}\<= 0$, and
 ({\small\bf2})~the mass of the field $\s$ is
  $\sqrt{9|x_2|^2 +|x_3|^2 +|x_4|^2}>0$.
 Since $\gcd[q(x_2),q(x_3),q(x_4)]=1$, the $\U(1)$ gauge symmetry is completely broken ($\U(1)\to\Ione$) at a generic point $(x_2,x_3,x_4)\neq(0,0,0)$, and
 $\U(1)\to\ZZ_3$ at $(x_2,x_3,x_4)=(1,0,0)$.
This latter point however does not belong to the zero-set
 $\{f(x)=0\}\<\subset \IP^2_{(3:1:1)}$, which justifies (a fortiori) the deferring of the desingularization of this singularity; see the comment just before \SS\:\ref{s:GndStoy}.

\paragraph{$r<0$:}
 The condition~\eqref{e:GLSM-D} now implies:
\begin{equation}
 0<3|x_2|^2 +|x_3|^2 +|x_4|^2 +|r| = 5|x_0|^2,
 \label{e:r<0}
\end{equation}
which forces $x_0\<\neq 0$. But then, Eqs.~\eqref{e:GLSM-2}--\eqref{e:GLSM-4} limit the field-space to the base-locus: $(x_3,x_4,x_2)\to(0,0,0)$. Contrary to appearances, this limit is unambiguous owing to the constraint~\eqref{e:GLSM-4}, which may be used first to eliminate $x_2\to\pm\sqrt{5}\,x_4\!^3$, reducing the system~\eqref{e:GLSM-V} to
\begin{subequations}
\begin{alignat}9
 \frac{\vd W}{\vd x_2}&:~~&
  0&\overset!=
    x_0\,\Big(\frac{\vd f}{\vd x_2}
              \to \pm2\sqrt{5}\,x_4\!^2\Big)\quad
    \To~~x_4=0,~~\To~x_2=0;\\
 \frac{\vd W}{\vd x_3}&:~~&
  0&\overset!=
    x_0\,\Big(\frac{\vd f}{\vd x_3} \to 5\,x_3\!^4\Big)\quad
    \To~~x_3=0,
\end{alignat}
\end{subequations}
whereupon also $f(x) \to x_3\!^5 +6x_4\!^5=0$.
This reduces the system to the base-locus of $f(x)$, the origin $(x_1,x_2,x_3)=(0,0,0)$, while~\eqref{e:GLSM-D} fixes $|x_0|=\sqrt{-r/5}>0$ and~\eqref{e:GLSM-M} continues to force $\vev{\s}\<= 0$ and sets the mass of the $\s$ field to $\sqrt{5|r|}$.
 Since $q(x_0)=-5$, the resulting value $\vev{x_0}\neq0$ breaks the gauge symmetry $\U(1)\to\ZZ_5$, turning the model into a $\ZZ_5$ Landau--Ginzburg orbifold\cite{rLGO0,rLGO}.

\paragraph{$r\<\in \IC$:} 
The simple toy-model GLSM~\eqref{e:GLSM-V} then interpolates between the $r>0$ and $r<0$ ``phases'' exactly as described in Refs.\cite{rPhases,rMP0}, passing ``around'' $r\<= 0$ by way of the complex-analytic extension provided by the full Fayet-Iliopoulos term, which complexifies the (``phase'') $r$-space.

\subsubsection{Quantum Field Theory}
 \label{s:LQFT}
The occurrence of division by the superfield $X_4$ may be seen as precluding an application of the Laurent polynomial~\eqref{e:LaurV0} as a superpotential in a fully  fledged (world-sheet) quantum field theory, since division by quantum (super)fields is generally ill-defined. The previous section shows that the ground state of the model is defined unambiguously, which then allows relying on  background field expansion around any such vacuum configuration: We expand the superfields, $X_i=\vev{x_i}+\Phi_i$, about their background values $\vev{x_i}$ (see footnote~\ref{n:vev}) and focus on the quantum fluctuations, $\Phi_i$.

In the chirality condition~\eqref{e:BL<infty}, the factors 
 $\bar{D}_{\dot\a}\,X_0$ and $\bar{D}_{\dot\a}\,X_i$ vanish 
 {\em\/a priori\/} --- i.e., before any restriction or constraint is imposed on the superfields and their component fields. It then follows that the chirality~\eqref{e:BL<infty} of the superpotential is preserved straightforwardly away from the location $\vev{x_4}=0$. At the location $\vev{x_4}=0$, the above analysis shows that the background (ground-state) value $\vev{x_4}$ is also constrained and preserves the conditions~\eqref{e:BL<infty} throughout the ground state. It then remains to discuss the quantum fluctuations about this background.

When $r\,{>}\,0$, $\vev{x_4}\,{\neq}\,0$, and this background field expansion of the superpotential produces a manifestly regular, well defined function of the (quantum) fluctuation superfields $\Phi_i$:
\begin{alignat}9
  W\big(\vev{x_i}+\F_i\big)\big|_{r>0}\to
  &\,\F_0 \bigg(\big(\!\vev{x_3}+\F_3\big)^5 +\big(\!\vev{x_4}+\F_4\big)^5
              +\frac{\big(\!\vev{x_2}+\F_2\big)^2}{\vev{x_4}+\F_4}\bigg),\\
  &=\F_0 \bigg(\big(\!\vev{x_3}+\F_3\big)^5 +\big(\!\vev{x_4}+\F_4\big)^5
              +\frac1{\vev{x_4}}\,\big(\!\vev{x_2}+\F_2\big)^2\,
               \sum_{k=0}^\infty\frac{(-1)^k}{\vev{x_3}^k}\,\F_3^{\,k}\bigg).
 \label{e:QW>0}
\end{alignat}
For $r\,{<}\,0$ in turn, the condition $\vev{x_2},\vev{x_3},\vev{x_4}\,{=}\,0$ reduces
\begin{equation}
  W\big(\vev{x_i}+\F_i\big)\to
   \big(\sqrt{-r/5}+\F_0\big)
    \bigg(\F_3^5 +\F_4^5 +\frac{\F_2\!^2}{\F_4}\bigg).
 \label{e:QW<0}
\end{equation}
For this to remain well defined, the (quantum) fluctuation superfields should correlate so that $\Phi_2\!^2/\Phi_4<\infty$ even when $\Phi_4\to0$. This correlated limitation of the (quantum) fluctuations in some ways reminds of to the momentum space limitation in the Wilsonian definition of an effective field theory, wherein (modes of) fields with masses/energies higher than a specified cutoff are (formally) integrated out or otherwise eliminated. Here, the fluctuation limitations are correlated in that the (quantum) fluctuations of one field are limited by the magnitude of the fluctuations of another rather than a fixed bound. In turn, this correlating limitation is an {\em\/inequality\/} and so cannot possibly be an equation of motion. Therefore, the (quantum) fluctuation superfields $\Phi_i$, albeit correlated, are still off-shell in the usual sense. They also define an {\em\/open\/} region of validity in the field space and so define a feasible framework.

In addition to the correlated limitation of the (quantum) superfield fluctuations, this description manifestly depends on the choice of the background. It may thus be desirable to find a ``background-independent'' formulation that does not depend on presuppositions about $\vev{x_i}$ and does not require correlations between the (quantum) fluctuations of the various superfields.
 There certainly exist other QFT methods of dealing with the divergence of~\eqref{e:QW<0}, as $W(\Phi_i)$ itself is a chiral superfield apparently subject to an infinite renormalization.
 It is also reasonable to regard a GLSM with a Laurent superpotential (containing rational monomials) such as~\eqref{e:LaurV0} as merely an effective limiting model, in which case it is clearly of interest to find the better-defined ``parent'' model and we briefly explore one logical possibility to this end.

\subsubsection{A Reciprocal Extension}
 \label{s:recip}
Since $X_4$ occurs in the superpotential~\eqref{e:LaurV0} {\em\/both\/} with positive and negative exponents, a simple field redefinition
 $X_4\<\to Z_4\!:=\!1/X_4$
would not eliminate the appearance of (putative) poles. It is however possible to only replace the instances of the fields in the denominators, leading us to consider a superpotential (with an additional ``Lagrange-Witten" superfields $X_{-1}$) where such a substitution is done only in the Laurent terms:
\begin{equation}
 \Tw{W} :=
  X_0\big(X_3\!^5 +X_4\!^5 +X_2\!^2 Z_4\big)
 +X_{-1}\big(Z_4 X_4-1\big).
 \label{e:LaurV1}
\end{equation}
 This admits a correspondingly extended action of the $\U(1)$ gauge symmetry with charges:
\begin{equation}
  q(X_{-1},X_0;\, X_2,X_3,X_4;\, Z_4)=(0,-5;3,1,1;-1).
 \label{e:LaurV2}
\end{equation}
 
In the original GLSM analyses\cite{rPhases,rMP0}, the lowest-component field $x_0$ in the superfield $X_0$ serves as the fiber-coordinate of the
 $\cO_{\mathscr{X}}(5)=\cK^*_{\mathscr{X}}$ bundle over
 $\mathscr{X}\<= \IP_{(3:1:1)}^2$ where $x_i$ serve as homogeneous (base) coordinates. Being uncharged, $X_{-1}$ has no gauge interactions, does not affect the $D$-term conditions~\eqref{e:GLSM-D}, but does have Yukawa interactions with the propagating fields $X_4,Z_4$. In turn, $X_3,Z_3$ may be regarded as providing a base-and-fiber coordinates for a $\cO_{\IP^0}(1)$ sub-sector in the full model.
 The precise dynamical and geometrical role of these extension variables ($Z_4,X_{-1}$) in the so-extended GLSM, and the possible lifting of ``intrinsic limit'' specifications required for the $\vev{x_4}\to0$ limit, will not be necessary here.

Treating $X_{-1}$ for the moment and formally as a Lagrange multiplier, its equations of motion\footnote{With kinetic (D-)terms $\sum_{i=0}^2\kappa_i\int\rd^4\q\,\bar{X}_iX_i$, the superspace/superfield {\em\/constrained\/} variation\cite[Eqs.~(3.8.10)--(3.8.11)]{r1001} and\cite[Eqs.~(3.1.30)--(3.1.48)]{rBK} produces the superfield equations of motion $\kappa_i\bar{D}^2\bar{X}_i=-\pd{W}{X_i}$. If kinetic terms are induced via quantum corrections as in Ref.\cite{rPhases} (regardless of $X_{-1}$ being uncharged) or otherwise included with $\kappa_{-1}\neq0$, na{\"\ii}vely treating the $X_{-1}$ as Lagrange multipliers is incorrect in principle, but is nevertheless an apt low-energy approximation\cite{rChaSM,rMargD}; foremost, it is consistent with the subsequent analysis of the ground-state configuration.} produces the algebraic equation $X_4\,Z_4=1$, which is manifestly a chiral equation. It may of course be solved for $Z_4=1/X_4$, which reproduces the original, Laurent superpotential~\eqref{e:LaurV0}, as intended. Thereby, the Laurent polynomial~\eqref{e:LaurV0} is said to be ``classically equivalent'' to the well defined and standard-looking superpotential~\eqref{e:LaurV1}. However, this manifestly holomorphic superpotential function~\eqref{e:LaurV1} {\em\/can\/} be used straightforwardly in any supersymmetric quantum field theory, since it is perfectly analytic in all of the involved (super)fields.

We may verify this ``classical equivalence'' also in a somewhat indirect way, which however more precisely identifies the ground-state field configuration and is more directly relevant for the GLSM, by analyzing its full potential. Also, this ``classical equivalence'' and its derivation is conceptually similar to the standard relation between the Nambu-Goto and the Polyakov action for strings, as well as the analogous equivalence for point-particles\cite{rJPS,rBZ-StrTh}.

Before we proceed however, notice that introducing the superfield $Z_4$ causes the sum of charges~\eqref{e:LaurV2} no longer to vanish. Consequently, the $\U(1)$ gauge symmetry is {\em\/anomalous\/} in the extended model~\eqs{e:LaurV1}{e:LaurV2}. To remedy this, we may introduce a ``spectator'' superfield $Z_0$ with $q(Z_0)=+1$ with a standard kinetic term and standard gauge interactions. This modifies the $D$-terms~\eqref{e:GLSM-4} and the mixed term~\eqref{e:GLSM-M}. Since the superpotential does not become renormalized, we may (and do) choose to not introduce $Z_0$ into the superpotential, so as to leave the $F$-terms in~\eqref{e:GndS} and the definition of the ground states unchanged. For example, the charges would permit adding the terms $\a_{k,l}\,X_0\,Z_0^{\,5-k-l}\,X_3^{\,k}\,X_4^{\,l}$ to the superpotential. The values of $\a_{k,l}$ are protected by the non-renormalization theorems, and we are free to choose $\a_{k,l}=0$; these terms are not (re)introduced by radiative/quantum corrections. Desirably so: including such couplings would put $Z_0$ on par with $X_3,X_4$, as another coordinate superfield, turning~\eqref{e:LaurV1} into a superpotential for
 $f(X,Z_0)\<\in \IP^3_{(3:1:1:1)}[5]$, which would no longer be Ricci-flat.
 At this preliminary stage, we need not worry about world-sheet gravity, i.e., interactions with world-sheet reparametrization ghosts.

\paragraph{The Full Potential:}
The full potential has three separate contributions:
The $F$-term is derived from the superpotential, now selected to be~\eqref{e:LaurV1}, and reads:
\begin{subequations}
 \label{e:xGLSM}
\begin{alignat}9
 \|\vec{F}\|^2
 &=\big|x_3{}^5 +x_4{}^5 +x_2{}^2\,z_4\big|^2
 \label{e:Fterm-1}\\
 &~~+\big|2\,x_0\,x_2\,z_4\big|^2
     +\big|5\,x_0\,x_3{}^4\big|^2
      +\big|5\,x_0\,x_4{}^4 +x_{-1}\,z_4\big|^2
 \label{e:Fterm-2}\\
 &~~+\Big|x_0\,x_2{}^2+x_{-1}\,x_4\Big|^2
    +|z_4\,x_4-1|^2.
 \label{e:Fterm-3}
\intertext{For the one $\U(1)$ and with the charges given in~\eqref{e:LaurV2}, the standard $D$-term contribution to the potential energy is:}
  \frac1{2e^2}D^2
 &=\frac{e^2}2\Big(3|x_2|^2+|x_3|^2+|x_4|^2
                  -|z_4|^2-5|x_0|^2-\Tw{r}\,\Big)^2.
 \label{e:Dterm}
\intertext{We introduce the effective ``radial'' variable $\Tw{r}:=r-|z_0|^2$ since $|z_0|$ does not appear in any of the previous conditions. Finally,~\eqref{e:GLSM-M} now becomes}
 \Vev{\s\big|Q^2(x)\big|\s}
 &=|\s|^2\Big(9|x_2|^2+|x_3|^2+|x_4|^2+|z_0|^2+|z_4|^2+25|x_0|^2\Big).
 \label{e:Mterm}
\end{alignat}
\end{subequations}

Substituting the vanishing locus of~\eqref{e:Fterm-3},
 $x_{-1}\<= {-}x_0\,x_2\!^2/x_4$ and $z_4\<= 1/x_4$, into~\eqref{e:Fterm-1}--\eqref{e:Fterm-2} reproduces the unextended $F$-terms~\eqref{e:GLSM-G}--\eqref{e:GLSM-4}. The constraint~\eqref{e:Fterm-1} may again be solved,
\begin{equation}
  f(x)=0\quad\To\quad x_2=\pm i\sqrt{x_4(x_3\!^5 +x_4\!^5)},
 \label{e:x2=34}
\end{equation}
assuming that $x_4\neq0$, and then using this to define the ``intrinsic limit'' of \SS\;\ref{s:SuS}. That is, the zero-locus of the $F$-terms are essentially unchanged.
 However, the $D$-term~\eqref{e:GLSM-D} and the mixed term~\eqref{e:GLSM-M} do change:
\begin{alignat}9
0&\overset!=\Big(3|x_2|^2 +|x_3|^2 +|x_4|^2 \Big)
            -\bigg(\frac1{|x_4|^2}+5|x_0|^2+\Tw{r}\bigg);
 \label{e:r}\\
0&\overset!=|\s|^2\bigg( 9|x_2|^2
                        +|x_3|^2
                        +|x_4|^2 +\frac1{|x_4|^2}
                        +|z_0|^2 +25|x_0|^2\bigg). \label{e:s}
\end{alignat}
An immediate consequence of the difference between~\eqref{e:s} and its unextended analogue~\eqref{e:GLSM-M} is that, owing to the pair of terms 
 $|x_3|^2{+}\frac1{|x_3|^2}$, the quantity in the round parentheses in~\eqref{e:s} has a nonzero minimum. Thereby, $\vev{\s}=0$ and the scalar field $\s$ is massive independently of $\Tw{r}$ and so throughout the phase-space of the model.

\paragraph{$\Tw{r}>0$:}
The condition~\eqref{e:r} implies
\begin{equation}
  0<\frac1{|x_4|^2}+5|x_0|^2+\Tw{r}
   =3|x_2|^2 +|x_3|^2 +|x_4|^2,
 \label{e:twr>0}
\end{equation}
and $x_3,x_4,x_2$ must not vanish simultaneously. Since $f(x)$ is transverse and~\eqref{e:Fterm-1} requires that $f(x)=0$, the derivatives $\frac{\vd f}{\vd x_i}$ for $i=1,2,3$ cannot all vanish. Then~\eqref{e:Fterm-2}--\eqref{e:Fterm-3} force $x_0\<= 0$. If $x_4$ were to vanish,~\eqref{e:x2=34} would imply that so does $x_2$, whereupon~\eqref{e:Fterm-1} would force $x_3=0$, implying $(x_2,x_3,x_4)\to(0,0,0)$, which is prohibited by~\eqref{e:twr>0}. Thus, the ground states automatically avoid the putative pole at $x_4\to0$.

\paragraph{$\Tw{r}<0$:}
The condition~\eqref{e:r} now implies
\begin{equation}
  0<|x_3|^2 +|x_4|^2 +3|x_2|^2 -\Tw{r}
   =\frac1{|x_4|^2}+5|x_0|^2,
 \label{e:twr<0}
\end{equation}
Unlike in~\eqref{e:r<0} and with $x_3\<<\infty$, the condition~\eqref{e:twr<0} does not force $x_0\neq0$ but is {\em\/consistent\/} with this; we thus have two possibilities:

\begin{enumerate}[itemsep=-1pt, topsep=0pt]
 \item{\boldmath$x_0=0$:} The condition~\eqref{e:p0} does not force the derivatives $\pd{f}{x_i}$ to vanish, whereby $x_3,x_4,x_2$ need not vanish simultaneously but do need to satisfy $f(x)=0$. This is the ``geometric'' branch of the ground states, with~\eqref{e:x2=34}.

\item{\boldmath$x_0\neq0$:} The condition~\eqref{e:p0} now does force the derivatives $\pd{f}{x_i}$ to vanish, whereby $x_3,x_4,x_2$ must vanish simultaneously in a limit that satisfies $f(x)=0$. This is the LGO branch of the ground states. In this limit,
\begin{alignat}9
  \lim_{x_i\to0} \Big(3|x_2|^2 +|x_3|^2 +|x_4|^2 -\Tw{r}\Big)
 &=\lim_{x_i\to0} \Big(\frac1{|x_3|^2}+5|x_0|^2\Big),\quad i=2,3,4,\nonumber\\
  \To\quad \Tw{r}
 &=-\lim_{x_4\to0} \Big(\frac1{|x_4|^2}+5|x_0|^2\Big)=-\infty.
\end{alignat}
\end{enumerate}
This extends the ``geometric'' branch of the ground states throughout the range\footnote{As explained in Ref.\cite{rPhases}, the semi-classical approximation employed in this analysis is really valid only for $|\Tw{r}|\to\infty$. For intermediate ---and especially small--- values of $|\Tw{r}|$, quantum corrections are expected to significantly affect analysis.} of $\Tw{r}$, and {\em\/limits\/} to the LGO only as $\Tw{r}\to-\infty$. 
Recalling that $\Tw{r}\!:=\!r{-}|z_0|^2$ and that $|z_0|$ remains arbitrary, the location (in $r\<= \Tw{r}{+}|z_0|^2$) of the transition from the ``geometric'' into the LGO phase as well as the location of the ``short-distance'' phenomena in the original $r$-space remains just as unspecified.

By avoiding the inclusion of the superfield $Z_0$ in the superpotential, its vev remains unbounded and parametrizes a non-compact direction, possibly interpretable as a fiber coordinate over the ground-state variety defined by the vevs of the other fields. Furthermore, although the superpotential~\eqref{e:LaurV1} remains unchanged owing to non-renormalization theorems, the mixing/mass-term~\eqref{e:Mterm} clearly involves an interaction of $Z_0$ with the twisted chiral superfield of which $\s$ is the lowest component. That is, $Z_0$ does not fully decouple from the dynamics of the rest of the model, and so complicates the analysis.

To summarize, the division by a (super)field in at least one of the terms in the Laurent monomial (e.g., with $X_4$ in the denominators) of the original superpotential~\eqref{e:LaurV0} causes the specification of the quantum field theory with such a potential require non-standard approach. As a remedy, we may:
\begin{enumerate}\itemsep=-3pt\vspace*{-1mm}

 \item Introduce additional superfields ($Z_4$ and $X_{-1}$) with the aid of which a Laurent superpotential such as~\eqref{e:LaurV0} becomes a regular, chiral/complex-analytic function such as~\eqref{e:LaurV1}.
 
 \item The additional fields (such as $Z_4$) however introduce a gauge anomaly, as the sum of at least one $\U(1)$ gauge symmetry charges no longer cancels out.

 \item The gauge anomaly may be canceled by introducing additional, ``spectator'' superfields (such as $Z_0$), for which however no (gauge-invariant and so permissible) terms are introduced into the superpotential. (This choice is admittedly {\em\/ad hoc\/}, but is protected by non-renormalization theorems.)

 \item These spectator superfields cause the ground state variety in the field space to become non-compact, while preventing a full decoupling of the non-compact directions.
\end{enumerate}
This procedure is designed to leave the superpotential and its base-locus essentially unchanged, but does end up changing (extending) the $D$-terms and mixing terms. This extends the geometry of the ground states into a bundle- or sheaf-like structure over the variety that was the intended ground-state algebraic variety of the original, Laurent model.

The lack of a straightforward distinction between the two options in the $\Tw{r}<0$ phase as well as the nontrivial changes~\eqref{e:Mterm} from the toy model~\eqref{e:LaurV0} to its ``regularizing'' extension~\eqref{e:LaurV1}, including the non-compact $Z_0$ included in~\eqref{e:Dterm}--\eqref{e:Mterm}, seem to indicate that
 either the model~\eqref{e:LaurV1} needs further remedial extension(s),
 or, more likely, the original model~\eqref{e:LaurV0} cannot be ``regularized'' by such extensions.
 After all, the non-convex and multi-layered VEX multitopes described herein specify toric varieties that are by design {\em\/excluded\/} from the standard toric geometry considerations\cite{rD-TV,rO-TV,rF-TV,rCox,rGE-CCAG,rCLS-TV,rCK} as well as from systematic studies leading to the most comprehensive database to date\cite{rKreSka00b}.
 
This substantial generalization from (reflexive) convex polytopes to VEX multitopes corresponds to a need for a substantial generalization of the GLSM and its QFT analysis.
 This generalization seems well motivated by the fascinating fact that the usual analysis can be employed nevertheless, with but minimal and most straightforward modifications, and we are gratified to find that this approach continues to produce self-consistent and coherent results for a very large class of significant generalizations of $(2,2)$-supersymmetric GLSMs and their $(0,2)$-supersymmetric generalizations\cite{rPhases, rPhasesMF, rMP0, Distler:1993mk, rMP1, Schafer-Nameki:2016cfr, Sharpe:2024dcd}. 
 Finally, the above description may be deemed ``semi-classical,'' and must be amended by incorporating world-sheet instanton effects and other consequences of the full quantum description of GLSMs. These topics remain to be explored.

\section{A Closer Surgical Insight}
\label{s:CSI}
As {\em\/sections\/} of the anticanonical bundle, $\cK^*_X$, the anticanonical monomials such as displayed in Figure~\ref{f:F1-2-3} are
{\em\/generalized functions\/} of specified $U(1;\IC)^n$-degrees,
defined over the underlying torus manifold, $X$, and so are {\em\/dual\/} to $X$
 ($\FF[2]1,\, \FF[2]2,\, \FF[2]3$ in Figure~\ref{f:F1-2-3}). Correspondingly, the so-called {\em\/Newton polytope,} $\pDN{X}$, formed by these monomials encodes the torus manifold $X$ over which such monomials are well defined by means of the {\em\/(trans)polar\/}\cite{rBH-gB, Berglund:2022dgb, Berglund:2024zuz}, consisting of lattice-normals to each convex {\em\/face\/} (vertex, edge, etc.):
\begin{subequations}
 \label{e:tP}
 \begin{gather}
  \Q \too{\sss\wtd}
  \Q^\wtd\<= 
   \q:=\{\n\in L^\vee: \vev{\n,\m}={-}1,~\m\in\Q\},\quad
  \forall\Q\subsetneq\pDN{X}\subsetneq L, \label{e:tPa}\\
  \bigg\{
  \begin{array}{@{\,}l@{}}
   \,\Q_1\subsetneq\Q_2
   \overset{\sss\wtd~}\To
    \q_2\subsetneq\q_1,   \quad\text{so that}\\
   \big[(\Q_4\cap\Q_5)=
        \Q_3\subsetneq\vd\Q_4,\vd\Q_5\big]
   \overset{\sss\wtd~}\To
   \vd\q_3\supseteq(\q_4\cup\q_5),
  \end{array}
  \label{e:tPb}
  \end{gather}
\end{subequations}
where $L$ is the lattice wherein the anticanonical sections are plotted in Figure~\ref{f:F1-2-3}, and $L^\vee$ is its dual.
 The {\em\/inclusion reversing\/} property~\eqref{e:tPb} encodes the assembling of the {\em\/(trans)polar\/} multitope, $\pDs{X}=(\pDN{X})^\wtd$, from its individually obtained faces, 
 $\q_i=\Q_i^\wtd$.

\begin{figure}[htb]
 \centering\unitlength=1mm
 \begin{picture}(135,80)(-4,2)
 \put(0,0){
  \TikZ{[scale=.9]\path[use as bounding box](-1,-2)--(1,4);
          \corner{(0,0)}{0}{90+atan(.5)}{1}{red};
          \corner{(0,0)}{90+atan(.5)}{225}{1}{yellow};
          \corner{(0,0)}{225}{315}{1}{green};
          \corner{(0,0)}{315}{360}{1}{blue};
          \draw[densely dotted](0,0)--(-1,2);
          \draw[densely dotted](0,0)--(-1,-1);
          \draw[densely dotted](0,0)--( 1,-1);
          \draw[densely dotted](0,0)--( 1,0);
         \draw[blue, very thick](-1,-1)--(1,-1)--(1,0)--(-1,2)--cycle;
          \path(-1,.5)node[left=-3pt, blue]{\footnotesize$\Q_1$};
           \path(-.8,-1)node[below left=-3pt, blue]
           {\footnotesize$\Q_{12}$};
          \path(0,-1)node[below right=-2pt, blue]{\footnotesize$\Q_2$};
           \path(1,-1)node[below right=-3pt, blue]
           {\footnotesize$\Q_{23}$};
          \path(1,-.5)node[right=-3pt, blue]{\footnotesize$\Q_3$};
           \path(1,.1)node[right=-3pt, blue]{\footnotesize$\Q_{34}$};
          \path(0,1)node[right, blue]{\footnotesize$\Q_4$};
           \path(-1,2)node[above left=-3pt, blue]{\footnotesize$\Q_{41}$};
         \draw[green!67!black, densely dashed, very thick, -stealth]
             (-1,0)--++(1,0);
          \path(-.67,0)node[above=-3pt, green!50!black]
          {\footnotesize$\q_1$};
         \draw[green!67!black, densely dashed, very thick, -stealth]
             (0,-1)--++(0,1);
          \path(0,-.67)node[right=-3pt, green!50!black]
          {\footnotesize$\q_2$};
         \draw[green!67!black, densely dashed, very thick, -stealth]
             (1,0)--++(-1,0);
          \path(.67,0)node[below=-3pt, green!50!black]
          {\footnotesize$\q_3$};
         \draw[green!67!black, densely dashed, very thick, -stealth]
             (45:.71)--++(-1,-1);
          \path(225:.6)node[below left=-3pt, green!50!black]
             {\footnotesize$\q_4$};
         \draw[very thick, densely dotted, -stealth](45:.8)to[out=45,in=-60]
           node[left=-3pt]{\footnotesize\eqref{e:tP}}++(0,3.7);
         \path(1.5,.75)node{$\pDN{\FF[2]1}$};
         \foreach\y in{-1,...,2}\filldraw(-1,\y)circle(.5mm);
         \foreach\y in{-1,...,1}\filldraw( 0,\y)circle(.5mm);
         \foreach\y in{-1,...,0}\filldraw( 1,\y)circle(.5mm);
         \filldraw[thick, fill=white](0,0)circle(.5mm);
            }}
 \put(0,45){
  \TikZ{[scale=.9]\path[use as bounding box](-1,-2)--(1,4);
          \corner{(0,0)}{0}{90}{.8}{orange};
          \corner{(0,0)}{90}{180}{.8}{yellow!50!green};
          \corner{(0,0)}{180}{225}{.8}{teal};
          \corner{(0,0)}{225}{360}{.8}{magenta};
         \draw[green!67!black, very thick]
             (1,0)--(0,1)--(-1,0)--(-1,-1)--cycle;
         \draw[green!67!black, very thick, -stealth](0,0)--(1,0);
          \path(1,0)node[right, green!50!black]
             {\footnotesize$\q_1$};
          \path(45:.71)node[above right=-3pt, green!50!black]
             {\footnotesize$\q_{12}$};
         \draw[green!67!black, very thick, -stealth](0,0)--(0,1);
          \path(0,1)node[above, green!50!black]{\footnotesize$\q_2$};
          \path(135:.71)node[above left=-3pt, green!50!black]
             {\footnotesize$\q_{23}$};
         \draw[green!67!black, very thick, -stealth](0,0)--(-1,0);
          \path(-1,0)node[left, green!50!black]{\footnotesize$\q_3$};
          \path(-1,-.5)node[left=-3pt, green!50!black]
             {\footnotesize$\q_{34}$};
         \draw[green!67!black, very thick, -stealth](0,0)--(-1,-1);
          \path(-1,-1)node[below left=-3pt, green!50!black]
             {\footnotesize$\q_4$};
          \path({-atan(2)}:.71)node[green!50!black]{\footnotesize$\q_{41}$};
         \draw[very thick, densely dotted, -stealth](250:.8)to[out=250,in=100]
           node[right=-3pt]{\footnotesize\eqref{e:tP}}++(-.05,-2.65);
         \path(.5,-.5)node[right]{$\pDs{\FF[2]1}$};
         \foreach\y in{-1,...,1}\filldraw(-1,\y)circle(.5mm);
         \foreach\y in{-1,...,1}\filldraw( 0,\y)circle(.5mm);
         \foreach\y in{ 0,...,1}\filldraw( 1,\y)circle(.5mm);
         \filldraw[thick, fill=white](0,0)circle(.5mm);
            }}
 \put(50,0){
  \TikZ{[scale=.9]\path[use as bounding box](-1,-2)--(1,4);
          \corner{(0,0)}{-45}{90+atan(.33)}{1}{red};
          \corner{(0,0)}{90+atan(.33)}{225}{1}{yellow};
          \corner{(0,0)}{225}{315}{1}{green};
          \draw[densely dotted](0,0)--(-1,3);
          \draw[densely dotted](0,0)--(-1,-1);
          \draw[densely dotted](0,0)--( 1,-1);
         \draw[blue, very thick, line join=round]
             (-1,-1)--(1,-1)--(-1,3)--cycle;
          \path(-1,.9)node[left=-3pt, blue]{\footnotesize$\Q_1$};
           \path(-.8,-1)node[below left=-3pt, blue]
           {\footnotesize$\Q_{12}$};
          \path(0,-1)node[below right=-2pt, blue]{\footnotesize$\Q_2$};
           \path(1,-1)node[below right=-3pt, blue]
           {\footnotesize$\Q_{23}$};
          \path(0,1)node[right, blue]{\footnotesize$\Q_3$};
           \path(-1,3)node[above left=-3pt, blue]{\footnotesize$\Q_{31}$};
         \draw[green!67!black, densely dashed, very thick, -stealth]
             (-1,0)--++(1,0);
          \path(-.67,0)node[above=-3pt, green!50!black]
          {\footnotesize$\q_1$};
         \draw[green!67!black, densely dashed, very thick, -stealth]
             (0,-1)--++(0,1);
          \path(0,-.67)node[right=-3pt, green!50!black]
          {\footnotesize$\q_2$};
         \draw[green!67!black, densely dashed, very thick, -stealth]
             (27:.4)--++(-2,-1);
          \path(205:1.2)node[above left=-3pt, green!50!black]
             {\footnotesize$\q_3$};
         \draw[very thick, densely dotted, -stealth](40:.6)to[out=45,in=-50]
           node[left=-3pt]{\footnotesize\eqref{e:tP}}++(-.1,4.55);
         \path(1.5,.5)node{$\pDN{\FF[2]2}$};
         \foreach\y in{-1,...,3}\filldraw(-1,\y)circle(.5mm);
         \foreach\y in{-1,...,1}\filldraw( 0,\y)circle(.5mm);
         \filldraw( 1,-1)circle(.5mm);
         \filldraw[thick, fill=white](0,0)circle(.5mm);
            }}
 \put(50,47.5){
  \TikZ{[scale=.9]\path[use as bounding box](-1,-2)--(1,4);
          \corner{(0,0)}{0}{90}{.8}{orange};
          \corner{(0,0)}{90}{180+atan(.5)}{.8}{yellow!50!green};
          \corner{(0,0)}{180+atan(.5)}{360}{.8}{magenta};
         \draw[green!67!black, very thick](1,0)--(0,1)--(-2,-1)--cycle;
         \draw[green!67!black, very thick, -stealth](0,0)--(1,0);
          \path(1,0)node[right, green!50!black]{\footnotesize$\q_1$};
          \path(45:.71)node[above right=-3pt, green!50!black]
             {\footnotesize$\q_{12}$};
         \draw[green!67!black, very thick, -stealth](0,0)--(0,1);
          \path(0,1)node[above, green!50!black]{\footnotesize$\q_2$};
          \path(-1,0)node[above left=-3pt, green!50!black]
             {\footnotesize$\q_{23}$};
         \draw[green!67!black, very thick, -stealth](0,0)--(-2,-1);
          \path(-2,-1)node[left=-3pt, green!50!black]
          {\footnotesize$\q_3$};
          \path(-85:.6)node[green!50!black]{\footnotesize$\q_{31}$};
         \draw[very thick, densely dotted, -stealth](240:.6)to[out=240,in=95]
           node[right=-3pt]{\footnotesize\eqref{e:tP}}++(-.2,-2.45);
         \path(105:.8)node[left]{$\pDs{\FF[2]2}$};
         \foreach\x in{-2,...,0}\filldraw(\x,-1)circle(.5mm);
         \foreach\x in{-1,...,1}\filldraw(\x, 0)circle(.5mm);
         \foreach\x in{ 0,...,1}\filldraw(\x, 1)circle(.5mm);
         \filldraw[thick, fill=white](0,0)circle(.5mm);
            }}
 \put(105,2){
  \TikZ{[scale=.9]\path[use as bounding box](-1,-2)--(1,4);
          \corner{(0,0)}{-atan(2)}{90+atan(.25)}{1.1}{yellow};
           \draw[orange, thick, densely dotted]
               ({-atan(2)}:.7)arc({-atan(2)}:{90+atan(.25)}:.7);
          \corner{(0,0)}{90+atan(.25)}{225}{1}{green};
           \draw[green!75!black, thick, densely dotted]
               ({90+atan(.25)}:.7)arc({90+atan(.25)}:225:.7);
          \corner{(0,0)}{-atan(2)}{-45}{1.5}{red};
           \draw[red, thick, densely dotted]
               ({-atan(2)}:1)arc({-atan(2)}:-45:1);
          \corner{(0,0)}{225}{315}{1}{blue};
           \draw[blue, thick, densely dotted](225:.8)arc(225:315:.8);
          \draw[densely dotted](0,0)--(-1,4);
          \draw[densely dotted](0,0)--(-1,-1);
          \draw[densely dotted](0,0)--( 1,-1);
          \draw[densely dotted](0,0)--( 1,-2);
         \draw[blue, very thick, line join=round]
             (1,-2)--(-1,4)--(-1,-1)--(1,-1);
         \draw[red, very thick](1,-1)--(1,-2);
          \path(-1,1.5)node[left=-3pt, blue]
          {\footnotesize$\Q_1$};
           \path(-.8,-1)node[below left=-3pt, blue]
           {\footnotesize$\Q_{12}$};
          \path(0,-1)node[below right=-2pt, blue]{\footnotesize$\Q_2$};
           \path(1,-.9)node[right=-2pt, red]{\footnotesize$\Q_{23}$};
          \path(1,-1.5)node[right=-2pt, red]{\footnotesize$\Q_3$};
           \path(1,-2.1)node[right=-2pt, red]{\footnotesize$\Q_{34}$};
          \path(0,1)node[right, blue]{\footnotesize$\Q_4$};
           \path(-1,4)node[above left=-3pt, blue]{\footnotesize$\Q_{41}$};
         \draw[green!67!black, densely dashed, very thick, -stealth]
             (-1,0)--++(1,0);
          \path(-.67,0)node[above=-3pt, green!50!black]
          {\footnotesize$\q_1$};
         \draw[green!67!black, densely dashed, very thick, -stealth]
             (0,-1)--++(0,1);
          \path(0,-.67)node[left=-3pt, green!50!black]
          {\footnotesize$\q_2$};
         \draw[red, densely dashed, very thick, -stealth](1,0)--++(-1,0);
          \draw[red, densely dotted](1,-1)--++(0,1.2);
          \path(.67,0)node[above=-3pt, red]{\footnotesize$\q_3$};
         \draw[green!67!black, densely dashed, very thick, -stealth]
             (22:.33)--++(-3,-1);
          \path(188:1.2)node[left=-3pt, green!50!black]
          {\footnotesize$\q_4$};
         \draw[very thick, densely dotted, -stealth](50:.5)to[out=45,in=-60]
           node[left=-3pt]{\footnotesize\eqref{e:tP}}++(-.1,4.6);
         \path(1.5,.5)node{$\pDN{\FF[2]3}$};
         \foreach\y in{-1,...,4}\filldraw(-1,\y)circle(.5mm);
         \foreach\y in{-1,...,1}\filldraw( 0,\y)circle(.5mm);
         \foreach\y in{-2,...,-1}\filldraw( 1,\y)circle(.5mm);
         \filldraw[thick, fill=white](0,0)circle(.5mm);
            }}
 \put(105,50){
  \TikZ{[scale=.9]\path[use as bounding box](-1,-2)--(1,4);
          \corner{(0,0)}{0}{90}{.8}{orange};
          \corner{(0,0)}{90}{180}{.8}{yellow!50!green};
          \corner{(0,0)}{180}{180+atan(1/3)}{1}{teal};
          \corner{(0,0)}{180+atan(1/3)}{360}{.8}{magenta};
         \draw[green!50!black, very thick, line join=round]
             (1,0)--(0,1)--(-1,0)--(-3,-1)--cycle;
         \draw[green!67!black, very thick, -stealth](0,0)--(1,0);
          \path(1,0)node[right, green!50!black]{\footnotesize$\q_1$};
          \path(45:.71)node[above right=-3pt, green!50!black]
             {\footnotesize$\q_{12}$};
         \draw[green!67!black, very thick, -stealth](0,0)--(0,1);
          \path(0,1)node[above, green!50!black]{\footnotesize$\q_2$};
          \path(135:.71)node[above left=-3pt, green!50!black]
             {\footnotesize$\q_{23}$};
         \draw[red, very thick, -stealth](0,0)--(-1,0);
          \path(-1,.15)node[left=-2pt, red]{\footnotesize$\q_3$};
          \path(-1.67,-.15)node[left=-3pt, green!50!black]
             {\footnotesize$\q_{34}$};
         \draw[green!67!black, very thick, -stealth](0,0)--(-3,-1);
          \path(-3,-1)node[below right=-3pt, green!50!black]
             {\footnotesize$\q_4$};
          \path(265:.55)node[green!50!black]{\footnotesize$\q_{41}$};
         \draw[very thick, densely dotted, -stealth](230:.6)to[out=240,in=90]
           node[right=-3pt]{\footnotesize\eqref{e:tP}}++(0,-2.5);
         \path(160:1.4)node[left]{$\pDs{\FF[2]3}$};
         \filldraw(-3,-1)circle(.5mm);
         \foreach\x in{-1,...,1}\filldraw(\x,0)circle(.5mm);
         \foreach\x in{-1,...,1}\filldraw(\x,1)circle(.5mm);
         \filldraw[thick, fill=white](0,0)circle(.5mm);
            }}
 \end{picture}
 \caption{The $m=1,2,3$ sequence of (trans)polar pairs $(\pDN{\FF[2]m},\pDs{\FF[2]m})$; note that $\sfa\Q_3=\sfa\Q_2\cap\sfa\Q_4$ accounts for the overlap --- this corresponds to the non-convexity of $\pDs{\FF[2]3}$ at the concave vertex $\q_3$}
 \label{f:tP}
\end{figure}
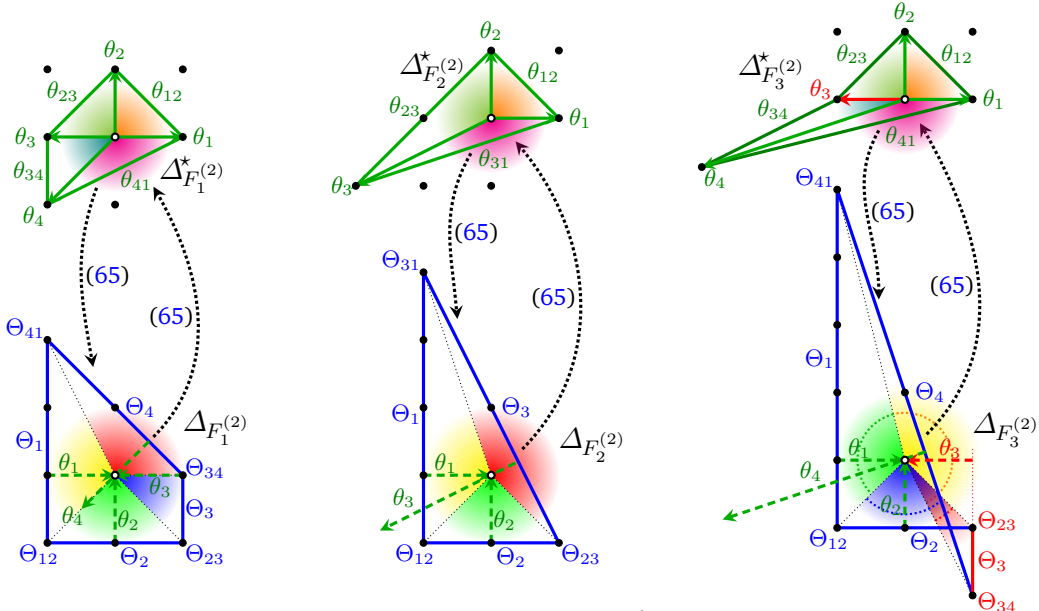

 Standard complex algebraic toric geometry restricts to {\em\/convex polytopes\/}\footnote{The 2-dimensional case of the operation~\eqref{e:tP} was called simply the {\em\/dual,} and 2-dimensional VEX multitopes are {\em\/generalized legal loops\/} in Ref.\cite{rP+RV-12}. The polar operation itself ``stretches back into the mists of time (David Gale, Max Dehn, Duncan Sommerville, etc.), but reflexive polytopes have been defined by Viktor Batyrev\cite{rBaty01},'' as communicated to me by Hal Schenck.} and the {\em\/fans\/} they span\cite{rD-TV, rO-TV, rF-TV, rGE-CCAG, rCLS-TV, rCK}. The rightmost plot in Figure~\ref{f:F1-2-3} however indicates the need for the flip-folding generalizations, called {\em\/multitopes,} and the {\em\/multifans\/} they span\cite{rM-MFans, Masuda:2000aa, rHM-MFs, Masuda:2006aa, rHM-EG+MF, rH-EG+MFs2, Nishimura:2006vs, Ishida:2013ab, rBH-gB, Berglund:2022dgb, Berglund:2024zuz}, as depicted by the sequence in Figure~\ref{f:tP}.
The fans spanned by the so-obtained polygons $\pDs{\FF[2]m}$ indeed encode the standard $m=1,2,3$ Hirzebruch surfaces, $\FF[2]m$. But the combinatorial operation~\eqref{e:tP} works equally well ``the other way around,'' so that the (multi)fan spanned by $\pDN{\FF[2]m}$ corresponds to a torus manifold, ${}^\wtd\!\FF[2]m$, on the right-hand (mirror) half in the diagram~\eqref{e:gMM}. As long as the polygons
 $(\pDN{X},\pDs{X})$ are convex and reflexive, both $X$ and $Y$ are Fano complex algebraic varieties and the two transpose anticanonical hypersurfaces
 $X[c_1]\ni Z_f \too{~~} Z_g\in Y[c_1]$ are each other's mirror\cite{rBaty01, rCK, rBH-gB}.

\subsection{VEX Multitopes}
\label{s:VEX}
The notions of {\em multifan\/}\cite{rM-MFans, Masuda:2000aa, rHM-MFs} spanned by a {\em\/VEX multitope\/}\cite{rBH-gB, Berglund:2022dgb, Berglund:2024zuz} generalize (and so include) the definitions of the standard fan\cite{rKKMS-TE1, rD-TV, rO-TV, rF-TV, rGE-CCAG, rCLS-TV, rCK} spanned by a reflexive convex polytope\cite{rBaty01}, respectively:
\begin{defn}
 Fans (polytopes) are {\em\/flat\/} multifans (multitopes), without mutually overlapping cones (faces). {\em\/VEX\/} multitopes (resp.\ {\em\/reflexive\/} polytopes) span (are star-triangulated by) {\em\/regular\/} multifans (resp.\ fans) --- which consist of unit-degree cones, the top-dimensional of which encode smooth coordinate charts covering a smooth torus manifold corresponding to the multifan and its spanning VEX multitope.
\end{defn}

The inclusion of non-convex polytopes and (self-overlapping and otherwise multilayered) multitopes (as exemplified on the far right of Figure~\ref{f:tP}) generalizes this to torus manifolds that are neither Fano nor necessarily complex algebraic varieties\cite{rM-MFans, Masuda:2000aa, rHM-MFs, Masuda:2006aa, rHM-EG+MF, rH-EG+MFs2, Nishimura:2006vs, Ishida:2013ab, rBH-gB, Berglund:2022dgb, Berglund:2024zuz}. Nevertheless, torus manifolds corresponding to VEX multitopes (star-triangulable by regular multifans) are smooth manifolds with maximal toric action, i.e., $U(1;\IC)^n$-gauge symmetry such as the projective $\IC^*$-scaling in Sections~\ref{s:FmXm}--\ref{s:TR}\ However, the complex (and mirror-symmetrically, symplectic) structure may become ill-defined at certain locations. A highly simplified example is the 4-sphere, which admits no global complex structure, while
 $(S^4\ssm\{\text{pt.}\})\approx\IR^4\approx\IC^2$ clearly does:
A single point obstructs extending the complex structure over all of $S^4$.
Instead, ``blowing up'' that one point (replacing it with $S^2\approx\IP^1$) produces
\begin{equation}
  S^4 ~\leadsto~
 (S^4\ssm\{\text{pt.}\})\cup\IP^1\approx\IC^2\cup\IP^1\approx\IP^2,
 \label{e:S4>P2}
\end{equation}
which of course {\em\/does\/} admit a global complex structure.
 Amusingly, this one-point obstruction to a global complex structure however does not preclude the construction of the Dolbeault complex (at heart of superstring applications), generated by the holomorphic exterior derivative and its Hermitian conjugate\cite{rS-S4notCplx}. I am not aware of any higher-dimensional generalization of this result nor of a generalization of this construction.

Polar pairs of convex polytopes such as depicted on the left-hand side of Figure~\ref{f:tP} both encode standard (complex algebraic) toric varieties.
 The top-left polytope, $\pDs{\FF[2]1}$, spans a simple fan consisting of four cones,
\begin{equation}
   \pDs{\FF[2]1}\<\smt \S_{\smash{\FF[2]1}} =
   \{\sfa(\q_{i,\,[i+1\:(\mathrm{mod}\:4)]})\},
\end{equation}
each of which have degree~1,\footnote{In two dimensions, this is seen easily by noting that the base-facets, $\overline{\q_i,\q_{i+1}}$ contain no subdividing lattice points. Also\cite{rBeast}, the matrix obtained by stacking the cone-generating vectors is unimodular,
 $\det\!\big[\!\big[\vec\q_i|\vec\q_{i+1}\big]\!\big]\<= 1$, which generalizes to higher dimensions.} and so encodes a smooth, $\IC^2$-like chart. These charts glue pairwise along the $\IP^1$ corresponding to the common $\vec\q_i$.
 In turn, the bottom-left polytope, $\pDN{\FF[2]1}$, spans a fan consisting of four cones of varying degrees:
\begin{equation}
  \deg\sfa(\Q_1)=3,\quad
  \deg\sfa(\Q_2)=2,\quad
  \deg\sfa(\Q_3)=1,\quad
  \deg\sfa(\Q_4)=2.
\end{equation}
These then encode, respectively, charts of the form
 $\IC^2/\ZZ_3$, $\IC^2/\ZZ_2$, $\IC^2$, $\IC^2/\ZZ_2$.
In order to glue together forming a {\em\/smooth\/} manifold, all but the third one require (``MPCP''\cite{rBaty01}) desingularizing blowups --- encoded by the subdividing lattice points along the base-edges, each of which corresponds to an {\em\/exceptional\/} (and isolated) $\IP^1$, four in total. The so-constructed smooth toric manifold is in fact a standard complex algebraic variety, which we denote $\tW\FF[2]1$.

The top-right polytope analogously encodes the Hirzebruch surface, $\FF[2]3$, via its fan, $\S_{\smash{\FF[2]3}}\<\lat \pDs{\FF[2]3}$. The transpolar operation~\eqref{e:tP} produces the bottom-right flip-folded multitope, $\pDN{\FF[2]3}$, where the orientation of the cone $\sfa(\Q_3)$ by continuity is retrograde (clockwise) as compared to that of the rest of the (counter-clockwise) cones: This retrograde cone $\sfa(\Q_3)$ is transpolar to the {\em\/convex\/} vertex $\q_3\in\pDs{\FF[2]3}$, and its lattice-points $\Q_{23}$ and $\Q_{34}$ correspond to the rational monomials, $X_2\!^2/X_4$ and $X_2\!^2/X_3$, respectively; see Figure~\ref{f:2F3F1}. The vertex generators $\vec\Q_{23}$ and $\vec\Q_{34}$ encode the glueing of this retrograde cone to the preceding and succeeding forward-oriented cones, and this cone-orientation reversal indicates that the local chart of 
 $\sfa(\Q_3)$ is glued in the the remaining patchwork in a ways that is not compatible with a global complex structure. This local chart thus contains an {\em\/obstruction\/} to a global complex structure, i.e., a {\em\/defect\/} in it --- perhaps not too different from~\eqref{e:S4>P2}, and so usable in string theory.

Explicit deformation families of generalized complete intersections such as
 $\XX3\<\in \ssK[{r||c|c}{\IP^n&1&n\\ \IP^1&3&-1}]\subset
  \ssK[{r||c}{\IP^n&1\\ \IP^1&3}]$ contain both models diffeomorphic to standard algebraic hypersurfaces in (semi-)Fano varieties as well as the exceptional (central, $\e\<= 0$) model $\XX3\<\subset \FF3$, where
 $\FF3$ is not Fano, and $\XX3$ is either reducible and so singular (Tyurin-degenerate), or (non-algebraically) Laurent-smoothed. Its transposition-mirror \`a la~\eqref{e:gMM} is an anticanonical hypersurface in the torus manifold, $\tW\FF3$, corresponding to the flip-folded multifan spanned by $\pDN{\FF3}$, as shown at the right in Figure~\ref{f:tP} for $n\<= 2$.
 The coordinate chart $\mathcal{U}_3\<\subset \tW\FF3$ encoded by the cone 
 $\sfa\Q_3$ glues to the adjacent charts
 ($\mathcal{U}_2$ and $\mathcal{U}_4$) 
 in a ``wrong'' way (corresponding to the flip-folding of
 $\sfa\Q_3$ between
 $\sfa\Q_2$ and $\sfa\Q_4$),
 which obstructs the complex structure that is well defined on 
 $\FF3\<\ssm \mathcal{U}_3$. Intersecting this $\mathcal{U}_3$ makes the mirror of $\XX3$ also {\em\/pre-complex\/} and its (putative) K\"ahler metric degenerate.
 Homological mirror symmetry then implies that the symplectic structure in $\XX3$ is analogously obstructed, making it {\em\/pre-symplectic,} defining a correspondingly degenerate (pre-)K\"ahler metric; see observation~\ref{o:CKS} in \SS\;\ref{s:org}. 
 This motivates:
\begin{conj}\label{C:pCS}
Each regular, complex-algebraic Calabi--Yau hypersurface $Z_f\<\subset X$ in a non-Fano variety is singular, while its Laurent-smoothing\cite{rBH-gB, Berglund:2022dgb, Berglund:2024zuz} is pre-symplectic and pre-K\"ahler: its (putative) symplectic form and K\"ahler metric degenerate at isolated loci. The \eqref{e:gMM0}-mirror, $Z_g\<\subset Y$, is analogously pre-complex and pre-K\"ahler, as is $Y$. The ``compatible triples'' of Laurent-smoothed Calabi--Yau hypersurfaces within torus manifolds corresponding to generic (non-convex and multi-layered) VEX multitopes are degenerate at isolated loci, i.e., have isolated $(J,\w,g_{\sss K})$-degenerating defects.
\end{conj}
\noindent
 Reid's original conjecture\cite{rReidK0} includes non-K\"ahler (complex-analytic, but not algebraic) Calabi--Yau 3-folds $(S^3\<\times S^3)^{\#N}$ {\em\/as the generic models,} where the successive conifold/extremal/geometric transitions and blowdown iterations render $h^{1,1}\<= b_2\<= 0$, so the symplectic 2-form (or its compatibility with the complex structure) also degenerates.
 The number, $N$, of connected copies of $(S^3\<\times S^3)$ as well as the dimension of the complex structure moduli space may become arbitrarily large --- precisely akin to Deligne--Mumford's ``universal curve.'' The foregoing discussion then aims to generalize this mirror-symmetrically, allowing the compatible triple, $(J,\w,g_{\sss K})$, to degenerate {\em\/arbitrarily far/often\/} in both directions, and transporting the mirror-symmetric framework~\eqref{e:gMM0} along.
 As these ``defective'' but smooth manifolds are $\e\<\to 0$ limits of models that are diffeomorphic to standard Calabi--Yau hypersurfaces in (semi-)Fano varieties, they should all be admissible in string theory\cite{McNamara:2019rup}.

The transpolar construction~\eqref{e:tP} corresponds concave regions in a VEX multitope to flip-folded regions in its transpolar VEX multitope.
 Now, a flip-folded region in a multitope encodes local charts in a corresponding torus manifold that contain isolated obstructions to the complex structure,
 e.g., $\mathcal{U}_3\<\subset \tW\FF3$, specified by the multifan spanned by the flip-folded $\pDN{\FF3}$; see right-hand side of Figure~\ref{f:tP} for the $n\<= 2$ case.
 It is then tempting to conclude that the transpolar relation corresponds this to isolated obstructions to the symplectic structure of the torus manifold corresponding to the transpolar multitope. This however cannot be true, since
 $\pDs{\tW\FF3}\!:=\!\pDN{\FF3}\!
   \xleftrightarrow{{}_\wtd}\pDs{\FF3}\<\smt \S_{\smash{\FF3}}$,
and $\FF{m}$ are known to be K\"ahler for all $(m,n)$, each of which then has a compatible triple that contains an unobstructed, non-degenerate symplectic structure.
 What exactly does the mirror-symmetric complex--symplectic correspondence at the ``ground floor'' of~\eqref{e:gMM} and~\eqref{e:gMM0} imply to the two ``higher floors'' in this framework, $(X\<\ssm Z_f)$ vs.\ $(Y\<\ssm Z_g)$ and then $\cK^*_X$ vs.\ $\cK_Y^*$, thus remain tantalizingly open.

\subsection{Surgery}
\label{s:Srg}
Toric --- or at least $(S^1)^n$-equivariant ``surgery'' operations can be used to relate complex algebraic toric varieties to their non-algebraic, non-complex torus manifold generalizations\cite{rM-MFans, Masuda:2000aa, rHM-MFs, Masuda:2006aa, rHM-EG+MF, rH-EG+MFs2, Nishimura:2006vs, Ishida:2013ab, rBH-gB, Berglund:2022dgb, Berglund:2024zuz}. This then equally applies also to the Calabi--Yau hypersurfaces therein, which may intersect/contain the so-``repaired'' such special locations.
 In turn, this strongly reminds of the algebraic suturing of two different solution sheets in the earliest known exact solutions to Einstein equations\cite{Kasner:1921Fin, rCF-BH}, as well as the time-honored ``repairing'' of singularities by replacing them with Eguchi-Gilkey patches and their generalizations\cite{Candelas:1989js, rBeast, rBBS, rD+G-DBrM, He:2018jtw}.

\begin{figure}[htbp]
 \centering\setlength{\unitlength}{1mm}
  \begin{picture}(110,52)
   \put(0,0){\TikZ{[thick]\path[use as bounding box](-1,-2);
              \fill[yellow!50](0,0)--(-1,3)--(-1,-2)--(1,0)--(0,0);
               \draw[brown, line join=round](-1,3)--(-1,-2)--(1,0);
              \fill[red!75, opacity=.85](0,0)--(1,0)--(1,-1)--(0,0);
               \draw[magenta]((1,0)--(1,-1);
               \draw[magenta, densely dashed, -Stealth](0,0)--(1,0);
              \fill[yellow!50, opacity=.85](0,0)--(1,-1)--(-1,3)--(0,0);
               \draw[brown, line join=round](1,-1)--(-1,3);
               \draw[-Stealth](0,0)--(-1,3);
               \draw[-Stealth](0,0)--(-1,-2);
               \draw[densely dotted, stealth-stealth](0,1)--(0,-1);
              \foreach\y in{-1,...,2}
               \draw[densely dotted, -stealth](0,0)--(-1,\y);
               \draw[magenta, densely dashed, -Stealth](0,0)--(1,-1);
              \foreach\y in{-2,...,3}
               \filldraw(-1,\y)circle(.3mm);
              \foreach\y in{-1,...,1}
               \filldraw(0,\y)circle(.3mm);
              \foreach\y in{-1,0}
               \filldraw[magenta](1,\y)circle(.4mm);
               \path(1,0)node[above, magenta]{\scriptsize$\Q_{23}$};
               \draw[magenta, densely dashed, <->](.9,.5)to[out=75, in=105]
                   node[above]{\footnotesize identify}++(4.5,0);
               \path(1,-1)node[right, magenta]{\scriptsize$\Q_{34}$};
               \draw[magenta, densely dashed, <->](1.4,-1.2)to[out=-60, in=170]
                   ++(1,-.8)to[out=-10, in=-120]
                   node[above left, rotate=-5]{\footnotesize identify}++(3,.6);
               \path(0,-1.5)node[right]{$\pDN{\FF[2]3}$};
              \filldraw[fill=white](0,0)circle(.5mm);
            }}
   \put(27,20){\LARGE$\biguplus_{\color{magenta}2\IP^1}$}
   \put(45,0){\TikZ{[thick]\path[use as bounding box](-1,-2);
              \fill[yellow!50](-2,1)--(1,-1)--(1,0);
               \draw[brown, line join=round](-2,1)--(1,-1)--(1,0)--cycle;
               \draw[-Stealth, blue](0,0)--(-2,1);
               \draw[magenta, densely dashed, -Stealth](0,0)--(1,0);
               \draw[magenta, densely dashed, -Stealth](0,0)--(1,-1);
               \filldraw[fill=white](0,0)circle(.5mm);
               \filldraw(-2,1)circle(.3mm);
                \path(-2,1)node[above right=-2pt, blue]
                {\scriptsize$\Q_*$};
               \filldraw(1,0)circle(.3mm);
               \filldraw(1,-1)circle(.3mm);
               \path(1,0)node[above, magenta]{\scriptsize$\Q_{23}$};
               \path(1,-1)node[below=-2pt, magenta]{\scriptsize$\Q_{34}$};
               \draw[red, densely dashed, double, <->](.8,-.5)--
                   node[below]{\footnotesize cancel}++(-4.5,0);
               \path(0,-1)node{$\pDN{{}^\wtd\!\IP^2}$};
            }}
   \put(70,22){\Huge$=$}
   \put(90,0){\TikZ{[thick]\path[use as bounding box](-1,-2);
              \fill[yellow!50](0,0)--(-2,1)--(1,0)--(0,0);
               \draw[brown, line join=round](-2,1)--(1,0);
              \fill[yellow!50, opacity=.85]
                  (0,0)--(-1,3)--(-1,-2)--(1,0)--(0,0);
               \draw[brown, line join=round](-1,3)--(-1,-2)--(1,0);
               \draw[-Stealth](0,0)--(1,0);
               \draw[-Stealth](0,0)--(-1,-2);
               \draw[densely dotted, -stealth](0,0)--(0,-1);
              \foreach\y in{-1,0}
               \draw[densely dotted, -stealth](0,0)--(-1,\y);
              \fill[yellow!50, opacity=.9]
                 (0,0)--(-2,1)--(1,-1)--(-1,3)--(0,0);
               \draw[brown, line join=round](-2,1)--(1,-1)--(-1,3);
               \draw[-Stealth](0,0)--(-1,3);
               \draw[-Stealth](0,0)--(1,-1);
               \draw[-Stealth, blue](0,0)--(-2,1);
               \draw[densely dotted, -stealth](0,0)--(0,1);
              \foreach\y in{1,2}\draw[densely dotted, -stealth](0,0)--(-1,\y);
               \filldraw(-2,1)circle(.3mm);
                \path(-2,1)node[above right=-2pt, blue]
                {\scriptsize$\Q_*$};
              \foreach\y in{-2,...,3}
               \filldraw(-1,\y)circle(.3mm);
              \foreach\y in{-1,...,1}
               \filldraw(0,\y)circle(.3mm);
              \foreach\y in{-1,0}
               \filldraw(1,\y)circle(.3mm);
               \path(1,0)node[above]{\scriptsize$\Q_{23}$};
               \path(1,-1)node[below]{\scriptsize$\Q_{34}$};
               \path(-1,-1.5)node[left]{$\pDN{\widehat{F}^{(2)}_3}$};
              \filldraw[fill=white](0,0)circle(.5mm);
            }}
  \end{picture}
 \caption{The flip-folded Newton multitope $\pDN{\FF[2]3}$ being ``repaired'' by splicing in the Newton polytope of 
 ${}^\wtd\!\IP^2\<= \mathrm{Bl}^\uA[\IP^2/\ZZ_3]$ along the two $\IP^1$s (dashed arrows), to produce the 2-sheeted (doubly winding) Newton multitope
 $\pDN{\widehat{F}^{(2)}_3}$}
 \label{f:FF}
\end{figure}
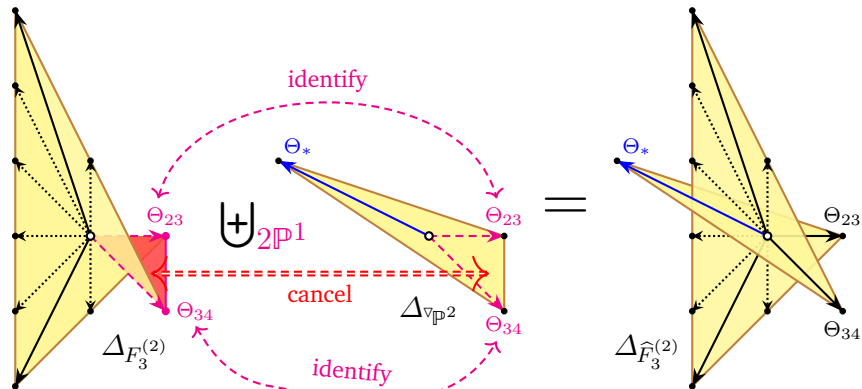

The multigon encoding of an example of such surgery is shown in Figure~\ref{f:FF}.
The flip-folded multigon at far left is the same as depicted on the right in Figures~\ref{f:F1-2-3} and~\ref{f:tP} and left of Figure~\ref{f:2F3F1}, up to a $\mathrm{GL}(2;\ZZ)$ ``skewing'' transformation to save some vertical space and better showcase the flip-fold on its right.
 Splicing in the Newton polygon (middle) of 
 ${}^\wtd\!\IP^2=\mathrm{Bl}^\uA[\IP^2/\ZZ_3]$ (a desingularizing ``MPCP-blowup''\cite{rBaty01} of $\IP^2/\ZZ_3$) along the two $\IP^1$s corresponding to the two 1-cones (dashed) produces the double-winding, 2-sheeted multigon on the far right. This spans a double-winding multifan that corresponds to a unitary torus manifold\cite{rHM-MFs}.
 Rather evidently, every flip-folded VEX multitope can be related in this way to a possibly multi-winding VEX multitope without flip-folds and at least one ordinary, plain (and reflexive) polytope.

Every (complex-algebraic) toric variety $X$ is precisely encoded by its Newton polytope, $\pDN{X}$\cite{rKKMS-TE1, rD-TV, rO-TV, rF-TV, rGE-CCAG, rCLS-TV, rCK}. If this lattice polytope has a single internal lattice point, $0$, then it also encodes a {\em\/polar\/} toric variety, $Y\!:=\!{}^\wtd\!\!X$, by means of the ({\em\/normal\/}) fan, $\S_Y$, of $0$-centered cones over the faces of $\pDN{X}$; $\S_Y\<\lat \pDN{X}$. The face-wise {\em\/transpolar\/} operation~\eqref{e:tP} generalizes this universally to all flip-folded or otherwise multi-layered VEX multitopes and multifans that they span\cite{rBH-gB, Berglund:2022dgb, Berglund:2024zuz}. By definition, VEX multitopes have a single {\em\/central\/} lattice point (even if flip-folding prevents them from having a strict interior), from which all faces are at unit lattice distance so their transpolar images are also lattice faces. 

 Alternatively, the surgery in Figure~\ref{f:FF} may be understood as a ``blowout'' of the $\IC^2$-like coordinate patch corresponding to the red-shaded
 $\sfa(\overline{\Q_{23},\Q_{34}})$, with the exceptional divisor corresponding to the lattice vector $\Q_*$.\footnote{The vector 
 $\Q_*\not\subset\sfa(\overline{\Q_{23},\Q_{34}})$ implies that this ``blowout'' is {\em\/not\/} the standard complex-algebraic {\em\/blowup,} but a variant that locally obstructs the complex structure.}
 The multitopes in Figure~\ref{f:FF} evidently span multifans, which then correspond to ``dual'' (transpolar) torus manifolds in this second way:
\begin{equation}
  \big((\pDN{\FF[2]3}{\smt}\S){:}\,\tW\FF[2]3\big),~~
  \big((\pDN{\tW\IP^2}{\smt}\S){:}\,\IP^2\big)
  ~~\text{and}~~
  \big((\pDN{\widehat{F}^{\sss(2)}_3}{\smt}\S){:}\,\tW\widehat{F}^{\sss(2)}_3\big),
\end{equation}
so the unitary torus manifold $\tW\widehat{F}^{\sss(2)}_3$ corresponding to the doubly-winding multifan spanned on the right-hand side of Figure~\ref{f:FF} is the non-complex-algebraic ``blowup'' of $\tW\FF[2]3$ corresponding to the flip-folded multifan spanned on the left-hand side of that figure.

Unlike complex-algebraic toric varieties, torus manifolds are not uniquely specified by a flip-folded or otherwise multi-layered VEX multitope. The various studies in the existing literature\cite{rM-MFans, Masuda:2000aa, rHM-MFs, Masuda:2006aa, rHM-EG+MF, rH-EG+MFs2, Nishimura:2006vs, Ishida:2013ab, Davis:1991uz, Ishida:2013aa, buchstaber2014toric, Jang:2023aa} have not been developed with mirror symmetry~\eqref{e:gMM} in mind, and so do not present any choice of additional data particularly suited to this end. A precise understanding of the surgery in Figure~\ref{f:FF} as well as the particular torus manifold corresponding to the flip-folded multifan spanned on the left-hand side then remain an open challenge for now.

\subsection{Multiple Mirrors}
\label{s:MM}
The original scope of the transposition mirror construction restricts to {\em\/Landau-Ginzburg orbifolds\/} (LGO), i.e., the zero-locus of so-called {\em\/invertible\/} polynomials, $f(x)$:
 These are linear combinations of $N$ monomials in $N$ variables, which are {\em\/transverse\/}: $\rd f(x)$ and $f(x)$ both vanish only at the origin, $x\<= 0$, and the matrix of exponents in $f(x)$ is regular\cite{rBH, rBH-LGO+EG, Krawitz:2009aa}. 
 Away from the origin, this is an $(N{-}2)$-dimensional hypersurface
 $\{f(x)\<= 0\}$ in the weighted-projective space,
 $\IP^{N-1}_{(q_1{:}\cdots{:}q_N)}$, where $q_i$ is the integral
 weight of the $i^{\text{th}}$ variable, equal to the $U(1)$-{\em\/charge\/} of the  corresponding worldsheet superfield. The field-space in such models (and their geometry!) has two distinct complete $U(1)$-orbits: the trivial origin (the LGO ``phase''), and the non-trivial complement (the CY ``phase'').
 GLSMs\cite{rPhases, rMP0, Distler:1993mk, Schafer-Nameki:2016cfr} generalize this to $U(1)^n$ gauge groups complexified by target-space supersymmetry to $(\IC^*)^n$ toric actions with a more diverse ``phase diagrams''; see for example\cite{rBH-gB} and~\eqref{e:2ndFn}.
 
In toric diagrams such as in Figure~\ref{f:tP}, transpose-mirror pairs of invertible defining polynomials are found by choosing a particular center-enclosing\footnote{Requiring transverseness and a few additional technical details seem to imply that both simplices, $(\Delta,\nabla)$, must {\em\/enclose\/} the center within their {\em\/strict interior\/}\cite{Berglund:2024zuz}.} {\em\/simplex\/} in each of the polytopes in a transpolar pair,
 $\Delta\<\subset \pDs{X}$ and $\nabla\<\subset \pDN{X}$:
 ({\small\bf1})~one to define the homogeneous coordinates,
 $x_\rho\<\iff \rho\<\in \Delta$,
 ({\small\bf2})~the other defines $f(x)$ as a linear combination of their anticanonical monomials,
 $f(x)=\sum_{\m\in\nabla}a_\m\prod_{\rho\in\Delta}x_\rho^{\vev{\m,\rho}+1}$.
Transposition mirror symmetry swaps the roles of these simplices:
 $(\Delta,\nabla)\<\iff (\nabla,\Delta)$.

A glance at Figure~\ref{f:tP} reveals that all $\pDs{\FF[2]m}$ have a single such simplex, $\Delta=\mathrm{Span}(\q_1,\q_2,\q_4)$, while there are multiple choices for $\nabla\<\subset \pDN{\FF[2]m}$.
 The unreduced pair $(\pDs{\FF[2]1},\pDN{\FF[2]1})$ specifies the transposed pair:
\begin{subequations}
 \label{e:2F1mm}
\begin{alignat}9
 f(x) &= a_1x_1\!^2x_3\!^3 +a_2x_1\!^2x_4\!^3 +a_3x_2\!^2x_4 +a_4x_2\!^2x_3,
 \label{e:2F1f}\\*
 g(x) &= b_1y_1\!^2y_2\!^2 +b_2y_3\!^2y_4\!^2 +b_3y_1\!^3y_4 +b_4y_2\!^3y_3,
 \label{e:2F1g}
 \qquad \smash{\raisebox{3mm}{$\mathbb{E}=
               \bigg[\begin{smallmatrix}
                             2 & 2 & 0 & 0 \\[1pt]
                             0 & 0 & 2 & 2 \\[1pt]
                             3 & 0 & 0 & 1 \\[1pt]
                             0 & 3 & 1 & 0 \\
                     \end{smallmatrix}\bigg],$}}
\end{alignat}
\end{subequations}
where $\mathbb{E}$ is the matrix of exponents: columns corresponding to the $f(x)$-monomials in row-indexed $x_i$, and vice versa for $g(y)$.
Analogously,
the unreduced pair $(\pDs{\FF[2]3},\pDN{\FF[2]3})$ specifies\cite{Berglund:2022dgb}:
\begin{subequations}
 \label{e:2F3mm}
\begin{alignat}9
 f(x) &= a_1x_1\!^2x_3\!^5 +a_2x_1\!^2x_4\!^5
        +a_3\tfrac{x_2\!^2}{x_4} +a_4\tfrac{x_2\!^2}{x_3},
 \label{e:2F3f}\\*
 g(x) &= b_1y_1\!^2y_2\!^2 +b_2y_3\!^2y_4\!^2
        +b_3\tfrac{y_1\!^5}{y_4} +b_4\tfrac{y_2\!^5}{y_3},
 \label{e:2F3g}
 \qquad \smash{\raisebox{3mm}{$\mathbb{E}=
               \bigg[\begin{smallmatrix}
                             2 & 2 & 0 & 0 \\[1pt]
                             0 & 0 & 2 & 2 \\[1pt]
                             5 & 0 & 0 &\!\!-1 \\[1pt]
                             0 & 5 &\!\!-1 & 0 \\
                     \end{smallmatrix}\bigg],$}}
\end{alignat}
\end{subequations}
All four hypersurfaces are smooth for generic choices of the $a_i,b_j$ coefficients, e.g., $a_1\,a_4\!^5\<\neq a_2\,a_3\!^5$ in~\eqref{e:2F3f} and
$b_1\!^5\<\neq b_2\,b_3\!^2\,b_4\!^2$ in~\eqref{e:2F3g}.
 In both cases, $\mathrm{rank}\,\mathbb{E}=3$ implies that the original transposition construction\cite{rBH, rBH-LGO+EG, Krawitz:2009aa, rLB-MirrBH} does not apply to the pairs~\eqref{e:2F1mm} and~\eqref{e:2F3mm}, but {\em\/does apply\/} to judicious reductions to ``$3{\times}3$'' subsystems, encoded by pairs of origin-enclosing simplices found within the original pair,
 $(\pDs{\FF[2]m},\pDN{\FF[2]m})$.

Suffice it here to compare two sample choices from each, where
 in~\eqref{e:2F1f} and~\eqref{e:2F3f} we simply set $x_1\<\to 1$ and $a_4\<\to 0$, and absorb the remaining $a_i,b_j$ coefficients into the coordinates. 
 In~\eqref{e:2F1g} and~\eqref{e:2F3g} we correspondingly set $b_1\<\to 0$ and $y_4\<\to 1$, and for a second simplex-reduction replace $(-1,-1)\<\to (-1,0)$, and correspondingly write $y_2\<\to \eta_2$:
\begin{subequations}
 \label{e:2F1a}
\begin{alignat}9
 \vC{\TikZ{[scale=.75, line join=round, thick]
            \path[use as bounding box](-1,1.5)--(1.5,2);
            \draw[orange, line width=3pt, opacity=.75]
                (1,0)--(0,1)--(-1,-1)--cycle;
            \draw[green!67!black](1,0)--(0,1)--(-1,0)--(-1,-1)--cycle;
            \foreach\y in{-1,...,1}
             \foreach\x in{-1,...,1}
              \fill(\x,\y)circle(.6mm);
            \filldraw[fill=white](0,0)circle(.6mm); 
            \path(.2,0)node[above]{$\Delta_1$};
            }}
 \qquad
 \vC{\TikZ{[scale=.75, line join=round, thick]
            \path[use as bounding box](-1,1.5)--(1.5,2);
            \draw[magenta, line width=3pt, opacity=.75]
                (-1,2)--(-1,-1)--(1,-1)--cycle;
            \draw[blue](-1,2)--(-1,-1)--(1,-1)--(1,0)--cycle;
            \foreach\y in{-1,...,2}
             \foreach\x in{-1,...,1}
              \fill(\x,\y)circle(.6mm);
            \filldraw[fill=white](0,0)circle(.6mm); 
            \path(0,0)node[above left]{$\nabla_{\!1}$};
            }}
 \qquad
 \{f_1(x)&=x_3\!^3 +x_4\!^3 +x_2\!^2x_4=0\}\in \IP^2[3],\\*
       &Q\<= \zZ3{\tfrac13{,}\tfrac13{,}\tfrac13},~~
        G\<= \zZ6{\tfrac16{,}0{,}\tfrac23};\\*
 \{g_1(y)&=y_3\!^2 +y_1\!^3 +y_2\!^3y_3=0\}\in \IP^2_{(2{:}1{:}3)}[6],\\*
       &\Tw{Q}\<= \zZ6{\tfrac13{,}\tfrac16{,}\tfrac12},~~
        \Tw{G}\<= \zZ3{\tfrac13{,}0{,}0}.
\end{alignat}
\end{subequations}
The ``quantum''-``geometric'' pair of symmetries, $(Q,G)$ and $(\Tw{Q},\Tw{G})$ for the mirror, clearly distinguish this from 
\begin{subequations}
 \label{e:2F1b}
\begin{alignat}9
 \vC{\TikZ{[scale=.75, line join=round, thick]
            \path[use as bounding box](-1,1.5)--(1.5,2);
            \draw[orange, line width=3pt, opacity=.75]
                (1,0)--(0,1)--(-1,-1)--cycle;
            \draw[green!67!black](1,0)--(0,1)--(-1,0)--(-1,-1)--cycle;
            \foreach\y in{-1,...,1}
             \foreach\x in{-1,...,1}
              \fill(\x,\y)circle(.6mm);
            \filldraw[fill=white](0,0)circle(.6mm); 
            \path(.2,0)node[above]{$\Delta_2$};
            }}
 \qquad
 \vC{\TikZ{[scale=.75, line join=round, thick]
            \path[use as bounding box](-1,1.5)--(1.5,2);
            \draw[magenta, line width=3pt, opacity=.75]
                (-1,2)--(-1,0)--(1,-1)--cycle;
            \draw[blue](-1,2)--(-1,-1)--(1,-1)--(1,0)--cycle;
            \foreach\y in{-1,...,2}
             \foreach\x in{-1,...,1}
              \fill(\x,\y)circle(.6mm);
            \filldraw[fill=white](0,0)circle(.6mm); 
            \path(0,0)node[above left]{$\nabla_{\!2}$};
            }}
 \qquad
 \{f_2(x)&=x_3\!^3 \!+\!x_3x_4\!^2 \!+\!x_2\!^2x_4 \<= 0\}\in \IP^2[3],\\*
       &Q\<= \zZ3{\tfrac13{,}\tfrac13{,}\tfrac13},~~
        G\<= \zZ4{\tfrac14{,}0{,}\tfrac12};\\*
 \{g_2(y)&=y_3\!^2 \!+\!\eta_2y_1\!^3 \!+\!\eta_2\!^2y_3 \<= 0\}
           \in \IP^2_{(1{:}1{:}2)}[4],\\*
       &\Tw{Q}\<= \zZ4{\tfrac14{,}\tfrac14{,}\tfrac12},~~
        \Tw{G}\<= \zZ3{\tfrac13{,}0{,}0}.
\end{alignat}
\end{subequations}
even though $f_1(x)$ is clearly a simple deformation of $f_2(x)$. Using the same two simplices within $\pDN{\FF[2]3}$ paired with the analogous one in $\pDs{\FF[2]3}$ requires starting with $a_1x_1\!^2x_3\!^5\<\to a_1x_1\!^2x_3\!^3x_4\!^2$ in~\eqref{e:2F3f} and using the Cox coordinate $y_1\<\to \eta_1$ corresponding to ``lowering'' $(-1,4)\<\in \pDN{\FF[2]3}$ to the non-vertex lattice point $(-1,2)$:
\begin{subequations}
 \label{e:2F3a}
\begin{alignat}9
 \vC{\TikZ{[scale=.5, line join=round, thick]
            \path[use as bounding box](-3,2.5)--(1.5,3);
            \draw[orange, line width=3pt, opacity=.75]
                (1,0)--(0,1)--(-3,-1)--cycle;
            \draw[green!67!black](1,0)--(0,1)--(-1,0)--(-3,-1)--cycle;
            \foreach\y in{-1,...,1}
             \foreach\x in{-3,...,1}
              \fill(\x,\y)circle(.6mm);
            \filldraw[fill=white](0,0)circle(.6mm); 
            \path(0,0)node[above]{\scriptsize$\Delta_3$};
            }}
 \qquad
 \vC{\TikZ{[scale=.5, line join=round, thick]
            \path[use as bounding box](-1,3)--(1.5,3.5);
            \draw[magenta, line width=3pt, opacity=.75]
                (-1,2)--(-1,-1)--(1,-1)--cycle;
            \draw[blue](-1,4)--(-1,-1)--(1,-1)--(1,-2)--cycle;
            \foreach\y in{-2,...,4}
             \foreach\x in{-1,...,1}
              \fill(\x,\y)circle(.6mm);
            \filldraw[fill=white](0,0)circle(.8mm); 
            \path(0,0)node[above left=-2pt]{\scriptsize$\nabla_{\!3}$};
            }}
 \qquad
 \{f_3(x)&=x_3\!^3x_4\!^2 +x_4\!^5 +\frac{x_2\!^2}{x_4}=0\}
            \in \IP^2_{(3{:}1{:}1)}[5],\\*
       &Q\<= \zZ5{\tfrac35{,}\tfrac15{,}\tfrac15},~~
        G\<= \zZ6{\tfrac12{,}\tfrac13{,}0};\\*
 \{g_3(y)&=y_3\!^2 +\eta_1\!^3 +\frac{\eta_1\!^2 y_2\!^5}{y_3}=0\}
            \in \IP^2_{(2{:}1{:}3)}[6],\\*
       &\Tw{Q}\<= \zZ6{\tfrac13{,}\tfrac16{,}\tfrac12},~~
        \Tw{G}\<= \zZ5{0{,}\tfrac15{,}0}.
\end{alignat}
\end{subequations}
as compared with\backUp[.5]
\begin{subequations}
 \label{e:2F3b}
\begin{alignat}9
 \vC{\TikZ{[scale=.5, line join=round, thick]
            \path[use as bounding box](-3,2.5)--(1.5,3);
            \draw[orange, line width=3pt, opacity=.75]
                (1,0)--(0,1)--(-3,-1)--cycle;
            \draw[green!67!black](1,0)--(0,1)--(-1,0)--(-3,-1)--cycle;
            \foreach\y in{-1,...,1}
             \foreach\x in{-3,...,1}
              \fill(\x,\y)circle(.6mm);
            \filldraw[fill=white](0,0)circle(.6mm); 
            \path(0,0)node[above]{\scriptsize$\Delta_4$};
            }}
 \qquad
 \vC{\TikZ{[scale=.5, line join=round, thick]
            \path[use as bounding box](-1,2.5)--(1.5,3);
            \draw[magenta, line width=3pt, opacity=.75]
                (-1,2)--(-1,0)--(1,-1)--cycle;
            \draw[blue](-1,4)--(-1,-1)--(1,-1)--(1,-2)--cycle;
            \foreach\y in{-2,...,4}
             \foreach\x in{-1,...,1}
              \fill(\x,\y)circle(.6mm);
            \filldraw[fill=white](0,0)circle(.8mm); 
            \path(0,0)node[above left=-2pt]{\scriptsize$\nabla_{\!4}$};
            }}
 \qquad
 \{f_4(x)&=x_3\!^3x_4\!^2 \!+\!x_3x_4^4 \!+\!\frac{x_2\!^2}{x_4}\<= 0\}
  \in \IP^2_{(3{:}1{:}1)}[5],\\*
       &Q\<= \zZ5{\tfrac35{,}\tfrac15{,}\tfrac15},~~
        G\<= \zZ4{\tfrac14{,}0{,}\tfrac12};\\*
 \{g_4(y)&=y_3\!^2 \!+\!\eta_1\!^3\eta_2
           \!+\!\frac{\eta_1\!^2\eta_2\!^4}{y_3} \<= 0\}
  \in \IP^2_{(1{:}1{:}2)}[4],\\*
       &\Tw{Q}\<= \zZ4{\tfrac14{,}\tfrac14{,}\tfrac12},~~
        \Tw{G}\<= \zZ5{\tfrac15{,}\tfrac25{,}0}.
\end{alignat}
\end{subequations}
The two distinct models, $\{g_1\<= 0\}\<\in \IP^2_{(2{:}1{:}3)}[6]$ and 
 $\{g_2\<= 0\}\<\in \IP^2_{(1{:}1{:}2)}[4]$, are dubbed ``multiple mirrors'' 
since they are mirror models of $\{f_1\<= 0\}$ and its explicit deformation,
 $\{f_2\<= 0\}\<\in \IP^2[3]$; this obviously includes all other center-enclosing sub-simplices in $\pDN{\FF[2]1}$. All center-enclosing sub-simplices in 
 $\pDN{\FF[2]3}$ then analogously specify multiple mirrors in the $m\<= 3$ case.
In all cases, $|Q|$ (i.e., $|\Tw{Q}|$) equals the degree of the simplex the vertices of which define the Cox coordinates:
\begin{equation}
  \vC{\TikZ{[scale=.75, line join=round, thick]
            \path[use as bounding box](-1.2,-1.5)--(1.2,2);
            \draw[orange, ultra thick, opacity=.75]
                (1,0)--(0,1)--(-1,-1)--cycle;
            \foreach\y in{-1,...,1}
             \foreach\x in{-1,...,1}
              \fill(\x,\y)circle(.6mm);
            \draw(0,1)--(0,0)--(1,0); \draw(0,0)--(-1,-1);
            \filldraw[fill=white](0,0)circle(.6mm); 
            \path(0,-1.5)node{$\deg\Delta_{1,2}\<=3$};
            }}
 \qquad\quad
  \vC{\TikZ{[scale=.75, line join=round, thick]
            \path[use as bounding box](-3.2,-1.5)--(1.2,2);
            \draw[orange, ultra thick, opacity=.75]
                (1,0)--(0,1)--(-3,-1)--cycle;
            \foreach\y in{-1,...,1}
             \foreach\x in{-3,...,1}
              \fill(\x,\y)circle(.6mm);
            \draw(-1,0)--(0,0)--(1,0); \draw(0,1)--(0,0)--(-3,-1)--(-1,0)--cycle;
            \filldraw[fill=white](0,0)circle(.6mm); 
            \path(-1,-1.5)node{$\deg\Delta_{3,4}\<=5$};
            }}
 \qquad\quad
  \vC{\TikZ{[scale=.75, line join=round, thick]
            \path[use as bounding box](-1.2,-1.5)--(1.2,2);
            \draw[magenta, ultra thick, opacity=.75]
                (-1,2)--(-1,-1)--(1,-1)--cycle;
            \foreach\y in{-1,...,2}
             \foreach\x in{-1,...,1}
              \fill(\x,\y)circle(.6mm);
            \draw(-1,2)--(0,0)--(-1,1);
            \draw(-1,0)--(0,0)--(-1,-1);
            \draw(0,-1)--(0,0)--(1,-1);
            \filldraw[fill=white](0,0)circle(.8mm); 
            \path(0,-1.5)node{$\deg\nabla_{\!1,3}\<=6$};
            }}
 \qquad\quad
  \vC{\TikZ{[scale=.75, line join=round, thick]
            \path[use as bounding box](-1.2,-1.5)--(1.2,2);
            \draw[magenta, ultra thick, opacity=.75]
                (-1,2)--(-1,0)--(1,-1)--cycle;
            \foreach\y in{-1,...,2}
             \foreach\x in{-1,...,1}
              \fill(\x,\y)circle(.6mm);
            \draw(-1,2)--(0,0)--(-1,1);
            \draw(-1,0)--(0,0)--(1,-1);
            \filldraw[fill=white](0,0)circle(.8mm); 
            \path(0,-1.5)node{$\deg\nabla_{\!2,4}\<=4$};
            }}
\end{equation}

Such constructions are clearly possible with any polar pair of reflexive polytopes (see also\cite{rA+P-MM}), as well as with their non-convex generalizations among the transpolar pairs of VEX multitopes\cite{Berglund:2022dgb, Berglund:2024zuz}:
\label{sc:nw}
\begin{cons}\label{c:nw}
Every transpolar pair of VEX multitopes,
 $(\pDs{X},\pDN{X})$, encodes a network of multiple mirror pairs given by all possible center-enclosing reductions to lattice simplices
 $(\Delta\<\subset \pDs{X},\nabla\<\subset \pDN{X})$.
\end{cons}
\noindent
The size and variety of the so-defined network of multiple mirror pairs grow combinatorially with the complexity of the original multitope pair. 
\label{sC:3}
\begin{conj}\label{C:3}
Multiple mirrors specified by center-enclosing sub-simplex reductions of a transpolar pair of multitopes, $(\pDs{X},\pDN{X})$, are special cases of the mirror mapping~\eqref{e:gMM} extended to the anticanonical hypersurfaces in the torus manifolds corresponding to the full multifans spanned by $(\pDs{X},\pDN{X})$.
\end{conj}
\noindent{\bf Notes:}
 ({\small\bf5})~The fact that the order of $Q\<\times G$ grows, unbounded, with increasing $m$ indicates that the infinite sequence of Hirzebruch scrolls, $\FF{m}$ and their transpolar pairs as in~\eqref{e:gMM}, harbor an infinite variety of Calabi--Yau mirror hypersurfaces and string compactifications, distinguished by the infinite variety of $(Q,G)$ symmetries.
 ({\small\bf6})~The simplex-reduced models such as~\eqref{e:2F1a}--\eqref{e:2F3b} all involve quasi-homogeneous hypersurfaces in weighted projective spaces, the projectivized top-dimensional orbit of the $Q$-encoded $\mathbb{C}^*\<= U(1;\mathbb{C})$-symmetry action in the $\mathbb{C}^3$-like affine space of the respective Cox coordinates. Its complement in this $\mathbb{C}^3$-like affine space is the origin, which is the only other orbit and also the {\em\/isolated\/} singularity of the affine hypersurface.
 By contrast, in the $\mathbb{C}^4$-like affine space of the respective Cox coordinates in~\eqref{e:2F1mm}--\eqref{e:2F3mm}, both the (toric/gauge) 
 $(\mathbb{C}^*)^2\,{=}\,U(1;\mathbb{C})^2$-symmetry orbit decomposition and the affine hypersurface singularity locus are more complicated, so the techniques employed to date in studying the transposition mirror construction require corresponding generalization.
 ({\small\bf7})~A flip-folded or otherwise multi-layered multifan corresponds to multiple torus manifolds, requiring additional data for disambiguation. Refs.\cite{rBH-gB, Berglund:2022dgb, Berglund:2024zuz} provide substantial computational evidence that a global, continuous orientation data should suffice within the mirror-framework~\eqref{e:gMM}. This however does not seem to match any of the structures on torus manifolds considered so far by mathematicians\cite{rM-MFans, Masuda:2000aa, rHM-MFs, Masuda:2006aa, rHM-EG+MF, rH-EG+MFs2, Nishimura:2006vs, Ishida:2013ab, Davis:1991uz, Ishida:2013aa, buchstaber2014toric, Jang:2023aa}. Finding the precise sub-class of torus manifolds that is usable in string theory then remains an open challenge.

\section{Conclusions and Outlook}
\label{s:Coda}
This article builds on the previous works\cite{rBH-Fm, rBH-gB, Berglund:2022dgb, Berglund:2024zuz} to argue for a general framework~\eqref{e:gMM}--\eqref{e:gMM0} for constructing mirror-pairs of compact (real $2n$-dim.)\ Calabi--Yau spaces,
 $(Z_f,Z_g)$, as hypersurfaces in (real $2(n{+}1)$-dim.)\ embedding spaces $(X,Y)$. The same then framework also suggests a tentative mirror symmetry (yet to be explored) for the pairs of complementary {\em\/non-compact\/} spaces, $(X\<\ssm Z_f,Y\<\ssm Z_g)$, as well as for the pair of non-compact (real $2(n{+}2)$-dim.)\ anticanonical bundle total spaces, $(\cK^*_X,\cK^*_Y)$; 
 see Conjecture~\ref{C:0} in \SS\;\ref{s:org}.
 
 Showcasing this framework with hypersurfaces in Hirzebruch scrolls reveals that an infinite and varied sequence of diverse toric varieties, $\FF{\sss\ora{m}}$, may be realized as (diffeomorphic to) specific hypersurfaces in products of projective spaces. This led to the sweeping conjecture that in fact all (complex algebraic) toric varieties may be realized as (complete intersections of) hypersurfaces in products of projective spaces, and then the Calabi--Yau hypersurface in those as (generalized) CICYs;
 see Conjecture~\ref{C:1} in \SS\;\ref{s:2C}\
 This motivates four subsequent conjectures:
 Conjecture~\ref{C:2}, that so-called (algebraically) ``unsmoothable'' Calabi--Yau hypersurfaces in non-Fano varieties can in fact be smoothed by Laurent deformations, which are themselves continuous limits of regular smoothings found elsewhere in the same explicit deformation family;
 Conjecture~\ref{C:01dS}, that general worldsheet $(0,1)$-supersymmetric GLSMs have stable target spaces that are not supersymmetric, and so may well include de~Sitter spacetime and no supersymmetry, and
 Conjecture~\ref{C:VEX}, that such ground-state spaces may be modeled by suitable hypersurfaces in torus manifolds specified by VEX multitopes.
 
 Section~\ref{s:CSI} then builds to
 Conjecture~\ref{C:pCS}, that generic such ground-state spaces have ``defects'' of positive codimension that obstruct supersymmetry, and the correlated ``compatible triple'' of complex, symplectic and K\"ahler structures of the compactification space --- somewhat akin to the description involving~\eqref{e:S4>P2}, and
 Conjecture~\ref{C:3}, that all transpolar pairs of VEX multitopes, $(\pDs{X},\pDN{X})$, define deformation spaces of transposition mirror-pair Calabi--Yau hypersurfaces in torus manifolds, amongst which the combinatorially diverse network of multiple transposition mirror-pairs constructed by various center-enclosing sub-simplex reductions of $(\pDs{X},\pDN{X})$ are but isolated, special cases.

 Together with the seven {\em\/observations\/} about the mirror-framework~\eqref{e:gMM} enumerated in \SS\:\ref{s:org} and seven {\em\/notes\/} (four in \SS\;\ref{s:org} and three more at end of \SS\;\ref{s:CSI}), these seven conjectures convey the main portend of this article.
 
In closing, one last development worth noting is the recent surge of various machine learning methods and techniques that are being vigorously advanced to compute the Ricci-flat metrics, Yukawa couplings and curvature on Calabi--Yau spaces\cite{He:2018jtw, Constantin:2018xkj, Butbaia:2024tje, Constantin:2024yxh, Berglund:2024uqv, Constantin:2024yaz, Berglund:2024psp, Fraser-Taliente:2024etl}, and most recently even the stringy, $\,\a'$-corrected metrics\cite{Fraser-Taliente:2024etl}. These novel and burgeoning developments are beginning to enable string theory to start delivering feasibly on the enthusiastic but by now already four decades old promise of string theory\cite{rHetStr1}.
 These methods also provide for a profoundly novel and fascinatingly promising way of analyzing the local and even global geometry of these vacuum solutions to Einstein equations, which both verifies and is helped by mirror symmetry\cite{Berglund:2022gvm, Berglund:2024psp}.

\paragraph{Acknowledgments:}
First and foremost, I should like to thank Branko Dragovich for the kind invitation and opportunity to present this work at the {\em 11th Mathematical Physics Meeting,} in Belgrade, Serbia, Sept.\;2--6, 2024.
 I am deeply grateful to Per Berglund for decades-long collaborations including on the topics reviewed here.
 I am also indebted to Mikiya Masuda for his guidance through the realm and literature of {\em\/unitary torus manifolds,}
 to Yong Cui, Amin Gholampour and Weikun Wang for many discussions on these topics,
 as well as Marco Aldi, Jonathan Rosenberg and Hal Schenck for insightful and instructive comments.
 I am grateful to
 the Mathematics Department of the University of Maryland, and 
 the Physics Department of the University of Novi Sad, Serbia,
for recurring hospitality and resources.

\begingroup
\footnotesize
\bibliographystyle{utphys}
\bibliography{RefsHR}
\endgroup

\end{document}